\documentclass[openany,envcountsame,envcountchap,sectref]{svmono}
\usepackage{hyperref}
\usepackage{mathptmx}       % selects Times Roman as basic font
\usepackage{helvet}         % selects Helvetica as sans-serif font
\usepackage{courier}        % selects Courier as typewriter font
\usepackage{type1cm}        % activate if the above 3 fonts are
\usepackage{makeidx}         % allows index generation

\usepackage{amscd,amssymb,amsmath,latexsym,bm}
\usepackage[mathcal,mathscr]{euscript}
\usepackage{lipsum}
\usepackage{amsfonts}
\usepackage{graphicx}
\usepackage{epstopdf}
\usepackage{algorithmic}

\usepackage{hyperref}
\usepackage{xcolor}
\usepackage{enumerate}
\usepackage{ulem}

\usepackage[small,nohug]{diagrams}
\diagramstyle[labelstyle=\scriptstyle]

\newtheorem{assumption}[theorem]{Assumption}
\numberwithin{equation}{chapter}
\numberwithin{figure}{chapter}
\renewcommand{\theenumi}{\roman{enumi}}

\newcommand{\BM}{{\mathbb B}}
\newcommand{\CM}{{\mathbb C}}
\newcommand{\NM}{{\mathbb N}}
\newcommand{\RM}{{\mathbb R}}

\newcommand{\TM}{{\mathbb T}}
\newcommand{\ZM}{{\mathbb Z}}
\newcommand{\PM}{{\mathbb P}}

\newcommand{\Aa}{{\mathcal A}}

\newcommand{\Dd}{{\mathcal D}}
\newcommand{\Ff}{{\mathcal F}}

\newcommand{\Ww}{{\mathcal W}}
\newcommand{\Uu}{{\mathcal U}}

\newcommand{\Ss}{{\mathcal S}}

\newcommand{\Tt}{{\mathcal T}}
\newcommand{\Rr}{{\mathcal R}}

\newcommand{\Cc}{{\mathcal C}}

\newcommand{\Ll}{{\mathcal L}}

\newcommand{\Hh}{{\mathcal H}}

\newcommand{\tr}{{\rm tr}}

\newcommand{\sgn}{{\rm sgn}} 
\newcommand{\sign}{{\rm sign}}

\newcommand{\PI}{{\rm \Pi}} 

\begin{document}

\title{A Computational Non-Commutative Geometry Program for Disordered Topological Insulators}

%\subtitle{From $K$-Theory to Physics}

\author{Emil Prodan}

\maketitle

\frontmatter

\noindent Emil Prodan \\
 Department of Physics \& \\
 Department of Mathematical Sciences \\ 
Yeshiva University\\
New York, NY, 10016, USA \\
E-mail: prodan@yu.edu

\include{Preface}

\setcounter{tocdepth}{2}

\tableofcontents

\mainmatter

\chapter{Disordered Topological Insulators: A Brief Introduction}
\label{Cha-TiIntro}

\abstract{Our main goal for this Chapter is to introduce the periodic table of topological insulators and superconductors (see Table~\ref{Table1}). Since the physical principles behind this table involve the robustness of certain physical properties against disorder, we will take a short detour and introduce first the class of homogeneous materials and then the sub-class of homogeneous disordered crystals. Our brief exposition introduces the main physical characteristics of these materials and puts forward a possible physical definition of the class of homogeneous materials. In parallel, it introduces general concepts from the mathematical program pioneered by Jean Bellissard and his collaborators \cite{1Bel1,1BeHZ,1Bel2}, which will become central to our exposition.}

\section{Homogeneous Materials}\label{Sec-HomMat}

Suppose one is given a stack of mesoscopic samples, all cut out from one big piece of a ``homogeneous" solid material prepared, for example, in a big furnace at an industrial facility. If the samples were to be examined under a powerful microscope, one will discover microscopic defects in the atomic configurations, which are unavoidable regardless of how careful the fabrication process was. The inescapable conclusion is that the samples are not exactly identical and one fundamental question can be immediately raised: Is it possible to characterize the homogeneous material produced in the furnace in any meaningful way, at the macroscopic level? In other words, does it make sense to place in a table the values of, let's say, the experimentally measured transport coefficients? Do these values have a meaning or do they differ from sample to sample? This is not just an academic question but a real practical issue because the factory has no way to measure and characterize the entire batch of solid material. Let us also recall that this type of questions represented a real issue when the quasi-crystals were first characterized by transport measurements. Now, if our character proceeds with the experiments anyway, he/she will witness the remarkable fact that all the samples return the same macroscopic physical parameters no matter how accurate the experiments, provided the size of the samples is large enough (typically a few microns or more).  We can then put forward the following physical definition of a homogenous materials: A material can be called homogeneous if the macroscopic physical parameters, such as transport coefficients, susceptibilities, dielectric constant, refractive index and so on, do not depend on the microscopic details of the macroscopic samples used for the measurements. The class of homogeneous materials includes the perfectly periodic crystals but there are many other distinct sub-classes such as the disordered crystals, quasi-crystals, amorphous solids, liquids and gases.  

\vspace{0.1cm}

The physical properties mentioned above can involve the electronic as well as the atomic degrees of freedom. In these notes, we focus entirely on those physical properties which involve only the electronic degrees of freedom, such as the linear dc-transport coefficients of a solid. Very often in the physics community, the defining property of the homogeneous solids is referred to as the self-averaging property of the intensive thermodynamic coefficients. The shear generality of the concept suggests that this can be framed in a simple and universal mathematics, which applies equally to all (quite distinct) sub-classes of homogeneous materials mentioned above. This has been recognized as early as 1980's by Jean Bellissard \cite{1Bel1} and the whole issue has indeed a simple and elegant resolution. Suppose we are given a sample, which is so large that can be treated in the thermodynamic limit.  Assume that the dynamics of the electrons is determined by the Hamiltonian $H$, assumed for simplicity to be a bounded operator. Physical translations of the sample, relative to the measuring devices, will modify the generator of the dynamics into $T_a H T_a^\ast$, where $T_a$ are the unitary operators implementing the translations on the Hilbert space $\mathcal H$ of the electrons. Let $\BM(\mathcal H)$ be the algebra of bounded linear operators over $\Hh$ and consider the subset defined by all the translates of $H$, namely:
\begin{equation}\label{Eq-Hull}
 \Omega = \big \{ T_a H T_a^\ast \in \BM(\Hh) \, | \, T_a = \mbox{translation} \big \} \subset \BM(\Hh)\; .
\end{equation} 
Then \cite{1Bel1} put forward the following mathematical definition: 

\begin{definition} The condensed matter system described by $H$ is homogeneous if $\Omega$ defined above has a compact closure in the strong topology of $\BM(\Hh)$.
\end{definition}

\begin{remark} {\rm A similar definition was given in \cite{1Bel1} for quasi-periodic systems, where the strong topology is replaced by the norm topology. For disordered crystals, however, the latter topology is not appropriate.
\hfill $\diamond$
}
\end{remark}

\begin{remark} {\rm The formalism can be developed for discrete as well as continuous translations. Since we will be dealing exclusively with the discrete case, we will assume that $a \in \ZM^d$ from now on, where $d$ is the dimension of the physical space.
\hfill $\diamond$
}
\end{remark}

The closure of \eqref{Eq-Hull}, denoted by the same $\Omega$ in the following, is called the hull of $H$. It follows from above that, for a homogeneous condensed matter system, $\Omega$ is a compact set and, since it is a norm-bounded subset of $\BM(\Hh)$, $\Omega$ is also metrizable. Furthermore, there is a natural continuous action of the space translations on $\Omega$, denoted by $\tau$ in the following and actually throughout. This action is topologically transitive because the orbit of $H$ is dense in $\Omega$, by definition. 

\begin{remark} {\rm Kellendonk in \cite{1Kel0} calls a system homogeneous if the hull of $H$ and the hull of $H'$ coincide for any $H'$ from the hull of $H$, without the compactness assumption on the hulls. These two definitions are not necessarily the same, tough they do overlap for many cases of interest. Note that always $\Omega_{H'} \subset \Omega_H$ if $H' \in \Omega_H$, which is a consequence of completness of $\Omega_H$. 
\hfill $\diamond$
}
\end{remark}

Things can be turned around and a homogeneous system can be equivalently defined by a family $\{H_\omega\}_{\omega \in \Omega} \subset \BM(\Hh)$ indexed by $\Omega$ and possessing the following covariant property:
\begin{equation}\label{Eq-FCovariantProp}
T_a H_\omega T_a^\ast = H_{\tau_a \omega} \; , \quad \forall \; a \in \ZM^d \; , \ \omega \in \Omega \; .
\end{equation}
While this move may seem innocent at first sight, it actually has far reaching consequences: 
\begin{enumerate}[\rm (i)]

\item It shows that at the core of a homogeneous material, figuratively speaking, stands a topological dynamical system $(\Omega,\tau,\ZM^d)$. This dynamical system is related to the underlying potential experienced by the electrons, which ultimately is set by the atomic configurations \cite{1BeHZ}. Then \eqref{Eq-Hull} tells us that, in principle, we can ``read" the space of atomic configurations from experiments involving only the electronic degrees of freedom. This can be viewed as an inverse problem. In fact, it defines one of the fundamental directions in materials science.

\item Any such topological dynamical system accepts at least one invariant Baire measure. We recall that $\Omega$ is actually metrizable, which has the important consequence that the Baire and Borel $\sigma$-algebras coincide. The latter is preferred in functional analysis but the former one is the natural one to use when dealing with topological dynamical systems, as nicely explained in \cite{1EFHN}. The invariant measures form a linear space, which coincides with the convex hull of its extreme points. The measures corresponding to these extreme points are ergodic w.r.t. the space translations (see \cite[p. 224]{1Dav} or \cite[Chapters~5-6]{1EFHN}). As such, the topological dynamical system can be always promoted to an ergodic dynamical system, to be denoted as $(\Omega,\tau,\ZM^d,{\rm d} \PM)$ in the following.

\item As long as the homogeneous phase is thermodynamically stable, the Gibbs measure describing the classical motion of the atoms must be translationally invariant, hence it belongs to the convex hull mentioned above. Based on physical grounds \cite{1BeHZ}, the Gibbs measure for the atomic degrees of freedom is expected to be an extreme point of the state space, in which case the natural choice of ${\rm d} \PM$ is precisely the physical Gibbs measure. This in fact has been conjectured by Ruelle \cite[Ch.~6]{1Rue} as early as 1969, and the progress on it is discussed in \cite{1DSS}. This brings us to another fundamental direction in materials science, namely, determining ${\rm d} \PM$ either from experiment or first-principles predictive methods such as Quantum Molecular Dynamics.

\end{enumerate}

Bellissard's program goes a whole lot further. Indeed, one is naturally lead to consider the algebra generated by all the translates of the Hamiltonian. At a first stage, one will like to close this algebra to the enveloping (separable) $C^\ast$-algebra, since the latter comes with a set of fantastic tools provided by its $K$-theory \cite{1Bel1}. The latter, for example, became one of the essential tools in the classification of topological insulators \cite{1Kit,1Thi1,1Thi2,1Bou,1BCR1,1KZ,1Kel1}, for the associated bulk-boundary correspondence program \cite{1KRS,1ProdanSpringer2016ds,1BCR2,1BKR,1MT} and for defining non-trivial topological invariants \cite{1LH,1Lor,1PS2,1Kel2,1BR,1BS,1AAnd}. In the bounded discrete case, this algebra coincides with the algebra of covariant physical observables (w.r.t. the underlining dynamical system):
\begin{equation}
A= \{A_\omega\}_{\omega \in \Omega} \subset \BM(\Hh)\;, \quad T_a A_\omega T_a^\ast = A_{\tau_a \omega} \quad \forall a \in \ZM^d \; ,
\end{equation}
equipped with the $C^\ast$-norm:
\begin{equation}
\| A \| = \sup_{\omega \in \Omega} \| A_\omega\|_{\BM(\Hh)} \; .
\end{equation}
It contains all the physical observables that lead to intensive thermodynamic variables that are independent of the atomic configuration $\omega$. This is a direct consequence of Birkhoff's ergodic theorem \cite{1Bir} (see Proposition~\ref{Prop-TracePerVolume}). For example, the current operator:
\begin{equation}
J_\omega = \I [H_\omega,X] \; , \quad X = \mbox{position operator}\; , 
\end{equation}
belongs to this algebra.
 
\vspace{0.1cm}
 
An important question is how to characterize the algebra of the covariant physical observables? Given the extremely generic scenario at the start of our discussion, the answer is astonishing since it provides a connection with one of the most studied algebras in the field of operator algebras, namely, the crossed product algebras \cite{1Wil}. Indeed, the topological dynamical system $(\Omega, \tau, \ZM^d)$ has a canonically associated dual $C^\ast$-dynamical system $(C(\Omega),\tau,\ZM^d)$ ($C(\Omega)$ = algebra of complex valued continuous function over $\Omega$). One then realizes that \eqref{Eq-FCovariantProp} implements a covariant representation of this dynamical system in $\BM(\Hh)$, and these covariant representations coincide with the representations of the crossed product algebra $C(\Omega) \rtimes \ZM^d$ (to be introduced in Section~\ref{Sec-PhysObs}). The gain here is that, not only these algebras accept a Fourier and a differential calculus, but in many cases their $K$-theories can be solved explicitly. For the case when $\Omega$ is contractible, which is relevant for the disordered crystals introduced next, the reader can find in \cite{1ProdanSpringer2016ds} a computation of the $K$-groups together with an explicit characterization of their generators. Calculations of the $K$-groups for quasi-crystals can be found in {\it e.g.} \cite{1Kel0}.

\section{Homogeneous Disordered Crystals}
\label{Sec-HomCrystals}

The picture of a crystalline material that most of us have in mind is that of a strictly periodic array of atoms. As such, it may appear at first sight that the two words in the title, disorder and crystals, make very little sense when used together like that. But is that true? After all, upon heating, a crystal remains a crystal until the melting occurs, i.e. it displays a sharp X-ray diffraction pattern (XRDP) \cite{1FDS}. Even for a chemically pure crystal, near the melting line, the atoms are, at any moment, quite far from the idealized periodic positions and they will appear disordered in an instantaneous snapshot. One can then put forward the physical definition of a disordered crystal as being a homogeneous material which returns sharp XRDPs consistent with the known crystal space groups (hence excluding the quasi-crystals). 

\vspace{0.1cm}

 One could object to the above definition by arguing that the sharp XRDPs seen experimentally all the way to the melting \cite{1FDS} capture only the time-average of the atomic configuration. In other words, the sharp XRDPs have little to do with the instantaneous disordered configurations and instead the diffraction patterns originate from the time-averaged atomic configuration, which presumably is periodic. One should recall, though, the fundamental principle of statistical mechanics, which asserts that, at thermodynamic equilibrium, the time averages and ensembles averages coincide. Due to the self-averaging feature of the homogeneous materials, it follows that the XRDPs coming from one single representative atomic configuration coincides with the average over the atomic configurations \cite[Theorem 1.12]{1BeHZ}. The striking conclusion is that, even if the XRDP were recorded instantaneously, something that might be possible in the near future, they will be identical with the time-averaged ones, hence sharp. 

\begin{remark} {\rm We brought up the subject of X-ray diffraction only to introduce the concept of disordered crystal and we warn the reader that XRDPs are not at all the focus of the present work. Let us point out that the mathematics of diffraction on aperiodic systems is quite well developed and the reader can read more about it in the monograph \cite{1BG1}. An insightful account of XPRDs in periodic vs disordered structures is given in \cite{1BG2}. Another useful source of information is the collection of works in \cite{1KLS}, which touch on various modern aspects of the aperiodic order.
\hfill $\diamond$
}
\end{remark}

\begin{remark}{\rm The disorder may have various sources besides the thermal fluctuations, such as substitutions with impurity atoms or microscopic lattice defects. All such defects can be treated by the same formalism.
\hfill $\diamond$
}
\end{remark}

Mathematically, a disordered crystal can be characterized by a transitive topological dynamical system $(\Omega,\tau,\ZM^d)$, as introduced in the previous section. What really makes the system into a disordered crystal, from the mathematical point of view, is the discrete action of the $\ZM^d$ group, together with the fact that $\Omega$ is compact, metrizable and contractible. The last condition is new, compared with {\it e.g.} \cite{1Bel1}, but we feel it is necessary because otherwise the $K$-groups of the disordered crystal may differ from that of the perfect crystal.  But this can not happen if the periodic and disordered crystals belong to the same thermodynamic macroscopic phase. In a more direct way, this condition will automatically exclude the quasi-crystals for which $\Omega$ is a totally disconnected space \cite{1Kel0}.

\begin{remark}{\rm A possible refinement of the above characterization of disordered crystals is to consider a transitive dynamical system under the full action of the crystalographic space group. 
\hfill $\diamond$
}
\end{remark}

\section{Classification of Homogenous Disordered Crystals}

 Table~\ref{Table1} represents the broadly accepted classification table of topological insulators and of the fermionic excitations in superconductors \cite{1SRFL,1Kit,1RSFL}. We assume familiarity of the reader with this table, whose explanation will be kept to minimum. The purpose of this section is rather to highlight the aspects related to disorder. 
 
\vspace{0.1cm}
 
 Let us start with the statement that, quite often, it is mistakenly assumed that Table~\ref{Table1} was generated based on a mathematical classification criterium. This cannot be further from the reality.  The value of this table is in that it singles out phases of condensed matter with extremely robust transport properties. Put it differently, the table is about unique materials and devices and not about an abstract mathematical classification problem. Specifically, the table states that the solid state phases listed there:

\begin{enumerate}[1.]

\item  Are distinct macroscopic phases of matter in the true sense of the words. That is, the phases are separated from one another by a phase transition, the Anderson localization-delocalization transition characterized by the divergence of Anderson's localization length. Its experimental signature will be discussed in great detail in Sections~\ref{Sec-IQHE} and \ref{Sec-ChernTI}.

\vspace{0.1cm}

\item Can be labeled uniquely by the values of a strong bulk invariant. The bulk invariants are specific to each class but in general they have physical interpretations as linear and non-linear transport coefficient, or as linear and non-linear magneto-electric response functions.

\vspace{0.1cm}

\item Conduct electricity even in the presence of relatively large disorder once a surface is cut into the crystals and the phase is topologically non-trivial.

\end{enumerate}

 \begin{table}\small
\begin{center}
\begin{tabular}{|c|c|c|c||c||c|c|c|c|c|c|c|c|}
\hline
$j$ & TRS & PHS & CHS & CAZ & $0,8$ & $1$ & $2$ & $3$ & $4$ & $5$ & $6$ & $7$
\\\hline\hline
$0$ & $0$ &$0$&$0$& A  & \textcolor{red}{$\mathbb Z$} &  & \textcolor{red}{$\mathbb Z$} &  & \textcolor{red}{$\mathbb Z$} &  & \textcolor{red}{$\mathbb Z$} &  
\\
$1$& $0$&$0$&$ 1$ & AIII & & \textcolor{red}{$\mathbb Z$} &  & \textcolor{red}{$\mathbb Z$}  &  & \textcolor{red}{$\mathbb Z$} &  & \textcolor{red}{$\mathbb Z$}
\\
\hline\hline
$0$ & $+1$&$0$&$0$ & AI &  \textcolor{red}{$\mathbb Z$} & &  & & \textcolor{red}{$2 \, \mathbb Z$} & & $\mathbb Z_2$ & ${\mathbb Z_2}$
\\
$1$ & $+1$&$+1$&$1$  & BDI & $\mathbb Z_2$ & \textcolor{red}{$\mathbb Z$}  & &  &  & \textcolor{red}{$2 \, \mathbb Z$} & & $\mathbb Z_2$
\\
$2$ & $0$ &$+1$&$0$ & D & $\mathbb Z_2$ & ${\mathbb Z_2}$ & \textcolor{red}{$\mathbb Z$} &  & & & \textcolor{red}{$2\,\mathbb Z$} &
\\
$3$ & $-1$&$+1$&$1$  & DIII &  & $\mathbb Z_2$  &  $\mathbb Z_2$ &  \textcolor{red}{$\mathbb Z$} &  & & & \textcolor{red}{$2\,\mathbb Z$}
\\
$4$ & $-1$&$0$&$0$ & AII & \textcolor{red}{$2 \, \mathbb Z$}  & &  $\mathbb Z_2$ & ${\mathbb Z_2}$ & \textcolor{red}{$\mathbb Z$} & & &
\\
$5$ & $-1$&$-1$&$1$  & CII & & \textcolor{red}{$2 \, \mathbb Z$} &  & $\mathbb Z_2$  & $\mathbb Z_2$ & \textcolor{red}{$\mathbb Z$} & &
\\
$6$ & $0$ &$-1$&$0$ & C&  &  & \textcolor{red}{$2\,\mathbb Z$} &  & $\mathbb Z_2$ & ${\mathbb Z_2}$ & \textcolor{red}{$\mathbb Z$} &
\\
$7$ & $+1$&$-1$&$1$  &  CI &  & &   & \textcolor{red}{$2 \, \mathbb Z$} &  & $\mathbb Z_2$ & $\mathbb Z_2$ & \textcolor{red}{$\mathbb Z$}
\\
[0.1cm]
\hline
\end{tabular}
\end{center}
\caption{{\small Classification table of strong topological insulator and superconductors \cite{1SRFL,1Kit,1RSFL}. Each row represents a universal symmetry class, defined by the presence ($1$ or $\pm 1$) or absence ($0$) of the three symmetries: time-reversal (TRS), particle-hole (PHS) and chiral (CHS), and by how TRS and PHS transformations square to either $+I$ or $-I$. Each universality class is identified by a Cartan-Altland-Zirnbauer (CAZ) label. The strong topological phases are organized by their corresponding symmetry class and space dimension $d=0,\ldots,8$. These phases are in one-to-one relation with the elements of the trivial, $\mathbb Z_2$, $\mathbb Z$ or $2\, \mathbb Z$ groups. The table is further divided into the complex classes A and AIII (top two rows), which are the object of the present study, and the real classes AI, \ldots, CI (the remaining 8 rows). The phases highlighted in red are covered by the present work.}}
\label{Table1}
\end{table}

\begin{remark}{\rm We should point out that the $K$-theories of the physical algebras cannot reproduce Table~\ref{Table1}, as the classifications based on such methods return also the so called weak topological phases. The conjecture put forward by Table~\ref{Table1} is that, in the normal laboratory conditions where disorder is inevitable, these weak phases can be connected with the trivial phases without crossing a phase transition. Only the strong topological phases listed in Table~\ref{Table1} remain distinct in the presence of disorder. For the phases classified by $\ZM$ or $2\ZM$ in Table~\ref{Table1}, a proof of the characteristics 1-3 can be found in the monograph \cite{1ProdanSpringer2016ds}.
\hfill $\diamond$
}
\end{remark}

As one can see, disorder is an integral component of the field. In fact, it is at the very heart of it since the classification table will look very different if the disorder is taken out of the picture. It was precisely this important aspect that has led us to the development of the numerical program reviewed within these pages. The examples and the explicit computations presented in this work are all for the topological phases classified by $\ZM$ or $2\ZM$ in Table~\ref{Table1}. The simple reason is that, despite concentrated efforts \cite{1GS,1Bou,1BCR1,1BKR,1KK}, the $\ZM_2$ bulk and boundary invariants are still lacking expressions which can be evaluated on a computer in the presence of strong disorder. An important development in this direction are the index pairings of \cite{1GS} which are indeed stable in the regime of strong disorder but, unfortunately, local formulas for the proposed $\ZM_2$ topological indices are missing. This is a critical issue because Fredholm operators are difficult to represent with the finite algebras handled by a computer. In contradistinction, a local index formula can be efficiently approximated using finite algerbas, as we shall see later. This is why the existing local index formulas \cite{1BES,1PLB,1PS,1Pro3,1ProdanSpringer2016ds,1BR} for the $\ZM$ topological indices are essential for our computational non-commutative program.

\chapter{Electron Dynamics: Concrete Physical Models}
\label{Cha-ElecDyn}

\abstract{This Chapter introduces and characterizes the so called lattice models, which integrate naturally in the general framework for homogeneous materials introduced in Chapter~\ref{Cha-TiIntro}. It also fixes the notations and the conventions used throughout. A model of disorder is introduced and the various regimes of disorder are discussed. The Chapter concludes with the introduction and characterization of the topological invariants corresponding to the phases classified by $\ZM$ and $2\ZM$ in Table~\ref{Table1}.
}

\section{Notations and Conventions} 

From now on, our exposition deals exclusively with discrete lattice models. These models give accurate representations of the low-energy excitations which are responsible for the physics observed in the typical experiments performed on solid state phases, such as the transport measurements. Given a concrete material, accurate lattice models can be generated from first principles \cite{2LQZDFZ,2ZYZDF,2WDF,2ORV}. One such technique is based on projecting the continuum models on a discrete, periodic and localized partial basis set which accurately represents the spectral subspace in a finite energy window around the Fermi level. Benchmarks of this method can be found in \cite{2MMY}. The models can be also generated empirically by fitting the available experimental data. One relevant example of this type is the lattice model of the HgTe quantum wells \cite{2BHZ,2KBM}, the prototypical topological insulator from class AII in dimension two.

\vspace{0.1cm}

The Hilbert spaces of the lattice models take the generic form $\Hh = \CM^N \otimes \ell^2(\ZM^d)$, where $d$ is the dimension of the physical space and $\CM^N$ is often referred to as the fiber. As before, the algebra of bounded linear operator over $
\Hh$ will be denoted by $\BM(\Hh)$. Let us specify from the beginning that $\ZM^d$ does not represent the position of the atoms but it rather labels the primitive cells. The dimension $d$ needs to be understood accordingly, {\it e.g.} $d=1$ for a molecular chain, $d=2$ for graphene and $d=3$ for an ordinary crystal. This dimension can be higher than three, since one can add virtual dimensions by varying the parameters of the models \cite{2Pro7}. The primitive cells of topological insulators contain many atoms, in many cases 10 or even more, hence they are quite complex. The dimension $N$ of the fiber represents the number of molecular orbitals per primitive cell included in the model. The larger this number the more precise the model is. Due to the complex primitive cell of the topological materials, this number can be quite large in the first principle calculations (within the hundreds!). This is to be contrasted with the simple analytic models used by theoretical physicists, which quite often assume $N=2$.

\vspace{0.1cm}

Throughout, we denote the algebra of $N \times N$ matrices with complex entries by $M_N(\CM)$.  Also, ${\rm tr}$ will denote the normalized trace of matrices, such as those from $M_N(\mathbb C)$, or more general the standard normalized traces over finite dimensional Hilbert spaces. The semi-finite traces over infinite Hilbert spaces, such as $\ell^2(\mathbb Z^d)$ or $\mathbb C^N \otimes \ell^2(\mathbb Z^d)$, will be denoted as usual by ${\rm Tr}$. We will adopt the Dirac notation $|\alpha,x\rangle$ for the standard basis of $\CM^N \otimes \ell^2(\ZM^d)$ and, throughout, $X$ will denote the position operator on $\ell^2(\ZM^d)$, $X|x\rangle = x|x \rangle$. For an operator $T$ on $\mathbb C^N \otimes \ell^2(\mathbb Z^d)$, the notation $\langle x | T |y \rangle$ stands for the $N \times N$ matrix with entries $\langle \alpha,x | T |\beta,y \rangle$, $\alpha,\beta=1,\ldots,N$. Another set of useful operators on $\ell^2(\ZM^d)$ are the shift operators $S_j|x\rangle = |x + e_j \rangle$, where $e_j$, $j=1,\ldots,d$, are the generators of $\ZM^d$. The following notation:
\begin{equation}
S_y = S_1^{y_1} \ldots, S_d^{y_d}\;, \quad S_y |x\rangle = |x+y\rangle \; , \quad y \in \ZM^d\;,
\end{equation}
will be adopted throughout. Lastly, if $V$ is a finite subset of $\ZM^d$, then we can consider the associated finite-dimensional Hilbert space $\CM^N \otimes \ell^2(V)$ and we denote the corresponding normalized trace by $\tr_V$, and the partial isometry from $\CM^N \otimes \ell^2(\ZM^d)$ to $\CM^N \otimes \ell^2(V)$ by $\Pi_V = 1 \otimes \sum_{x \in V} |x\rangle \langle x|$.

\vspace{0.1cm}

Another convention we adopt throughout concerns the notation for the norms. While, the norms on different algebras will be defined explicitly, to simplify the notation, we will use most of the time the same symbol $\| \ \|$ for all of them. There should be no confusion in the calculations, because in all such instances the algebras can be easily identified. For example, the norm appearing in $\|\langle x | T | y \rangle \|$ is the one on $M_N(\CM)$.

\section{Physical Models} 

In the periodic or translational invariant case, the most general Hamiltonian that can be defined on a lattice takes the form:
\begin{align}
\label{Eq-GenericTrModel}
H: \CM^N \otimes \ell^2(\ZM^d)   \rightarrow \CM^N \otimes \ell^2(\ZM^d)
\;,
\qquad
H \, =\,\sum_{y\in \mathbb Z^d} w_y \otimes S_y\;,
\end{align}
where the hopping matrices $w_y \in M_N(\CM)$ satisfy the basic constraint $w_y^* \;=\; w_{-y}$ ensuring that $H$ is self-adjoint. When symmetries are present, the hopping matrices acquire additional structure. In particular, when the chiral symmetry is present and the dimension of the fiber is even, the hopping matrices have the structure:
\begin{equation}
w_{y} = \begin{pmatrix} 0 & g_y \\ g_y^\ast & 0 \end{pmatrix},
\end{equation}
which manifests in the chiral symmetry of the Hamiltonian \cite{2RSFL}:
\begin{equation}
\label{Eq-ChiralSymm}
J \, H \, J^\ast = -H\;, \quad J = \begin{pmatrix} I_N & 0 \\ 0 & -I_N \end{pmatrix}\;.
\end{equation} 
By examining the classification table~\ref{Table1}, we can see that the chiral symmetry is present for all phases classified by $\ZM$ or $2\ZM$ in odd dimensions.

\vspace{0.2cm}

As it is well known, the energy spectrum of the translational invariant Hamiltonians is continuous and, since we want to deal only with insulators, the only way to accommodate that is to assume a gap in the energy spectrum and that the Fermi level $\epsilon_F$ is somehow pinned inside this spectral gap. In the independent electron approximation, assumed throughout, the ground state of the Hamiltonian is encoded in the Fermi projection, that is, the spectral projection onto the energy spectrum below the Fermi level $\epsilon_F$:
\begin{equation}
P_F = \chi(H \leq \epsilon_F) \in \BM(\Hh)\; ,
\end{equation} 
where $\chi$ stands for the indicator function. For all topological phases in even dimensions that are classified by  $\ZM$ or $2\ZM$ in Table~\ref{Table1}, the topology is encoded in $P_F$. When the chiral symmetry is present, the Fermi projection has a particular structure:
\begin{equation}
P_F = \tfrac{1}{2} \begin{pmatrix} 1 & -U_F^\ast \\ -U_F & 1 \end{pmatrix}, \quad U_F^\ast U_F = U_F U_F^\ast =I \; ,
\end{equation}
and the topology is encoded in the unitary operator $U_F$ rather than $P_F$ itself. We refer to $U_F$ as the unitary Fermi operator. The strong topological invariants for the phases classified by $\ZM$ or $2\ZM$ in Table~\ref{Table1} are given by the top even Chern number applied to $P_F$, for even space dimensions, and by the top odd Chern number applied to $U_F$, for odd dimensions \cite{2RSFL}. They will be further elaborated in Section~\ref{Sec-TopInv} and, for now, let us present two examples of topological models, which can be solved explicitly in the bulk as well as with a boundary. The explicit calculations can be found in \cite{2ProdanSpringer2016ds}.

\begin{example}[A topological insulator from class A] {\rm Assume the space dimension  even and let $\{\gamma_j\}_{j=\overline{1,d}}$ be the irreducible representation of the even complex Clifford algebra ${\mathcal Cl}_d$ on $\CM^{2^\frac{d}{2}}$ and $\gamma_0$ be its standard inner grading. Consider the following Hamiltonian on the Hilbert space $\CM^{2^\frac{d}{2}} \otimes \ell^2(\ZM^d)$:
\begin{eqnarray}\label{Eq-Model1}
H \;=\; 
\tfrac{1}{2\I} \sum_{j=1}^d \gamma_j \otimes (S_j-S_j^\ast) 
\;+\; 
\gamma_0 \otimes \Big(m+\tfrac{1}{2}\sum_{j=1}^d  (S_j+S_j^\ast)\Big)
\;,
\end{eqnarray} 
with Fermi level fixed at the origin. Then $H$ has a spectral gap at the Fermi level, except when $m\in \{-d,-d+2,\ldots, d-2,d\}$. If $m \in (-d +2n, -d +2n+2)$, with $n=0, \ldots,d-1$, the model is in a topological insulating phase labeled by its top even Chern number which takes the value $(-1)^n \binom{d-1}{n}$.
\hfill $\diamond$
}
\end{example}

\begin{remark}{\rm If $d=4$, the above model is in fact time-reversal symmetric, hence it can be as well considered from the class AII.
\hfill $\diamond$
}
\end{remark} 

\begin{example}[A topological insulator from class AIII]\label{Ex-ChiralModel} {\rm Assume the space dimension odd and let $\{\gamma_j\}_{j=\overline{1,d+1}}$ be the irreducible representation of the even complex Clifford algebra ${\mathcal Cl}_{d+1}$ on $\CM^{2^\frac{d+1}{2}}$. Consider the following Hamiltonian on the Hilbert space $\CM^{2^\frac{d+1}{2}} \otimes \ell^2(\ZM^d)$:
\begin{eqnarray}\label{Eq-Model2}
H \;=\; 
\tfrac{1}{2\I} \sum_{j=1}^d \gamma_j \otimes (S_j-S_j^\ast) 
\;+\; 
\gamma_{d+1} \otimes \Big(m+\tfrac{1}{2}\sum_{j=1}^d  (S_j+S_j^\ast)\Big)
\;,
\end{eqnarray} 
with Fermi level fixed at the origin. Then $H$ has the chiral symmetry $\gamma_0 H \gamma_0^\ast = -H$ and the model displays a spectral gap at the Fermi level, except when $m\in \{-d,-d+2,\ldots, d-2,d\}$. If $m \in (-d +2n, -d +2n+2)$, with $n=0,\ldots,d-1$, the model is in a topological phase labeled by its top odd Chern number which takes the value $(-1)^n \binom{d-1}{n}$.
\hfill $\diamond$
}
\end{example}

The presence of a uniform magnetic field is incorporated in the lattice models by using the Peierls substitution \cite{2Pei}, which amounts to replacing the ordinary shift operators with the dual magnetic translations:
\begin{equation}
\label{DualMagTransSymmetric}
S_y  \;\;\mapsto \;\; U_y \;=\; e^{\I \, y \wedge X}S_y \;=\; S_y \, e^{\I \, y \wedge X}
\;.
\end{equation}
Throughout:
\begin{equation}
x \wedge x' = \tfrac{1}{2} \sum_{i,j=1}^d \phi_{ij} x_i x_j' \; , \quad \forall \ x, x' \in \RM^d \; ,
\end{equation}
where $\phi$ is a real anti-symmetric $d\times d$ matrix with entries from $[0,2 \pi)$, encoding the magnetic fluxes through the facets of the primitive cell, in units of flux quantum $\phi_0 = h/e$. When the dependence on the magnetic flux needs to be emphasized we will use $\wedge_\phi$. With this substitution, the lattice Hamiltonians become:
\begin{equation}
\label{Eq-GenericBModel}
H
\;=\;
\sum_{y\in \ZM^d}  w_y \otimes U_y 
\;=\;
\sum_{x,y \in \mathbb Z^d} e^{\I \, y \wedge x } \, w_y \otimes |x\rangle \langle x -y| 
\;.
\end{equation}
They are no longer invariant w.r.t. the ordinary shifts but they are invariant w.r.t. the magnetic translations:
\begin{equation}
\label{MagTransSymmetric}
V_y \,H \, V_y^\ast \, = \,H \; , 
\qquad  V_y \,= \, e^{-\I \, y \wedge X}S_y \,= \, S_y e^{-\I \, y \wedge X}
\;.
\end{equation}
It is important to note that the unitary classes of topological insulators, namely the classes A and AIII, can incorporate magnetic fields because time-reversal symmetry is not required.

\vspace{0.1cm}

The next step incorporates the disorder. As we have seen in the previous chapter, a homogeneous disorder is described by a measure-preserving ergodic dynamical system $(\Omega,\tau,\ZM^d,{\rm d} \PM)$. Below is an important example.

\begin{example}[Disorder induced by thermal fluctuations]\label{Ex-PhysDis}{\rm Assume we are dealing with a chemically pure crystal with the atoms arranged in a perfectly periodic lattice at zero temperature. At finite temperature, though, the atoms wander around the equilibrium positions and the crystal is in a disordered state. If $R_x^\alpha$ is the displacement from equilibrium position of atom $\alpha$ in the primitive cell $x$ at an instantaneous time, then the disorder configuration at that time is encoded in $\omega = \{R_x^\alpha\}_{x \in \ZM^d}^{\alpha = \overline{1,N_a}}$, where $N_a$ is the number of atoms in the repeating cell. One assumption is that each atom wanders only in a finite neighborhood $\Omega_x^\alpha$ of its equilibrium position. Then homogeneity requires that $\Omega_x^\alpha = \Omega_0^\alpha$, and the space of disorder configurations and the $\ZM^d$-action are given by:
\begin{equation}
\Omega = \prod_{x \in \ZM^d} \prod_{\alpha=1}^{N_a} \; \Omega_0^\alpha = \prod_{x \in \ZM^d} \; \Omega_0 \; , \quad \tau_y \omega = \tau_y\{R_x^\alpha\} = \{R_{x-y}^\alpha\}\; . 
\end{equation}
Note that $\Omega$ is a Tychononov space, hence compact and metrizable, and that $\tau$ is continuous and invertible hence $(\Omega, \tau, \ZM^d)$ is indeed a topological classical dynamical system. The Gibbs measure governing the classical statistical mechanics of the atoms is defined as the thermodynamic limit of the finite volume probability measures:
\begin{equation}\label{Eq-GibbsMeasure}
{\rm d}\PM_V (\omega) = {\mathcal Z}_V^{-1} e^{-\beta {\mathcal V}_V(\omega)}\prod_{x \in V \subset \ZM^d}\prod_{\alpha = 1}^{N_a} {\rm d} \omega_x^\alpha \; ,
\end{equation}
where ${\mathcal V}_V$ is a classical potential for the atomic degrees of freedom and $\beta = \frac{1}{kT}$, where $k$ is Boltzmann's constant and $T$ is the temperature. When the electronic and atomic degrees of freedom decouple as in the Born-Oppenheimer regime, the potential $\mathcal V_V$ can be mapped explicitly from the ground state of the electrons.
\hfill $\diamond$
}
\end{example}

As we have already seen, central to the theory of homogeneous aperiodic solids is the notion of covariance of physical observables, which in the presence of a uniform magnetic field takes the form:
\begin{equation}
\label{Eq-CovariantProp}
V_y F_{\omega} V_y^\ast = F_{\tau_y \omega} \;.
\end{equation}
We will always assume that the matrix elements $\langle x | F_\omega |y \rangle $ are measurable functions of $\omega$, which is indeed the case for the applications considered here.

\begin{proposition}\label{Prop-TracePerVolume} Let $F_\omega, F'_\omega, \ldots$ be covariant operators. Then $\PM$-almost surely:
\begin{equation}
\label{Eq-TrPerVolume}
\lim_{|V| \rightarrow \infty} \tr_V \big ( \Pi_V F_\omega F'_\omega \ldots \Pi_V^\ast \big ) = \int_\Omega {\rm d} \PM(\omega) \, \tr \big ( \langle 0 |F_\omega F'_\omega \ldots | 0 \rangle \big ) \; ,
\end{equation}
where $V$ is a cube from $\ZM^d$ and $|V|$ is its cardinality. The functional on the righthand side defines the trace per volume, which will be denoted by $\Tt( \ )$ throughout.
\end{proposition}

\proof We have:
\begin{equation}
\tr_V \big ( \Pi_V (F_\omega F'_\omega \ldots )\Pi_V \big ) = \frac{1}{|V|}\sum_{x \in V} \tr \big ( \langle x | F_\omega F'_\omega \ldots | x \rangle \big ) \; ,
\end{equation}
and the last term can be written equivalently as:
\begin{equation}
\frac{1}{|V|}\sum_{x \in V} \tr \big ( \langle 0 | V_x^\ast (F_\omega F'_\omega \ldots )V_x | 0 \rangle \big ) =  \frac{1}{|V|}\sum_{x \in V} \tr \big ( \langle 0 | F_{\tau_{-x}\omega} F'_{\tau_{-x}\omega} \ldots | 0 \rangle \big ) \; .
\end{equation}
Then we are in the conditions of Birkhoff ergodic theorem \cite{2Bir} and the statement follows.
\qed

\begin{remark}{\rm One important implication of the above statement is that the intensive thermodynamic coefficients connected to the covariant physical observables do not fluctuate from one disorder configuration to another or, in other words, it has the self-averaging property in the thermodynamic limit.
\hfill $\diamond$
}
\end{remark}

When disorder is present, all the coefficients of the Hamiltonians develop a random component. The most general form of a covariant Hamiltonian on a lattice takes the form:
\begin{equation}
\label{Eq-GenericDisModel}
H_\omega = \sum_{x,y \in \mathbb Z^d} \,w_y(\tau_x \omega) \otimes |x \rangle \langle x|U_y 
= \sum_{x,y \in \mathbb Z^d} \,e^{\I \, y \wedge x } \;w_y(\tau_x \omega) \otimes |x \rangle \langle x -y |
\; . 
\end{equation}
We will assume that the hopping matrices $w_y$ are continuous functions over $\Omega$ with values in $M_N(\mathbb C)$. We will also assume a certain rapid decay of $w_y$ with $y$, to be precisely quantified in Section~\ref{Sec-Assumptions}. These assumptions are quite natural and fulfilled by most model Hamiltonians.

\begin{example}[``Linearized" disordered models]\label{Ex-LinDM} {\rm  
Most of the numerical simulations found in the physics literature consider linearizations of the coefficients: 
\begin{equation}
w_y(\omega) = w_y + \sum_\alpha \omega_0^\alpha \lambda_y^\alpha,
\end{equation}
where $w_y$ and $\lambda_y^\alpha$ are non-stochastic $N\times N$ matrices. Furthermore, $\omega_x^\alpha$ are drawn randomly and independently from the interval $[-\frac{1}{2},\frac{1}{2}]$. Hence, the disorder configuration space is just the Hilbert cube whose topology is discussed in great detail in {\it e.g.} \cite[Ch.~5]{2IM}. Inserting these expressions in \eqref{Eq-GenericDisModel} leads to:
\begin{equation}
\label{Eq-LinDisHam}
H_\omega = \sum_{x,y \in \mathbb Z^d} \, \big ( w_y+ \sum_\alpha \omega_x^\alpha \lambda_y^\alpha \big) \otimes |x \rangle \langle x|U_y 
\; . 
\end{equation}
In the physicists' jargon, the random fluctuations in the coefficients $w_{y \neq 0}$ are often referred to as bond disorder, while to those in $w_{y=0}$ as on-site disorder.  
\hfill $\diamond$
}
\end{example}

\section{Disorder Regimes} 
\label{Sec-DisRegimes}

Consider the following simplified Hamiltonian on the Hilbert space $\ell^2(\ZM^d)$: 
\begin{equation}\label{Eq-ExRandHam}
H_\omega = \lambda_0 \, H_0 + \lambda_1 \sum_{x \in \ZM^d} \omega_x |x \rangle \langle x| \; ,
\end{equation} 
where $H_0$ is a non-stochastic periodic Hamiltonian and $\lambda$'s are just real numbers. There are two extreme limits where the spectral properties of $H_\omega$ can be readily understood. One is $\lambda_1=0$ when Bloch-Floquet theory can be applied, in which case the spectrum is absolutely continuous and can be graphed as energy bands versus quasi-momentum. The other extreme is $\lambda_0=0$, when the Hamiltonian is already diagonal and the spectrum is pure-point, with eigenvalues $\{ \lambda_1 \omega_x \}_{x \in \ZM^d}$ and corresponding eigenvectors localized on the lattice's sites $\{ |x\rangle\}_{x \in \ZM^d}$. Note that the set of eigenvalues is a countable subset of the real axis, hence of zero Lebesque measure, yet the set is $\PM$-almost sure dense in its interval. These are the trades of the Anderson localized spectrum.

\begin{definition}{\rm The part of the {\it essential} spectrum which is pure-point and has corresponding eigenvectors that are exponentially localized in space is called the Anderson localized spectrum. The complement part of the essential spectrum will be called the Anderson delocalized spectrum.
}
\end{definition} 

In the intermediate regimes where both randomly fluctuating and non-fluctuating components are present in a Hamiltonian, both Anderson localized and delocalized spectra are expected to be present. If the space dimension is low, specifically for $d=1$ and 2, Anderson and collaborators \cite{2AALR} found that sometimes even small random fluctuations can localize the entire energy spectrum. In fact, this was believed to be the rule rather than the exception because, until the discovery of the topological insulators, there were only a handful of known exceptions. The situation is somewhat different in higher dimensions, where typically only the regions around the spectral edges are localized while the spectral regions deep inside the bands remain delocalized, though, the spectrum is expected to become entirely localized for extremely strong disorder. Below we briefly discuss the regimes of weak and strong disorder. For an in-depth treatment of the subject, the reader can consult the monograph \cite{2AW}.

\begin{remark}{\rm Perhaps now it becomes more clear how extraordinary are the claims contained in the classification table. If the topology is nontrivial, then, even in lower dimensions, regions of extended spectrum can exist in the presence of large random fluctuations and the entire boundary spectrum can avoid the Anderson localization phenomenon.
\hfill $\diamond$
}
\end{remark}  

 Let us now assume that, in the absence of disorder, we are dealing with an insulator, hence the Hamiltonian has a spectral gap in its spectrum and the Fermi level is pinned in the middle of this gap. Note that the spectrum of covariant Hamiltonians is non-random and, quite generally, the edges of the energy spectrum depend continuously on the amplitudes of the random fluctuations. Hence, if the random fluctuations are small, such as when $\lambda_1 \ll \lambda_0$ in the example \eqref{Eq-ExRandHam}, the spectral gap persists and the Fermi level continues to be located in a region free of spectrum. One refers to this regime as the regime of weak disorder. In this regime, the matrix elements $\langle x | P_F(\omega)|y \rangle $ of the Fermi projection are rapidly decaying with the separation $|x-y|$, uniformly in $\omega$. If the Hamiltonian is of finite-range, {\it i.e.} $w_y =0$ outside a finite neighborhood of the origin, or if $w_y$ decays with $y$ exponentially fast, then:
\begin{equation}
\sup_{\omega \in \Omega} \big \| \langle x | P_F(\omega)|y \rangle \big \| \leq A e^{-\beta |x-y|} \; , \quad 0<A,\beta < \infty \; .
\end{equation}
The inequality can be established by a straightforward application of the Combes-Thomas technique \cite{2CT}. The ordinary perturbation theory is available in this regime and can be used to investigate the behavior of the Fermi projection under various deformations of the models. In particular, to establish that $P_F(\omega)$ varies continuously in the topology of $\BM(\Hh)$ induced by the operator norm. 

\vspace{0.1cm}

When the amplitudes of the random fluctuations become large, the spectral gap closes but a region of Anderson localized spectrum may still survive. This region is typically sandwitched between two regions of Anderson delocalized spectrum and in such situations one speaks of a mobility gap. One refers to this regime as the regime of strong disorder. Note that the Fermi level is now located in the essential spectrum but the quantum diffusion is absent. In this regime the ordinary perturbation theory does not apply. Also, the location of the eigenvalues in the Anderson localized spectrum fluctuate from one disorder configuration to another, and when a model is deformed continuously, these eigenvalues shift and inherently cross the Fermi level. These crossings can come simultaneously from below and above, hence they cannot be avoided, {\it e.g.} by shifting the Fermi level. As a result, the Fermi projection does not vary continuously under the deformations of the models, in neither of the natural topologies of $\BM(\Hh)$. 

\begin{remark}{\rm The above phenomenon happens also when one tries to compute the bulk invariants using twisted boundary conditions \cite{2LPp}. As a result, the numerical values of the invariants, as computed by such methods, fluctuate from one disorder configuration to another and the quantization can be seen only after the disorder average is taken. As we shall see, this is not the case for the computational methods reviewed here, where the self-averaging property manifests explicitly.
\hfill $\diamond$
}
\end{remark}

A convenient and effective way to describe the Anderson localized spectrum is through the Aizenman-Molchanov bound on the resolvent \cite{2AM}:
\begin{equation}
\label{Eq-AizenmannMolchanov}
\int_\Omega {\rm d} \PM(\omega) \; \big \| \langle x | (\epsilon + \I \delta - H_\omega )^{-1}|y\rangle \big \|^s 
\;\leq \;
A_s \,e^{-\beta_s |x-y|} \; , \quad s \in (0,1) \; ,
\end{equation}
assumed to hold uniformly in $\delta \in (0,\infty)$. Above, $A_s$ and $\beta_s$ are strictly positive and finite parameters. The Aizenman-Molchanov bound automatically implies all characteristics of the Anderson localization \cite{2Aiz,2AENSS,2AW}, such as absence of diffusion, pure-point nature of the spectrum, exponential decay of the eigenvectors at infinity and exponential decay of the averaged matrix elements of the Fermi projection:
\begin{equation}\label{Eq-FermiLoc}
\int_\Omega {\rm d} \PM(\omega) \, \big \| \langle x | P_F(\omega)|y \rangle \big \| \leq A e^{-\gamma |x-y|} \; , \quad 0<A,\gamma < \infty \;.
\end{equation} 
Note that without the average, the inequality \eqref{Eq-FermiLoc} fails for $\omega$ in a set of strictly positive measure, no matter how large we pick the constant $A$. The bound \eqref{Eq-AizenmannMolchanov} is known to hold for finite-range Hamiltonians in all situations where Anderson localization was established with mathematical rigor \cite{2AW}.

\section{Topological Invariants}
\label{Sec-TopInv} 

The theory of the topological bulk invariants for the phases classified by $\ZM$ or $2\ZM$ in Table~\ref{Table1} was developed in \cite{2BES,2PLB,2PS} for the regime of strong disorder. The recent work \cite{2BR} by Bourne and Rennie extended these result to continuum models. Below we briefly state the main results for the discrete case, which are covered extensively in \cite{2ProdanSpringer2016ds}. A brisk review of the methods involved by these works can be found in \cite{2Pro3}.

\vspace{0.1cm}

 For periodic models, the Bloch-Floquet transformation $\Ff$ over the $d$-dimensional Brillouin torus $\mathbb T^d$ gives:
\begin{equation}
\label{eq-BlochFloquet}
\Ff H\Ff^*\;=\;
\int^\oplus_{\TM^d} {\rm d} k \; H_k \; , \quad \Ff P_F\Ff^*\;=\;
\int^\oplus_{\TM^d} {\rm d} k \; P_F(k) \; .
\end{equation}
The analysis is then reduced to that of smooth families of  $N \times N$ matrices:
$$
H_k : \mathbb C^N \rightarrow \mathbb C^N\;, 
\qquad 
H_k \,=\, \sum_{y\in \ZM} e^{\I \, y \cdot k} w_y
\;, \quad P_F(k) = \chi(H_k \leq \epsilon_F) \; .
$$
Throughout, $x \cdot x'$ will denote the Euclidean scalar product. In this simplified setting, the top even Chern number takes the familiar form \cite{2ASSS} ($d=$ even):
\begin{equation}\label{Eq-EvenChernK}
\mathrm{Ch}_{d}(P_F)
\;=\;
\tfrac{(2 \pi \I)^\frac{d}{2}}{ (\frac{d}{2})!} N \sum_{\rho \in \Ss_d} (-1)^\rho  \int_{\mathbb{T}^d} \frac{{\rm d}k}{(2\pi)^d} \ \tr \Big ( P_F(k) \prod_{j=1}^d \frac{\partial P_F (k)}{\partial k_{\rho_j}} \Big )
\;.
\end{equation}
Similarly, the top odd Chern number takes the familiar form \cite{2RSFL} ($d=$ odd):
\begin{equation}
\label{Eq-OddChernK}
{\mathrm{Ch}}_d  (U_F)
\;=\; 
\tfrac{\I (\I \pi)^\frac{d-1}{2}}{ d!!} N \sum_{\rho \in \Ss_d} (-1)^{\rho}\int_{\mathbb{T}^d}\frac{{\rm d}k}{(2 \pi)^d} \  
\tr \Big ( \prod_{j=1}^d U^\ast_F(k) \frac{\partial U_F(k)}{\partial k_{\rho_j}}  \Big )
\;.
\end{equation}
Throughout, $\Ss_I$ will denote the group of permutations group of the finite set $I$. In particular $\Ss_d$ denotes the permutation group of $\{1,\ldots,d\}$. 

\vspace{0.1cm}

For the generic models \eqref{Eq-GenericDisModel}, the top even Chern number can be formulated \cite{2BES,2PLB} using a real-space representation of the operators and the trace per volume $\Tt( \ ) = \lim_{V \rightarrow \infty} \tr_V ( \ )$ from Proposition~\ref{Prop-TracePerVolume}:
\begin{equation}
\label{Eq-EvenChernNR}
\mathrm{Ch}_{d}(P_F)
\;=\;
\tfrac{(2 \pi \I)^\frac{d}{2}}{(\frac{d}{2})!} N
\;
\sum_{\rho \in \Ss_d}(-1)^\rho\; \Tt \Big ( P_F(\omega) \prod_{j=1}^{d} \big (\I[P_F(\omega),X_{\rho_j}]\big ) \Big )
\;.
\end{equation}
When the disorder and the magnetic fields are turned off, this expression become identically with \eqref{Eq-EvenChernK}. In the general case, the quantization of \eqref{Eq-EvenChernNR} follows from: 

\begin{theorem}[\cite{2BES,2PLB}]
\label{Th-EvenChernIndex}
Let $d$ be even and let $\{P_\omega\}_{\omega \in \Omega}$ be a family of covariant projections over $\CM^N \otimes \ell^2(\ZM^d)$ such that:
\begin{equation}
\sum_{x \in \ZM^d} (1+|x|)^{d+1}\int_\Omega {\rm d} \mathbb P(\omega) \ \tr \Big ( \big | \langle 0 \big | P_\omega |x \rangle \big |^{d+1} \Big ) <  \infty \; .
\end{equation}
Consider the family of operators on $\CM^{2^\frac{d}{2}} \otimes \CM^N \otimes \ell^2(\ZM^d)$:
\begin{equation}\label{Eq-FredOp} 
F_{\omega,x_0} = I \otimes P_\omega \left (\frac{\Gamma \cdot (X+x_0)}{|X+x_0|} \right ) I \otimes P_\omega = \begin{pmatrix} 0 & G^\ast_{\omega,x_0} \\ G_{\omega,x_0} & 0 \end{pmatrix}, \quad x_0 \in (0,1)^d \; ,
\end{equation}
where $\Gamma=(\Gamma_1, \ldots,\Gamma_d)$ is an irreducible representation of the even $d$-dimensional complex Clifford algebra ${\mathcal Cl}_d$ and $\Gamma \cdot (X+x_0)$ is a shorthand for $\sum_{j=1}^d \Gamma_j \otimes I \otimes (X+x_0)_j$. Also, the second equality in \eqref{Eq-FredOp} gives the decomposition of $F_{\omega,x_0}$ w.r.t. to the natural inner grading of ${\mathcal Cl}_d$. Then $G_{\omega,x_0}$ is $\PM$-almost surely a Fredholm operator on the range of $I \otimes P_\omega$. Its almost sure Fredholm index is independent of $x_0$ and $\PM$-almost surely independent of $\omega \in \Omega$, and is given by the formula:
\begin{equation}
\label{Eq-EvenChernIndex}
{\rm Index} \, G_{\omega,x_0} = \tfrac{(2\I \pi)^{\frac{d}{2}}}{(\frac{d}{2})!} N\sum_{\rho \in S_d} (-1)^\rho \Tt \Big ( P_\omega \prod_{j=1}^d \I \big [P_\omega,X_{\rho_j} \big] \Big ) \; .
\end{equation}
\end{theorem}

\begin{remark}\label{Re-PhysEvenChern} {\rm In the regime of strong disorder when the Fermi level lies in the essential spectrum, the physical interpretation of the top even Chern number in dimension $d=2$ has been elucidated in \cite{2BES}. There, the reader can find a derivation of the finite-temperature Kubo-formula for the conductivity tensor in the presence of disorder and dissipation, together with a proof of the following limit:
\begin{equation}\label{Eq-ChernLimit}
\lim_{T \rightarrow 0} \sigma_{12}(T, \epsilon_F) = {\rm Ch}_2(P_F) \; ,
\end{equation}
provided the conditions of the Theorem~\ref{Th-EvenChernIndex} are satisfied. Here, $\sigma_{12}$ denotes the off-diagonal or the Hall component of the linear conductivity tensor $\{\sigma_{ij}\}_{i,j=\overline{1,d}}$. The latter provides the linear link between the current-density $J$ set in motion by an externally applied electric field $E$:
\begin{equation}
J_i = \sum_{j =1}^d \sigma_{ij} E_j \; .
\end{equation}
In higher dimensions, it was shown in \cite{2ProdanSpringer2016ds} that, up to an un-interesting constant:
\begin{equation}\label{Eq-EvenChernPhys}
{\rm Ch}_d(P_F) = \frac{\partial^{\frac{d}{2}-1} \sigma_{i_{d-1} i_d}}{\partial \phi_{i_1 i_2} \ldots \partial \phi_{i_{d-3} i_{d-2}} } \;, 
\end{equation}
where the set of indices at the righthand side is such that $\{i_1, \ldots i_d\} = \{1,\ldots,d\}$. According to this result, the top even Chern numbers can be interpreted as non-linear magneto-electric transport coefficients.
\hfill $\diamond$
}
\end{remark}  

Similarly, in odd dimensions, the top odd Chern number can be formulated for generic disordered models with chiral symmetry as \cite{2MSHP}:
\begin{equation}
\label{Eq-OddChernNR}
{\mathrm{Ch}}_{d}(U_F) 
\;=\; 
\tfrac{\I (\I \pi)^\frac{d-1}{2}}{ d!!} N \sum_{\rho \in \mathcal S_d} (-1)^\rho \;\mathcal{T}\Big ( \prod_{j=1}^{d}   U_F^\ast(\omega) \, \I [U_F(\omega),X_{\rho_j}] \Big )
\;.
\end{equation}
When the disorder and the magnetic fields are turned off, this expression become identically with \eqref{Eq-OddChernK}. In the general case, the quantization of \eqref{Eq-OddChernNR} follows from: 

\begin{theorem}[\cite{2PS}]
\label{Th-OddChernIndex}
Let $d$ be odd and let $\{U_\omega\}_{\omega \in \Omega}$ be a family of covariant unitary operators over $\CM^N \otimes \ell^2(\ZM^d)$ such that:
\begin{equation}
\sum_{x \in \ZM^d} (1+|x|)^{d+1}\int_\Omega {\rm d} \PM (\omega) \ \tr \Big ( \big | \langle 0 \big | U_\omega |x \rangle \big |^{d+1} \Big ) <  \infty \; .
\end{equation}
Consider the family of operators on $\CM^{2^\frac{d-1}{2}} \otimes \CM^N \otimes \ell^2(\ZM^d)$:
\begin{equation} 
F_{\omega,x_0} = \frac{1}{4}\left (I+\frac{\Sigma \cdot (X+x_0)}{|X+x_0|} \right ) (I\otimes U_\omega) \left (I+\frac{\Sigma \cdot (X+x_0)}{|X+x_0|} \right )\; , \quad x_0 \in (0,1)^d \; ,
\end{equation}
where $\Sigma=(\Sigma_1, \ldots,\Sigma_d)$ is an irreducible representation of the odd $d$-dimensional complex Clifford algebra $\mathcal Cl_d$ and $\Sigma \cdot (X+x_0)$ is a shorthand for $\sum_{j=1}^d \Sigma_j \otimes I \otimes (X+x_0)_j$. Then $F_{\omega,x_0}$ is $\PM$-almost surely a Fredholm operator. Its almost sure Fredholm index is independent of $x_0$ and $\PM$-almost surely independent of $\omega \in \Omega$, and is given by the formula:
\begin{equation}
\label{Eq-OddChernIndex}
{\rm Index} \, F_{\omega,x_0} = \tfrac{\I (\I \pi)^\frac{d-1}{2}}{ d!!} N \sum_\rho (-1)^\rho \;\mathcal{T}\Big ( \prod_{j=1}^{d}   U_\omega^\ast \, \I [U_\omega,X_{\rho_j}] \Big ) \; .
\end{equation}
\end{theorem}

\begin{remark}\label{Re-PhysOddChern}{\rm The physical interpretation of the top odd Chern numbers has been elucidated in \cite{2ProdanSpringer2016ds}. If $P^c$ denotes the vector of chiral polarization, that is, the difference between the electric polarizations corresponding to the two chiral sectors, then, for $d$ odd:
\begin{equation}\label{Eq-OddChernPhys}
{\rm Ch}_d(P_F) = \frac{\partial^\frac{d-1}{2} P^c_{i_d}}{\partial \phi_{i_1 i_2} \ldots \partial \phi_{i_{d-2} i_{d-1}} } \;, 
\end{equation}
where the set of indices at the righthand side is such that $\{i_1, \ldots i_d\} = \{1,\ldots,d\}$. According to this result, the top odd Chern numbers can be interpreted as non-linear magneto-electric response coefficients.
\hfill $\diamond$
}
\end{remark} 

\begin{remark}{\rm Note that the Dirac operators and the appended Hilbert space in the index theorems for both even and odd Chern numbers are just auxiliary mathematical constructs and have no direct connection with the physics. 
\hfill $\diamond$
}
\end{remark} 

As we already mentioned, in the regime of strong disorder, the Fermi projection, hence also the Fermi unitary operator when chiral symmetry is present,  do not vary continuously in norm under the deformations of the models. As such, the invariance of the Fredholm index w.r.t. norm-deformations cannot be used in this context. However, under the Aizenman-Molchanov condition, the righthand side of \eqref{Eq-EvenChernIndex} (or of \eqref{Eq-OddChernIndex}) varies continuously with the deformations of the models \cite{2PLB,2PS}. As such, using both sides of \eqref{Eq-EvenChernIndex} and \eqref{Eq-OddChernIndex}, one can indeed establish the quantization and invariance of the bulk topological numbers under the Anderson localization assumption. Given all the above, we can conclude at once that the only way to change the bulk topological invariants is by violating the Aizenman-Molchanov condition, that is, through an Anderson localization-delocalization transition.

\vspace{0.1cm}

The derivations of Theorems~\ref{Th-EvenChernIndex} and \ref{Th-OddChernIndex} together with the context in which they naturally occur, which is Alain Connes' non-commutative geometry \cite{2Con}, is the subject of the monograph \cite{2ProdanSpringer2016ds}. In the present work, we are primarily concerned with how to compute the topological invariants once we are given a concrete disordered model. It turns out that the algebraic framework introduced by Jean Bellissard \cite{2Bel1}, and used in \cite{2ProdanSpringer2016ds} for quite different purposes, provides just the right framework for the computer assisted calculations as well as for the analysis of the error bounds.

\chapter{Non-Commutative Brillouin Torus}
\label{Cha-NCBrillouin}

\abstract{In this Chapter, we review the fundamental theoretical tools, starting with the space of disordered configurations and its associated dynamical systems, the $C^\ast$-algebra $\Aa_d$ of the physical observables, together with its Fourier and differential calculus. The latter is provided by a set of commuting derivations $\partial$ and a trace $\Tt$. The triple $(\Aa_d,\partial,\Tt)$ defines a non-commutative manifold known as the non-commutative Brillouin torus. We reformulate the topological invariants and other response functions in this new framework. We also introduce the magnetic derivations and investigate the behavior of the correlation functions w.r.t. the magnetic fields. This Chapter also fixes the notation and defines the precise settings for the rest of our calculations.}

\section{Disorder Configurations and Associated Dynamical Systems}\label{Sec-PhysObs}

The space of disorder configurations was briefly introduced in Chapter~\ref{Cha-ElecDyn}. Here we formulate it under very precise assumptions. 

\begin{assumption}\label{Def-DisorderSpace} {\rm Throughout, the disorder configurations are characterized by a classical topological dynamical systems $(\Omega,\tau, \ZM^d)$, where:
\begin{enumerate}[$\ \circ$]

\item The configuration space is assumed to be a Tychonov space: 
\begin{equation}
\Omega = \prod_{x \in \ZM^d} \Omega_0 \; ,
\end{equation}
where $\Omega_0$ is a compact metrizable topological space.  

\item The action of $\ZM^d$ is given by the homeomorphisms: 
\begin{equation}
\ZM^d \ni y \; \mapsto \; \tau_y \omega = \tau_y\{\omega_x\}_{x \in \ZM^d} = \{\omega_{x-y} \}_{x \in \ZM^d} \; , \quad \forall \, \omega \in \Omega \; .
\end{equation}

\end{enumerate}
}
\end{assumption}

\begin{remark} {\rm We recall that $\Omega$ is equipped with the product topology, defined to be the coarsest topology in which the projection maps $\omega \mapsto \omega_x$ are continuous, for all $x \in \ZM^d$. The standard metric for the product space is:
\begin{equation}
d(\omega,\omega') = \sum_{n \in \NM} \frac{d_0 \big (\omega_{\eta(n)},\omega'_{\eta(n)} \big )}{2^{n}} \; ,
\end{equation}
where $\eta : \NM \rightarrow \ZM^d$ is a bijection and $d_0(,)$ denotes the metric on $\Omega_0$.
\hfill $\diamond$
}
\end{remark}

Assumption~\ref{Def-DisorderSpace} is obviously in line with Example~\ref{Ex-PhysDis}. It also includes the classical Bernoulli shifts which are quite well understood, see {\it e.g.} \cite{3EFHN}. For example, it is well known that the Bernoulli shifts, hence the above generalization too, admit many natural ergodic measures. The statement below provides a large class of such ergodic measures. 

\begin{proposition}\label{Pro-ProdMeasure} Let ${\rm d} \PM_0$ be a Baire probability measure with full topological support on $\Omega_0$. Then the product measure:
\begin{equation}\label{Eq-ProdMeasure}
{\rm d} \PM(\omega) = \prod_{x \in \ZM^d} {\rm d} \PM_0(\omega_x) \; , \quad \omega = \{\omega_x\}_{x \in \ZM^d} \in \Omega \; ,
\end{equation}
is a well defined probability Baire measure, which is invariant and ergodic w.r.t. the $\tau$-action of $\ZM^d$ and has full topological support.
\end{proposition}

\proof It follows from a direct adaptation of Proposition~6.16 from \cite{3EFHN}. \qed

\vspace{0.2cm}

In \cite{3ProdanSpringer2016ds}, we have opted to work with such product measures solely to make the exposition more concrete and the conclusions derived there do not depend on this choice. The concrete numerical applications reported in the following chapters also assume product measures. However, in the near future, we hope to combine the algorithms with first-principle simulations and obtain more accurate representations of the physical Gibbs measures \ref{Eq-GibbsMeasure}, which are obviously not product measures. For this reason, we will try to remove any un-necessary assumptions on the probability measure.

\begin{assumption} The disorder configurations are also characterized by a probability Baire measure ${\rm d} \PM$ on $\Omega$, which is invariant and ergodic w.r.t. the $\tau$-action of $\ZM^d$ and has full topological support. 
\end{assumption}

The above assumption is entirely justifiable based on physical grounds \cite[Ch.~6]{3Rue} but, unfortunately, we need to introduce one more assumption. The product measures \eqref{Eq-ProdMeasure} have the special property that are invariant not only w.r.t. the translations but also w.r.t. any permutation of the coordinates. This property is essential for our numerical program, at least in its present form (see Remark~\ref{Re-PerTransInv}). 

\begin{assumption}\label{Ass-PermInv} The probability measure ${\rm d} \PM$ is assumed to be invariant against any permutation of the coordinates, that is, the push-forward of ${\rm d} \PM$ by any map: 
\begin{equation}
\{\omega_x\}_{x \in \ZM} \mapsto \{\omega_{r(x)}\}_{x \in \ZM^d} \; , \quad r: \ZM^d \rightarrow \ZM^d \ \mbox{a bijection} \; , 
\end{equation}
coincides with ${\rm d} \PM$.
\end{assumption}

We want to point out that the class of measures which satisfy the two assumption above is much larger than the class of product measures \eqref{Eq-ProdMeasure}. For example the Gibbs measures \eqref{Eq-GibbsMeasure} with symmetric classical potentials falls within this class.

\begin{definition}{\rm The dual $C^\ast$-dynamical system $(C_N(\Omega), \tau,\ZM^d, \Tt_0)$, $N \in \NM$, canonically associated to the ergodic dynamical system $(\Omega,\tau,\ZM^d,{\rm d}\PM)$, is defined by:

\begin{enumerate}[$\ \circ$]

\item The unital $C^\ast$-algebra $C_N(\Omega)$ of continuous functions from $\Omega$ to $M_N(\CM)$, equipped with the $C^\ast$-norm:
\begin{equation}
\|f\| = \sup_{\omega \in \Omega} \|f(\omega)\| \; .
\end{equation}

\vspace{0.1cm}

\item The action of $\ZM^d$ on $C_N(\Omega)$, given by the automorphisms $\tau_y(f) = f \circ \tau_y$, where on the right we have the old action of $\ZM^d$ on $\Omega$. To simplify, we will use the same notation for these two actions.

\vspace{0.1cm}

\item The trace: 
\begin{equation}
\Tt_0(f) = \int_\Omega {\rm d} \PM(\omega)\ \tr\big ( f(\omega) \big ) \; , \quad f \in C_N(\Omega) \; .
\end{equation} 
This trace is faithful (because $\PM$ has full topological support), normalized and invariant w.r.t. the $\tau$-action.

\end{enumerate}
}
\end{definition}

\section{Algebra of Covariant Physical Observables}

As already mentioned at the end Section~\ref{Sec-HomMat}, the classical dynamical system $(\Omega,\tau,\ZM^d)$ has a canonical crossed product algebra $C_N(\Omega) \rtimes_{\tau} \ZM^d$ associated to it, whose representations coincide with the covariant representations of the $C^\ast$-dynamical system defined above \cite{3Ped}. The algebra of physical observables for disordered crystals in magnetic fields is given precisely by such crossed product algebras, but with the action twisted by the magnetic field. Below we introduce this twisted crossed product algebra in a form taken from \cite{3NS}, which we found to be preferred by physicists.

\begin{definition}
\label{Def-BulkAlgebra}
{\rm The algebra of bulk physical observables is defined as the universal $C^\ast$-algebra:
$$
\Aa_d=C^\ast \big ( C_N(\Omega), u_1,\ldots,u_d \big )
$$
generated by the elements of $C_N(\Omega)$ and by additional $d$ elements, satisfying the commutation relations:
\begin{align*}
&  u_j u_j^\ast\; = \; u_j^\ast u_j\; = 1, \quad & j=1,\ldots,d\; , \\
& u_i u_j\;=\;e^{\I \, \phi_{ij}} u_j u_i \;,  \qquad&  i,j=1,\ldots,d\; ,\\
& f \, u_j \; =\; u_j \, \tau_{e_j}(f) \; =\; u_j (f \circ \tau_{e_j})\, , \qquad & \forall \ f \in C_N(\Omega),  \ \ j=1,\ldots,d\; .
\end{align*}
}
\end{definition}

A generic element of $\Aa_d$ can be presented as: 
\begin{equation}\label{Eq-GenericElement}
a\;=\; \sum_{x\in\mathbb Z^d} a_x \, u_x \; , \quad a_x \in C_N(\Omega)\; , \quad u_x = e^{\frac{\I}{2} \sum_{i<j} \Phi_{ij}x_i x_j} \, u_1^{x_1} \ldots u_d^{x_d} \; ,
\end{equation}
where the coefficients $a_x$ have a certain decay with $x$ and the infinite sum must be interpreted as explained below. The monomials $u_x$ in \eqref{Eq-GenericElement} obey the following commutations relations:
\begin{equation}
u_x u_y = e^{2\I \, x \wedge y} u_y u_x = e^{\I \, x \wedge y} u_{x+y} \; , \quad \forall \ x,y \in \ZM^d \; ,
\end{equation}
and
\begin{equation}
u_x^\ast = u_{-x} \; , \quad \forall \ x \in \ZM^d \;.
\end{equation}
The presentation of $\Aa_d$ given in \eqref{Eq-GenericElement} was called the symmetric presentation in \cite{3ProdanSpringer2016ds} because it is associated with the magnetic operators written in the symmetric gauge. A Landau presentation associated with the Landau gauge representation of the magnetic operators can be introduced too \cite{3ProdanSpringer2016ds}, by replacing the monomial $u_x$ with $u_1^{x_1} \ldots u_d^{x_d}$. This presentation is preferred when dealing with systems with a boundary but, since here we deal exclusively with bulk systems, the Landau presentation will not be used at all.

\vspace{0.1cm}

If $a,a' \in \Aa_d$, then their product is explicitly given by:
\begin{equation}\label{Eq-GenericProd}
a\,  a' = \sum_{x \in \ZM^d} \Big ( \sum_{y \in \ZM^d} e^{\I \, y \wedge x} \, a_y \, (a'_{x-y} \circ \tau_{-y}) \Big ) u_x \; ,
\end{equation}
which can be derived directly from the commutation relations. This is another way to see that $\Aa_d$ is indeed the twisted crossed product $C_N(\Omega) \rtimes_\tau^\phi \ZM^d$ \cite{3Dav}. The conjugate of a generic element can be computed as follows:
\begin{equation}\label{Eq-GenericStar}
a^\ast = \sum_{x \in \ZM^d} u_x^\ast \, a_x^\ast = \sum_{x \in \ZM^d} (a_x^\ast \circ \tau_x) u_x^\ast =\sum_{x \in \ZM^d} (a_x^\ast \circ \tau_x) u_{-x}= \sum_{x \in \ZM^d} (a_{-x}^\ast \circ \tau_{-x}) u_x \; .  
\end{equation}

\begin{proposition}[Canonical Representation \cite{3ProdanSpringer2016ds}]\label{Pro-CanRep} The following relations define a field $\{\pi_\omega\}_{\omega \in \Omega}$ of continuous $\PM$-almost surely faithful $\ast$-representations of $\Aa_d$ on the Hilbert space $\CM^N \otimes \ell^2(\mathbb Z^d)$: 
\begin{align}
& \pi_\omega(u_j)
\;=\; I \otimes U_j
\;, 
& \qquad j=1,\ldots,d\;, \\
& \pi_\omega(f) 
\;= \;
\sum_{x \in \mathbb Z^d} f(\tau_x \omega) \otimes |x \rangle \langle x|
\;, 
\quad & \forall \ f \in C_N(\Omega)
\;.
\end{align}
\end{proposition}

\proof It is fairly elementary to check that $\pi_\omega$ preserves the commutation relations \cite{3ProdanSpringer2016ds}. We would like to comment, however, on the faithfulness of the representation. Assume that $\pi_\omega(f)=0$, which can happen if and only if $f(\tau_x \omega) =0$ for all $x \in \ZM^d$. Given the ergodic nature of the action, for $\PM$-almost all $\omega$, this implies that $f$ cancels on the entire $\Omega$, excepting a possible subset of zero measure. Since the measure has full topological support, $f$ must be identically zero. Note that there are $\omega$'s for which $\pi_\omega$ is not faithful. Indeed, let $\omega=\{\omega_0\}_{x \in \ZM^d}$, for some $\omega_0 \in \Omega_0$. Then there are non-constant continuous functions on $\Omega$ such that $f(\omega)=0$. Then $\pi_\omega(f)=0$ and $f\neq 0$. Note that the orbit of this particular $\omega$ is not dense in $\Omega$.  \qed

\vspace{0.2cm}

If an element is presented as in \eqref{Eq-GenericElement}, then:
\begin{equation}\label{Eq-Rep1}
\pi_\omega(a) = \sum_{x,y \in \ZM^d} a_y(\tau_x \omega) \otimes |x\rangle \langle x| U_y \; .
\end{equation}
Comparing this expression with the generic models \eqref{Eq-GenericDisModel}, one can see that any homogeneous disorder crystal model with coefficients from $C_N(\Omega)$ can be generated as the representation of some element from $\Aa_d$.

\begin{example}\label{Ex-HamRep} The generic Hamiltonian \eqref{Eq-GenericDisModel} is generated by the element:
\begin{equation}
\label{Eq-HGenerator1}
h = \sum_{y \in \ZM^d} w_y \, u_y \in \Aa_d \; ,
\end{equation}
while the Hamiltonian in \eqref{Eq-LinDisHam} is generated by:
\begin{equation}
\label{Eq-HGenerator2}
h = \sum_{y \in \ZM^d} (w_y + \omega_0^\alpha \lambda_y^\alpha) \, u_y \in \Aa_d \; ,
\end{equation}
where $\omega_0^\alpha$ should be understood as the function on $\Omega$ which returns the component $\omega_0^\alpha$ corresponding to $x=0$ of $\omega$. 
\hfill $\diamond$
\end{example}

The universal algebras are definitely remarkable in many respects but we should recall that their $C^\ast$-norm is defined in a very abstract manner \cite[p.~158]{3Bla}. In the present context, due to amenability of the $\ZM^d$ group, the norm of $\Aa_d$ can be characterized quite concretely as:
\begin{equation}
\| a \| = \sup_{\omega \in \Omega} \| \pi_\omega(a) \| \;,
\end{equation}
where the norm appearing on the right is the operator norm over $\CM^N \otimes \ell^2(\ZM^d)$. One of the simplest consequences of this observation is the fact that:
\begin{equation}\label{Eq-CNorm}
\| f \|_{\Aa_d} = \| f\|_{C_N(\Omega)} \;, \quad \forall \ f \in C_N(\Omega) \subset \Aa_d \; .
\end{equation}
In other words, the imbedding of $C_N(\Omega)$ in $\Aa_d$ is isometric.

\section{Fourier Calculus}\label{Sec-Fourier} One can exploit the fact that the commutation relations in Definition~\ref{Def-BulkAlgebra} are invariant w.r.t. multiplications by phase factors. This together with universality of the algebra has far reaching consequences, as explained below.

\begin{proposition}[\cite{3Dav}] Let $\lambda = (\lambda_1, \ldots, \lambda_d) \in \TM^d$ and consider the following action of the torus on the unitary generators:
\begin{equation}
u_j \mapsto \lambda_j^{-1} u_j \; , \quad j = 1,\ldots,d \; .
\end{equation}
Then this torus action can be extended to a continuous group $\{\rho_\lambda\}_{\lambda \in \TM^d}$ of automorphism on the whole $\Aa_d$.
\end{proposition}

\proof The twisted unitary elements satisfy the same commutations relations as the ones in Definition~\ref{Def-BulkAlgebra}. Let $\Aa_d(\lambda)$ be the algebra generated by $C_N(\Omega)$ and the twisted $u_j$'s. From the universality property, there exists a homomorphism $\rho_\lambda$ between $\Aa_d$ and $\Aa_d(\lambda)$ such that $\rho_\lambda(u_j) = \lambda_j^{-1}u_j$. But the two algebras are in fact identical. Reversing the role between the twisted and un-twisted $u_j$'s, we obtain a homomorphism from $\Aa_d(\lambda)$ to $\Aa_d$ which is the inverse of $\rho_\lambda$. Hence, $\rho_\lambda$ is an automorphism. The group property and continuity follow from similar considerations.\qed

\begin{remark}{\rm The group of automorphisms introduced above are in fact $U(1)$-gauge transformations, and the latter are associated with the electric charge. For this reason, the calculus developed on this basis (see below) is relevant for the charge transport physics. In a theory where other charges are conserved, one will have to go beyond what is presented here.
\hfill $\diamond$
}
\end{remark}

\begin{definition}[\cite{3Dav}] The Fourier coefficients of an element $a \in \Aa_d$ are define by the Riemann integral:
\begin{equation}\label{Def-FourierCoeff}
\Phi_x(a) = \Big [ \int_{\TM^d} {\rm d} \mu(\lambda) \, \lambda^x \rho_\lambda(a) \Big ] u_x^\ast \; , \quad \lambda^x = \lambda_1^{x_1} \ldots \lambda_d^{x_d}\;, \quad x \in \ZM^d\; ,
\end{equation}
where $\mu$ is the normalized Haar measure of the torus.
\end{definition}

All Fourier coefficients take values in $C_N(\Omega)$. In particular, $\Phi_0$ has the following properties:
\begin{align}
& \Phi_0(f a f') = f \Phi_0(a) f' \; , \ & \forall \ f,f' \in C_N(\Omega)\; ; \\
& \Phi_0(f) =f \; , \ & \forall \ f \in C_N(\Omega)\;.
\end{align}
In other words, $\Phi_0$ is a $C_N(\Omega)$-bimodule map which reduces to the identity map over $C_N(\Omega)$, hence $\Phi_0$ is an expectation from $\Aa_d$ to $C_N(\Omega)$. This expectation can be shown to be continuous and faithful \cite{3Dav}. This observation will play a central role later.

\begin{proposition}[\cite{3Dav}] For a generic element $a \in \Aa_d$, the Ces\`{a}ro sums:
\begin{equation}\label{Cesaro}
a^{(L)}
\;=\;
\sum_{x \in V_L} \prod_{j=1}^d \left (1-\frac{|x_j|}{L+1} \right ) \Phi_{x}(a) u^x
\;,
\end{equation}
with $V_L = \{-L,\ldots,L\}^{d}$, converge in norm to $a$ as $L \rightarrow \infty$. In particular, two elements of the algebra with identical Fourier coefficients coincide.
\end{proposition}

The above statement enables us to present the elements of $\Aa_d$ by the formal infinite series \eqref{Eq-GenericElement}. It now becomes evident that $a_x$ are nothing but the Fourier coefficients and that the formal infinite sums should be interpreted as a Fourier expansion. From their very definition \eqref{Def-FourierCoeff} and from relation \eqref{Eq-CNorm}, it follows that:
\begin{equation}\label{Eq-FourierBound}
\|a_x\| \; \leq \; \| a\| \;, \quad \forall \ x \in \ZM^d \; .
\end{equation}
This simple observation will be used later.

Let us conclude the Section with a remark intended to highlight the generality of the concepts introduced here. The Fourier calculus is available whenever there is a continuous action of the torus on a $C^\ast$-algebra, sometime also called a gauge action. This is amply discussed and exemplified in \cite{3CGRS}. For example, such an action exists when one is dealing with an universal algebra and the defining commutation relations are invariant to $U(1)$ twists. In this more general context, $C_N(\Omega)$ is replaced by the fixed point algebra, {\it i.e.} the sub-algebra which is left invariant by the torus action. Note that the fixed point algebra can be non-commutative. The Fourier coefficients induce a grading on the original algebra and this can be exploited to build a differential calculus as explained next. 

\section{Differential Calculus}
\label{Sec-DiffCalculus}

The Fourier calculus over the algebra of bulk observables generates a system of unbounded closed $\ast$-derivations $\partial =(\partial_1,\ldots,\partial_d)$ and a faithful continuous trace. They will replace the classical differential calculus over the classical Brillouin torus. In the light of the last remark, such non-commutative differential calculus is always available for algebras which accept continuous torus actions. Applications along this line can be found in \cite{3Pro6,3PS2,3BR,3BS} for crossed product algebras, in  \cite{3PR,3PRS1,3PRS2} for graph algebras, in \cite{3CPR} for Cuntz algebras, in \cite{3CRT} for the SU$_q(2)$ algebra and in \cite{3BMS,3Su} for high-energy physics. An overarching study of these aspects can be found in \cite[Ch.~5]{3CGRS}. It is then our hope that the numerical algorithms presented here will find further generalizations and applications.

\begin{definition}{\rm In the following, $C^n(\Aa_d) \subset \Aa_d$ will denote the linear subspaces spanned by those elements $a\in \Aa_d$ for which the automorphisms $\rho_\lambda(a)$ are $n$-times differentiable of $\lambda$. 
}
\end{definition}

\begin{remark}{\rm These linear subspaces are non-empty and in fact are dense in $\Aa_d$. For example, they all contain the sub-algebra of polynomials $\sum_{|x|<R} a_x u_x$, $R<\infty$.
\hfill $\diamond$
}
\end{remark}

\begin{definition} {\rm The non-commutative derivations are defined over $C^1(\Aa_d)$ as the generators of the automorphisms $\rho_\lambda$. Their actions are given explicitly by:
\begin{equation}
\label{BulkDerivation}
\partial_j a \;=\; \partial_j \sum_{x \in \mathbb Z^d} a_x \, u_x 
\;=\; 
-\,\I \sum_{x \in \mathbb Z^d} x_j\, a_x \, u_x
\;, \quad a \in C^1(\Aa_d) \; .
\end{equation}
}
\end{definition}

The derivations commute with each other and they satisfy the Leibniz rule:
\begin{equation}
\partial_j (aa')\;=\; (\partial_j a)a' \; + \; a (\partial_j a')
\;,
\quad
a, \, a'\in C^1(\Aa_d)
\;, \quad j=1,\ldots,d \; .
\end{equation}
Furthermore, their canonical representations on $\CM^N \otimes \ell^2(\ZM^d)$ are nothing but the commutators with the components of the position operator:
\begin{equation}
\label{Eq-BulkDerivationRep}
\pi_{\omega}(\partial_j a)
\;=\;
\I \big [ \pi_{\omega}(a),I \otimes X_j \big ]
\; , \quad j = 1,\ldots,d \; .
\end{equation}
If the disorder and the magnetic field are turned off, then $\partial$ reduces to $\partial_k$, the derivations w.r.t. quasi-momentum encountered in Section~\ref{Sec-TopInv}. Below we showcase several simple but useful calculations.

\begin{exercise}\label{Ex-InvDer1} The Leibniz rule can be used to derive useful identities. One of them is: 
\begin{equation}
\partial_j a^{-1} = - a^{-1} (\partial_j a) a^{-1} \;,
\end{equation} 
whenever $a$ is invertible and $a \in C^1(\Aa_d)$. The identity shows that $a^{-1} \in C^1(\Aa_d)$ whenever $a \in C^1(\Aa_d)$ is invertible. It follows formally from:
\begin{equation}
0=\partial_j 1 = \partial_j (a a^{-1}) = (\partial_j a) a^{-1} + a (\partial_j a^{-1}) \; ,
\end{equation}
after solving for $\partial_j a^{-1}$. This identity comes in handy when one needs to compute the derivative of the resolvent $(z-h)^{-1}$ of a Hamiltonian $h \in C^1(\Aa_d)$.
\hfill $\diamond$
\end{exercise}

\begin{exercise} As we have already seen, the topological invariants involve derivatives of spectral projectors. Let $h \in \Aa_d$ be of finite range and assume $h$ has gap in its spectrum. Then the projection onto the lower part of the spectrum belongs to $C^1(\Aa_d)$ and can be computed by Riesz formula:
\begin{equation}
p = \tfrac{1}{2 \pi \I} \int_{\mathcal C} {\rm d} z \ (z-h)^{-1} \;.
\end{equation}
Then the above identity gives:
\begin{equation}
\partial_j p = \tfrac{1}{2 \pi \I} \int_{\mathcal C} {\rm d} z \ (z-h)^{-1}(\partial_j h) (z-h)^{-1} \; .
\end{equation}
Such formula is useful in theory but no so much in applications because the contour integrals are difficult to evaluate numerically. Other useful identities involving projections are:
\begin{equation}
p(\partial_j p ) p = (1-p) (\partial_j p) (1-p) =0, \quad j =1 , \ldots d\;,
\end{equation}
which follow from $\partial_j\big (p(1-p) \big ) = 0$ after applying the Leibniz rule.
\hfill $\diamond$
\end{exercise} 

\begin{exercise}\label{Ex-HigherDer} Another useful identity is the binomial formula for higher derivations:
\begin{equation}
\partial_j^n(aa') = \sum_{k=0}^n \binom{n}{k} (\partial_j^k a) (\partial_j^{n-k} a') \; ,
\end{equation}
which is a direct consequence of the Leibniz rule. As in the classical case, this identity can be straightforwardly generalized to multi-indices. \hfill $\diamond$
\end{exercise}

The Fourier calculus also defines a faithful continuous trace over $\Aa_d$. Indeed, if we combine the faithful and continuous expectation $\Phi_0$ from $\Aa_d$ to $C_N(\Omega)$ with the continuous and normalized trace over $C_N(\Omega)$, we obtain:

\begin{proposition} {\rm The algebra $\Aa_d$ accepts a canonical continuous and faithful trace:
\begin{equation}
\label{Eq-BulkTrace}
\Tt(a) = \Tt_0 \big ( \Phi_0(a) \big ) = \int_\Omega {\rm d} \PM(\omega) \, \tr \big ( \Phi_0(a)(\omega) \big ) \;.
\end{equation}
If the element is already decomposed as in \eqref{Eq-GenericElement}, then its trace is just:
\begin{equation}
\Tt(a) 
\;=\; 
 \int_\Omega {\rm d} \PM(\omega) \, \tr \big (a_0(\omega) \big ) \;.
\end{equation}
}
\end{proposition}

The trace is normalized, $\Tt(1)=1$, and invariant w.r.t. the automorphisms $\rho_\lambda$. Its physical meaning can be understood from  Birkhoff's ergodic theorem. Indeed, it follows directly from Proposition~\ref{Prop-TracePerVolume} that $\Tt$ is actually equal to the trace per unit volume:
\begin{equation}
\label{Eq-TracePerVolume}
\Tt(a)
\;=\;
\lim_{V \rightarrow \ZM^d} \tr_V \big (\PI_V \, \pi_\omega (a) \, \PI_V^\ast \big )\;.
\end{equation}
This equality holds for $\PM$-almost all $\omega$. Furthermore, if the magnetic field and disorder are turned off, then $\Tt$ becomes the usual integration over the classical Brilllouin torus, $\int_{\TM^d}\frac{{\rm d} k}{(2\pi)^d} \tr ( \ )$.

\vspace{0.1cm}

The triple $(\Aa_d, \partial, \Tt)$ is often referred to as a non-commutative differential manifold. It represents the natural replacement of the classical Brillouin torus hence the name non-commutative Brillouin torus given by Bellissard. Since the trace is invariant w.r.t. the automorphisms $\rho_\lambda$, it follows automatically that $\Tt(\partial_j a)=0$ for all $a \in C^1(\Aa_d)$ and $j=1,\ldots,d$, which implies that the non-commutative Brillouin torus has no boundary.

\begin{exercise}\label{Ex-TraceFaith} A trace is said to be faithful if $\Tt(a) = 0$ implies $a=0$ for any positive element $a \in \Aa_d$. The latter means $a$ is self-adjoint with non-negative spectrum or, equivalently, that $a = b b^\ast$ for some $b \in \Aa_d$ \cite{3Bla}. Then we can compute the $0$-th Fourier coefficient explicitly using \eqref{Eq-GenericProd} and \eqref{Eq-GenericStar}, $a_0 = \sum_{y \in \ZM^d} b_y b_y^\ast$, and:
\begin{equation}
\Tt(a) =  \sum_{y \in \ZM^d} \int_\Omega {\rm d}\PM(\omega) \, \tr \big ( b_y(\omega) b_y^\ast(\omega) \big ) = \sum_{y \in \ZM^d} \Tt_0 \big ( |b_y|^2 \big ) \;.
\end{equation}
Since the terms inside the sum are all positive, $\Tt(a)=0$ implies that every term of the last sum is zero and, since $\Tt_0$ is faithful, we conclude that each $b_y$ is zero. The conclusion is that indeed $\Tt$ is faithful. 
\hfill $\diamond$
\end{exercise}

\section{Smooth Sub-Algebra}
\label{Sec-SmoothAlg} 

The following sub-algebra plays a central role in our numerical program.

\begin{proposition}
\label{Prop-SmoothAlgebra}
The space of infinitely differentiable elements:
\begin{equation}
\Aa^\infty_d
\;=\;
C^\infty(\Aa_d)
\;=\;
\bigcap_{n\geq 1} C^n(\Aa_d)
\;,
\end{equation}
when endowed with the topology induced by the seminorms: 
\begin{equation}\label{Eq-SemiNorm}
\|a \|_\alpha \;=\; \|\partial^\alpha a \|\;, 
\qquad 
\partial^\alpha = \partial_1^{\alpha_1} \cdots \partial_d^{\alpha_d}
\;,\;\;
\alpha \;=\; 
(\alpha_1, \ldots \alpha_d)
\;,
\end{equation}
becomes a dense Fr\'{e}chet sub-algebra of $\Aa_d$, which is stable under holomorphic calculus. The norm appearing on the right in \eqref{Eq-SemiNorm} is the $C^\ast$-norm of $\Aa_d$.
\end{proposition}

Stability w.r.t. the holomorphic calculus refers to the property that if $a \in \Aa_d^\infty$ is invertible in $\Aa_d$, then its inverse $a^{-1}$ actually belongs to $\Aa_d^\infty$. This property plays an important role in Alain Connes' non-commutative geometry because the cyclic cohomology of $C^\ast$-algebras is more naturally defined over locally convex sub-algebras \cite[Ch.~3]{3Con}. In our context, this is relevant because the pairing between cyclic cohomology and $K$-theory provides all (weak and strong) topological invariants \cite{3ProdanSpringer2016ds}. The holomorphic stability ensures that the $K$-theories of the Frech\'et sub-algebras coincide with those of the original $C^\ast$-algebras \cite{3Con}. As it will become more clear later, the holomorphic stability of the smooth sub-algebra can be exploited quite efficiently inside our computational program.

\begin{proposition}[\cite{3ProdanSpringer2016ds}]\label{Pro-SmoothCriterion} The smooth algebra can be characterized as follows. If $a \in \Aa_d^\infty$, then its Fourier coefficients have the fast decay property:
\begin{equation}\label{Eq-SmoothDecay}
|x^\alpha|\, \|a_x\|_{C_N(\Omega)} \leq \| \partial^\alpha a \| < \infty, \quad x^\alpha = x^{\alpha_1} \ldots x^{\alpha_d}.
\end{equation}
Conversely, if for any multi-index $\alpha$:
\begin{equation}
|x^\alpha|\, \|a_x\|_{C_N(\Omega)} < \infty \; ,
\end{equation}
uniformly in $x \in \ZM^d$, then $a \in \Aa_d^\infty$.
\end{proposition} 

The value for our computational program of all the above is in that, if we take $h$ from $\Aa_d^\infty$ and if $G$ is any holomorphic function in a neighborhood of its spectrum, then $G(h)$ belongs automatically to $\Aa^\infty_d$ and its Fourier coefficients have the rapid decay property described in Proposition~\ref{Pro-SmoothCriterion}. Furthermore, note that $\Aa_d^\infty$ is invariant w.r.t. the derivations $\partial$, hence we can conclude at once that generic products:
\begin{equation}
\label{Eq-HolProd}
\partial^\alpha G_\alpha(h) \, \partial^{\beta}G_{\beta}(h) \ldots 
\end{equation}
belong to $\Aa_d^\infty$, hence their Fourier coefficients have the rapid decay property. We can actually show that all of the above are valid not only under the holomorphic but also under the $C^\infty$-functional calculus (with self-adjoint elements). While the former is standard in non-commutative geometry, the latter is not and has to be treated with care. The key to this extension is a functional calculus with smooth functions introduced by Dynkin \cite{3Dyn} and rediscovered by Helfffer and Shostrand \cite{3HS} (see also \cite{3Davis}).

\begin{proposition}[\cite{3Dyn,3HS,3Davis}]\label{Pro-SmoothFC} Let $G: \RM \rightarrow \CM $ be a smooth function with support in an interval and let $\Dd$ be a finite open neighborhood in the complex plane of this interval. Then for any $K \in \NM$, one can construct a function $G_K : \Dd \rightarrow \CM$ such that:
\begin{enumerate}[\rm (i)]
\item $G_K(z,\bar z) = G(z)$ when $z \in \RM$;
\item $|\partial_{\bar z} \, G_K(z,\bar z) | \leq A_K |{\rm Im} \; z|^K$, $A_K < \infty$. 
\end{enumerate}
Furthermore:
\begin{equation}\label{Eq-SmoothFC}
G(h) = \tfrac{1}{2\pi} \int_\Dd {\rm d}^2 z \; \partial_{\bar z} \, G_K(z,\bar z) (h -z)^{-1} \; ,
\end{equation}
for any self-adjoint element $h \in \Aa_d$.
\end{proposition}  

\begin{proposition} The sub-algebra $\Aa_d^\infty$ is stable under the smooth functional calculus with self-adjoint operators.
\end{proposition}

\proof Let $h \in \Aa_d^\infty$ be self-adjoint and $G$ be a smooth function on $\sigma(h)$. Our task is to show that $G(h) \in \Aa_d^\infty$, that is, $\|\partial^\alpha G(h)\| < \infty$ for all multi-indices $\alpha$. Let $K = |\alpha|+1$ and choose $G_K$ as in Proposition~\ref{Pro-SmoothFC}. By applying the rules of calculus from Exercises~\ref{Ex-InvDer1} and \ref{Ex-HigherDer} on \eqref{Eq-SmoothFC}, we can see that $\partial^\alpha G(h)$ can be expressed as a finite sum of terms like:
\begin{equation}
\frac{1}{2\pi} \int_\Dd {\rm d}^2 z \; \partial_{\bar z} \, G_K(z,\bar z) (h -z)^{-\beta_1} \, (\partial^{\gamma_1} h) \, (h -z)^{-\beta_2} \, (\partial^{\gamma_2} h) \ldots \; ,
\end{equation}
where $|\beta_1|+|\beta_2| \ldots \leq |\alpha|+1$ and $|\gamma_1| + |\gamma_2| \ldots = |\alpha|$. By using the elementary estimate $\|(h-z)^{-1}\| \leq |{\rm Im}\; z |^{-1}$, we obtain at once:
\begin{equation}
\|\partial^\alpha G(h) \| \leq {\rm ct.} \int_\Dd {\rm d}^2 z \; |G_K(z,\bar z)| |{\rm Im}\, z |^{-K} \; .
\end{equation}
The integrand is finite due to property (ii) in Proposition~\ref{Pro-SmoothFC} and the affirmation follows. \qed

\vspace{0.2cm}

Recall that the Fermi projection is defined by the functional calculus with a discontinuous function, $p_F = \chi(h \leq \epsilon_F)$. However, if the Fermi energy happens to fall in a spectral gap of $h$, than we can deform the discontinuous function in a smooth one and, as a consequence, the Fermi projection belongs to $\Aa_d^\infty$.

\section{Sobolev Spaces}
\label{Sec-Sobolev} 

Here we introduced the non-commutative Sobolev spaces and define the associated Frech\'et algebra $\bar \Aa_d^\infty$. As opposed to the smooth algebra, $\bar \Aa_d^\infty$ is stable w.r.t. the Borel functional calculus with Anderson localized self-adjoint elements. Here, Anderson localization will be synonymous with the Aizenman-Molchanov bound \cite{3AM,3AW} stated in \eqref{Eq-AizenmannMolchanov}. Let us first introduce and characterize the non-commutative $L^p$-spaces. 

\begin{definition} {\rm Let $|a|=\sqrt{a^\ast a}$ denote the absolute value of an element $a\in \Aa_d$. Then the completion of $\Aa_d$ under the norm:
\begin{equation}
\label{LPNorm}
\|a\|_p \;=\; 
\Tt \left ( |a|^p\right)^{\frac{1}{p}} \; , 
\qquad p\in [1,\infty]\;, 
\end{equation}
defines the non-commutative $L^p(\Aa_d,\Tt)$ Banach space.
}
\end{definition}

For $f \in C_N(\Omega) \subset \Aa_d$, we have:
\begin{equation}
\|f\|_p = \Tt_0 \big ( |f|^p \big )^\frac{1}{p} \; ,
\end{equation}
hence the closure of $C_N(\Omega) \subset \Aa_d$ in $L^p(\Aa_d,\Tt)$ is precisely $L^p(C_N(\Omega),\Tt_0)$. This suggests that the non-commutative $L^p$-spaces are generated by $L^p(C_N(\Omega),\Tt_0)$ and by the unitary elements $u_j$, $j =1,\ldots d$. That this is indeed the case can be seen as follows. First, the linear maps $\rho_\lambda$ from Section~\ref{Sec-Fourier} can be extended by continuity over the non-commutative $L^p$-spaces. Furthermore, note that $|au| = u^\ast |a| u$, hence $|au|^p = u^\ast |a|^p u$ and, consequently, $\| a u\|_p = \| a \|_p$ for any $a,u\in \Aa_d$ with $u$ unitary. Using \eqref{Def-FourierCoeff}:
\begin{align}\label{Eq-LpFourierBound}
\Tt_0 \big ( |\Phi_x(a)|^p \big )^\frac{1}{p} & = \Big \| \Big [ \int_{\TM^d} {\rm d}\mu(\lambda) \, \lambda^x \rho_\lambda(a) \Big ] u_x^\ast \Big \|_p  \\
& \leq \int_{\TM^d} | {\rm d} \mu(\lambda)| \,  \|\rho_\lambda(a)\|_p \leq {\rm ct.}\, \|a\|_p \;. \nonumber
\end{align}
Hence the linear map $\Phi_x$ can be extended by continuity from $\Aa_d$ to $L^p(\Aa,\Tt)$, and this extension indeed takes values in $L^p(C_N(\Omega),\Tt_0)$. As before, it follows that two elements with same Fourier coefficients necessarily coincide. Among other things, this provide the following useful upper bound on the $L^p$-norms.

\begin{proposition}[\cite{3ProdanSpringer2016ds}] 
\label{prop-lsbulk}
Let $a \in L^p(\Aa_d, \Tt)$ and assume $p$ is integer. Then
\begin{equation}
\label{eq-lsbound0}
\|a\|_p 
\;\leq \;
2\, \sum_{x \in \ZM^d} \Tt_0 \big ( |a_x|^p \big )^\frac{1}{p}
\;,
\end{equation}
where $a_x$ is the Fourier coefficient at $x$.
\end{proposition}

In general, the $L^p$-spaces are not closed under multiplication, hence they are not algebras, but only Banach spaces. The non-commutative version of H\"older's inequality is a useful tool in this respect:
\begin{equation}
\label{Holder}
 \|a_1 \cdots a_k\|_p 
 \;\leq \;
 \|a_1\|_{p_1} \cdots \| a_k \|_{p_k}
\;, 
\qquad \frac{1}{p_1} \,+\, \ldots \,+ \,\frac{1}{p_k} \,=\, \frac{1}{p} 
\;.
\end{equation}
As one can see, \eqref{Holder} enables one to make sense of the products of elements from different $L^p$-spaces, such as the one on the l.h.s. of  \eqref{Holder}, as elements of lower $L^p$-spaces. Taking some of the $f$'s in \eqref{Holder} to be the identity in $\Aa_d$, the following sequence of inclusions can be derived: 
\begin{equation}
\label{eq-sequence}
L^\infty ({\Aa_d},{\mathcal T})  \subset \ldots \subset L^p({\Aa_d},{\mathcal T}) \subset \ldots \subset L^1({\Aa_d},{\mathcal T})
\;.
\end{equation}
One should be aware that this chain of inclusions does not apply to $C^\ast$-algebras without a unit.

\vspace{0.1cm}

There is one important exception to the above rule. Indeed, the space $L^\infty({\Aa_d},{\mathcal T})$ is closed under the multiplication. It represents the weak von Neumann closure of $\Aa_d$ when the elements of the latter are viewed as multiplication operators on the Hilbert space $L^2(\Aa_d,\Tt)$. This weak closure can be characterized more concretely \cite{3Con3} as the closure of $\Aa_d$ under the norm: 
\begin{equation}
\|a\|_\infty\; =\; 
\mathbb P\!-\!\operatorname*{\mathrm{esssup}}\limits_{\omega \in \Omega} \|\pi_\omega (a)\|
\;.
\end{equation}
Since von Neumann algebras are stable under the Borel functional calculus, the Fermi projections and Fermi unitary operators belong to $L^\infty({\Aa_d},{\mathcal T})$, regardless of the existence of spectral or mobility gaps at the Fermi level. The topology of $L^\infty({\Aa_d},{\mathcal T})$ is, however, too strong to be useful just by itself, in the strong disorder regime. For example, the Fermi projections do not vary continuously w.r.t. $\|\cdot \|_{L^\infty}$ when the coefficients of the models are deformed continuously in $C_N(\Omega)$, even when the Fermi level is located in a region of Anderson localized spectrum. This is another reason for introducing the non-commutative Sobolev spaces.

\begin{definition}[\cite{3ProdanSpringer2016ds}]\label{Def-SobolevSpace} {\rm The Sobolev spaces $\mathcal W_{r,p}(\mathcal A,\mathcal T_0)$ are defined as the Banach spaces obtained as the closure of $\Aa_d^\infty \subset L^\infty({\Aa_d},{\mathcal T})$ under the norms:
\begin{equation}
\label{Eq-SobolevNorm}
\|a\|_{r,p}
\;=\; 
\sum_{\bm x \in \mathbb Z^d} \big (1+|x|\big )^r \, \Tt_0 \Big (|a_x|^p \Big )^\frac{1}{p}
\;, 
\quad 
r \in \NM \;, \;\; p \in \NM_+
\;. 
\end{equation}
}
\end{definition}

\begin{definition} {\rm The algebra $\bar \Aa_d^\infty$ is defined as the Frech\'et algebra defined by the closure of $\Aa_d^\infty \subset L^\infty({\Aa_d},{\mathcal T})$ in the topology defined by the norms $\| \ \|_{r,p}$, $r \in \NM$, $p \in \NM_+$.
}
\end{definition}

As we already anticipated, the Sobolev spaces are an essential tool for the regime of strong disorder. But first a definition.

\begin{definition}\label{Def-LocElement} {\rm Let $h$ be a self-adjoint element of $\Aa_d$. The spectrum in an open interval $\Delta \subset \sigma(h)$ is said to be Anderson localized if: 
\begin{equation}\label{Eq-MA}
\int_\Omega {\rm d} \PM(\omega) \, \big \|(h-z)^{-1}_x(\omega) \big \|^s \leq A_s(\delta) e^{-\gamma_s(\delta) |x|} \; , \quad s \in (0,1), \ \delta >0 \;,
\end{equation}
uniformly for all $z \in \CM \setminus \sigma(h)$ with ${\rm dist}(z , \sigma(h) \setminus \Delta) \geq \delta.$ Above, $A_s(\delta)$ and $\gamma_s(\delta)$ are finite positive constants which depend parametrically on $s$ and $\delta$ but are independent of $x$.
}
\end{definition}

The above definition is a convenient reformulation of the Aizenman-Molchanov bound \cite{3Aiz,3AM,3AW} stated in \eqref{Eq-AizenmannMolchanov}. In the physics jargon, such an interval is called a mobility gap because the diagonal transport coefficients vanish whenever the Fermi level resides in such interval \cite{3AW}. Note that the bound \eqref{Eq-MA}, at least for finite range Hamiltonians ({\it i.e.} those with $h_x =0$ if $x$ outside a compact subset), applies automatically to any interval $\Delta$ which is outside the spectrum. Hence, we require specifically that $\Delta \subset \sigma(h)$ in order to distinguish a mobility gap from an ordinary gap in the spectrum.

\begin{proposition}\label{Pro-BorelFC} Let $h \in \Aa_d$ be a self-adjoint element with a mobility gap $\Delta$. Then, $G(h) \in \bar \Aa_d^\infty$ for any Borel function $G$ with support in $\Delta$.
\end{proposition}

\proof We will show that the step function $\sgn(h-\mu)$ belongs to $\bar \Aa_d^\infty$ for all $\mu \in \Delta$. Since such step functions generate the Borel calculus, the affirmation will follow. Let us begin by proving the useful relation:
\begin{equation}\label{Eq-BasicTool}
\Big \| \int_\Cc {\rm d} z \; G(z) (h - z)^{-1} \Big \|_{r,p}  \leq \int_\Cc |{\rm d} z| \; \frac{|G(z)|}{|{\rm Im}\, z|^{1-s}} \; \big \|(h-z)^{-1} \big \|_{r,sp}^s \; ,
\end{equation}
where $\Cc$ is any contour in the complex plane such that $\Gamma \cap \sigma(h) = \emptyset$. In \eqref{Eq-BasicTool}, $s$ can be taken arbitrarily small, hence $sp$ is not restricted to $[1,\infty)$ as in \ref{Eq-SobolevNorm}. The proof of \eqref{Eq-BasicTool} starts from the observation that $\|(h-z)^{-1}\| \leq \frac{1}{|{\rm Im}\, z|}$, which automatically implies that all Fourier coefficients obey:
\begin{equation}
\big \|(h-z)^{-1}_x(\omega)  \big \| \leq \big \|(h-z)^{-1} \big \| \leq \frac{1}{|{\rm Im}\, z|} \; .
\end{equation}
Then:
\begin{align}
 \big \|(h-z)^{-1}_x(\omega) \big \| \leq \frac{1}{|{\rm Im} \, z|^{1-s}} \big \|(h-z)^{-1}_x(\omega) \big \|^{s} \; ,
\end{align}
and \eqref{Eq-BasicTool} follows. 

\vspace{0.1cm}

Next, we consider two contours in the complex plane, $\Cc_\epsilon^\pm \in \CM \setminus \sigma(h)$, which start at $\mu \mp \I \epsilon$ and end at $\mu \pm \I \epsilon$, respectively, and ${\rm Re}(z-\mu) \geq 0$ for $z \in \Cc^+_\epsilon$ while ${\rm Re}(z-\mu) \leq 0$ for $z \in \Cc^-_\epsilon$. We will show that: 
\begin{equation}
\lim_{\epsilon \searrow 0} \tfrac{\I}{2\pi} \int_{\Cc^\pm_\epsilon} {\rm d} z \; (h-z)^{-1} \in \Ww_{r,p}(\Aa_d,\Tt_0) \; .
\end{equation}
It will be convenient to introduce the notation:
\begin{equation}
a_\pm(\epsilon) = \tfrac{\I}{2\pi}\int_{\Cc^\pm_\epsilon} dz \; (h-z)^{-1} \in \mathscr A_d^\infty \; ,
\end{equation}
and let us consider $0<\epsilon_1 < \epsilon_2 < \epsilon$. Then:
\begin{align}
a_\pm(\epsilon_1) - a_\pm(\epsilon_2) =  \mp \, \frac{\I}{2 \pi} \int_{\epsilon_1}^{\epsilon_2} dw \; \Big ((h-\mu-\I w)^{-1} -(h-\mu+\I w)^{-1} \Big ) \; . 
\end{align}
By applying \eqref{Eq-BasicTool}, we obtain:
\begin{align}
\Big \| \int_{\epsilon_1}^{\epsilon_2} dw \; (h-\mu \pm \I w)^{-1} \Big \|_{r,p}  \leq \int_{\epsilon_1}^{\epsilon_2} \frac{dw}{w^{1-s}} \big \| (h-\mu \pm \I w)^{-1} \big \|_{r,sp}^s \; ,
\end{align}
and we take $s < \frac{1}{p}$ so that we can apply \eqref{Eq-MA}. Then: 
\begin{align}
\Big \|\int_{\epsilon_1}^{\epsilon_2} dw & \; (h-\mu \pm \I w )^{-1} dz \Big \|_{r,p} \\
\nonumber & \leq s(\epsilon_2^s -\epsilon_1^s) \sqrt[p]{A_{sp}(\delta) } \sum_{q \in \ZM^d} (1+ |q|)^r e^{-\frac{1}{p}\beta_{sp}(\delta) |q|},
\end{align}
where $\delta = {\rm dist}(\mu, \sigma(h) \setminus \Delta)$. The remaining sum is convergent for any $r\in \NM$ and $p \in [1,\infty)$, hence we established that: 
\begin{equation}
\|a_\pm(\epsilon_1) - a_\pm(\epsilon_2)\|_{r,p} \leq ct. (\epsilon_2^s -\epsilon_1^s) \rightarrow 0 \ \ {\rm as} \ \ \epsilon \searrow 0,
\end{equation}
therefore $a_\pm(\epsilon)$ defines Cauchy sequences  w.r.t. the Sobolev norms and, consequently, $a_\pm(\epsilon)$ have a limit in $\Ww_{r,p}(\Aa_d,\Tt_0)$. 

\vspace{0.1cm}

In the second step of the proof, we consider an approximate sign function:
\begin{equation}
\sgn_\epsilon(t) = \tanh(t/\epsilon) \; , \quad \lim_{\epsilon \searrow 0} \sign_\epsilon(t) = \sign(t)\;,
\end{equation} 
and show that:
\begin{equation}\label{Eq-Diff}
\sgn_\epsilon(h-\mu) -a_+\big(\tfrac{\pi \epsilon}{4} \big) + a_-\big(\tfrac{\pi \epsilon}{4} \big) \rightarrow 0 \ \ {\rm as} \ \ \epsilon \searrow 0 \; ,
\end{equation}
where the limit is w.r.t. any of the Sobolev norms. We will express $\sgn_\epsilon(h-\mu)$ using the holomorphic calculus. For this, we consider the contour $\Cc_\epsilon = \Cc_\frac{\pi \epsilon}{4}^+ \cup \Cc_\frac{\pi \epsilon}{4}^-$, which encircles $\sigma(h)$ and avoids the poles of $\tanh\big ((z-\mu)/\epsilon\big)$, located at $\mu+\epsilon(n+\frac{1}{2})\pi$, $n \in \ZM$, and write:
\begin{equation}
\sgn_\epsilon(h-\mu) = \tfrac{\I}{2\pi} \int_{\Cc_\epsilon} {\rm d} z \, \tanh\Big(\frac{z-\mu}{\epsilon}\Big)(h-z)^{-1} \; .
\end{equation}
Then \eqref{Eq-Diff} becomes:
\begin{align}
& \tfrac{\I}{2\pi} \int_{\Cc^+_\frac{\pi \epsilon}{4}} {\rm d} z \, \Big(\tanh\Big(\tfrac{z-\mu}{\epsilon}\Big)-1\Big)(h-z)^{-1} \\
\nonumber & + \tfrac{\I}{2\pi} \int_{\Cc^-_\frac{\pi \epsilon}{4}} {\rm d} z \, \Big(\tanh\Big(\tfrac{z-\mu}{\epsilon}\Big)+1\Big)(h-z)^{-1}.
\end{align}
By applying \eqref{Eq-BasicTool} and choosing again $s < \frac{1}{p}$:
\begin{align}
& \Big \| \int_{\Cc^\pm_\frac{\pi \epsilon}{4}} {\rm d} z \, \Big(\tanh\Big(\frac{z-\mu}{\epsilon}\Big) \mp 1\Big )(h-z)^{-1} \Big \|_{r,p}  \\
\nonumber & \quad \leq \sqrt[p]{ A_{sp}(\delta^\pm_\epsilon) }  \Big ( \sum_{q \in \ZM^k} (1+|q|)^r e^{-\frac{1}{p}\beta_{sp}(\delta^\pm_\epsilon) |q|}  \Big ) \\
\nonumber & \qquad \times \int_{\Cc^\pm_\frac{\pi \epsilon}{4}} |dz| \; \frac{|\tanh\big((z-\mu)/\epsilon\big)\mp 1|}{ |{\rm Im} \, z|^{1-s}}\; ,
\end{align}
where $\delta^\pm_\epsilon = {\rm dist} \big (\Cc^\pm_\frac{\pi \epsilon}{4}, \sigma(h) \setminus \Delta \big)$. These $\delta$'s converge to a finite value as $\epsilon \searrow 0$, hence the exponential factor inside the sum is present for all $\epsilon >0$. Furthermore, $\tanh(z/\epsilon) \rightarrow \pm 1$ point-wisely in the right/left complex semi-planes, respectively, and the singularity $|{\rm Im} \, z|^{s -1}$ is integrable. The immediate conclusion is that, indeed, the Sobolev norm of \eqref{Eq-Diff} goes to zero as $\epsilon \searrow 0$. As a consequence, $\sgn(h-\mu) \in \Ww_{r,p}(\Aa,\Tt_0)$. \qed

\begin{corollary}\label{Cor-BoreFC} The Borel functional calculus with self-adjoint elements from $\Aa_d^\infty$ lands in $\bar \Aa_d^\infty$, provided the Borel functions are smooth away from the mobility gaps of the self-adjoint elements.
\end{corollary}

\proof The statement follows by combining Propositions~\ref{Pro-BorelFC} and \ref{Pro-SmoothFC}. \qed

\vspace{0.1cm}

Let us conclude this section by noticing that non-commutative Sobolev spaces can be defined by closing the sub-algebra $\Aa_d^\infty$ in the topology induced by the semi-norms $\|\partial^\alpha a\|_p$. The result is a closely related but nevertheless different family of Sobolev spaces (see \cite{3ProdanSpringer2016ds}). We will not use them at all here because the new definition formulated in \ref{Def-SobolevSpace} presents certain advantages.

\section{Magnetic Derivations}\label{Sec-MagDer} 

In this Section we focus on the dependence of the calculations on the magnetic field. One goal is to introduce the Ito-like derivations introduced by Bellissard \cite{3Bel3,3Bel33}, which already proved to be quite a formidable tool for the analysis of the energy spectra \cite{3RB} and computation of the linear and non-linear magneto-electric response functions \cite{3LP, 3ST,3ProdanSpringer2016ds}. Our goal here is to provide the tools by which one can diagnose how smooth is a correlation functions as a function of the magnetic field. The issue is relevant for the following reason. As it is well known, the entries in the magnetic flux matrix $\phi$ become quantized when working on finite volumes (see Section~\ref{Sec-FiniteAlg}), hence the correlation functions can be computed in practice only for discrete values of the magnetic field. Hence, continuity of these functions w.r.t. the magnetic field is of central importance. The natural framework for this type of problems is that of continuous fields of $C^\ast$-algebras, which we introduce next, by following Dixmier's monograph \cite{3Dix}. See also \cite{3Wil} for a more modern approach and a slightly broader context.

\begin{definition}[\cite{3Dix}, p.~212]\label{Def-FieldAlg1} 
{\rm Let $\Sigma$ be a compact topological space. A continuous field $\mathcal E = (\{E(t)\}_{t \in \Sigma}, \Gamma)$ of Banach spaces over $\Sigma$ is a family $\{E(t)\}_{t \in \Sigma}$ of Banach spaces, together with a set $\Gamma \subset \prod_{t \in \Sigma} E(t)$ of sections such that:

\begin{enumerate}[\rm (i)] 

\vspace{0.1cm}

\item $\Gamma$ is a complex linear subspace of $\prod_{t \in \Sigma} E(t)$;

\vspace{0.1cm}

\item For every $t \in \Sigma$, the set $\mathfrak s(t)$ for $\mathfrak s \in \Gamma$ is dense in $E(t)$;

\vspace{0.1cm}

\item For every section $\mathfrak s \in \Gamma$, the function $t \rightarrow \|\mathfrak s (t)\|$ is continuous;

\vspace{0.1cm}

\item Let $\mathfrak s \in \prod_{t \in \Sigma} E(t)$ be a section; if for every $t \in \Sigma$ and every $\epsilon >0$, there exists an $\mathfrak s' \in \Gamma$ such that $\|\mathfrak s (t) - \mathfrak s'(t)\| \leq \epsilon$ throughout some neighborhood of $t$, then $\mathfrak s \in \Gamma$.

\end{enumerate}
}
\end{definition}

\begin{definition}[\cite{3Dix}, p.~218]\label{Def-FieldAlg1}
{\rm Let $\Sigma$ be a compact topological space. A continuous field of $C^\ast$-algebras over $\Sigma$ is a continuous field $(\{A(t)\}_{t \in \Sigma}, \Gamma )$ of Banach spaces, such that each $A(t)$ is being endowed with a multiplication and an involution with which it is a $C^\ast$-algebra, and the space of sections $\Gamma$ being closed under multiplication and involution.
}
\end{definition}

The definition can be easily specialized to our context \cite{3Bel3,3Bel33}, by considering the family of $C^\ast$-algebras $A_d(\phi)$ over the torus $\TM^{n_d}$, $n_d=\frac{d(d-1)}{2}$, where the anti-symmetric matrix $\phi$ lives. To be more precise, we will write the dependence of $u_j$'s on $\phi$ explicitly. As for the set of sections, we can take:
\begin{equation}
\Gamma = \Big \{ \mathfrak s \in \prod_{\phi \in \TM^{n_d} } \Aa_d(\phi), \  \mathfrak s (\phi)=\sum_{x \in \ZM^d} a_x(\phi) u_x(\phi) \ \big | \ a_x \in C \big (\TM^{n_d}, C_N(\Omega) \big ) \Big \}\; .
\end{equation} 
Then $\big (\{\Aa_d(\phi)\}_{\phi \in \TM^{n_d} }, \Gamma \big )$ defines our field of $C^\ast$-algebras. Our next goal is to define an action of a torus and construct the derivations w.r.t. $\phi$ as the generators of this action. The construction is very natural but first we need one more standard result. 

\begin{proposition}[\cite{3Dix}, p.~219] The set $\Gamma$ becomes a $C^\ast$-algebra when endowed with the natural algebraic operations and the norm:
\begin{equation}
\|\mathfrak s\|_\Gamma = \sup_{\phi \in \TM^{n_d} } \|\mathfrak s(\phi)\|_{\Aa_d(\phi)} \; .
\end{equation}
\end{proposition} 

\begin{proposition} Let $\sigma_\xi$ be the natural action of $\TM^{n_d}$ on $C \big (\TM^{n_d},C_N(\Omega) \big )$:
\begin{equation}
\TM^{n_d} \times C \big (\TM^{n_d},C_N(\Omega) \big ) \ni (\xi,f) \mapsto \sigma_\xi(f)\;, \ \  (\sigma_\xi(f))(\phi) = f(\phi+\xi)\; .
\end{equation} 
Then this action can be naturally lifted to a continuous group of automorphisms on $\Gamma$ by:
\begin{equation}
\sigma_\xi(\mathfrak s) = \sum_{x \in \ZM^d} \sigma_\xi (a_x) u_x \; , \quad \xi \in \TM^{n_d} \; .
\end{equation}
Furthermore, the generators of this action are Bellissard's derivations \cite{3Bel3} w.r.t. the magnetic field:
\begin{equation}
\delta_{\phi_{ij}} \mathfrak s = \sum_{x \in \ZM^d} (\partial_{\phi_{ij}} a_x) \, u_x \; .
\end{equation}
The derivations are defined over $C^1(\Gamma)$, the subspace of $\Gamma$ on which $\sigma_\xi$ is first order differentiable. This space becomes a Banach space when endowed with the norm:
 \begin{equation}
\|\mathfrak s\|_1 = \|\mathfrak s\| + \sum_{j=1}^d \| \partial_j \mathfrak s\| + \sum_{i<j} \|\delta_{\phi_{ij}} \mathfrak s \| \; .
\end{equation}
\end{proposition}

\begin{proposition}[Rules of calculus]\label{Pro-MagCalculus} The $\delta$-derivations
satisfy the following identities:

\begin{enumerate}[\rm (i)]

\item $\delta_{\phi_{ij}} \partial_k = \partial_k \delta_{\phi_{ij}}$;

\vspace{0.1cm}

\item $(\delta_{\phi_{ij}} \mathfrak s )^\ast = \delta_{\phi_{ij}} \mathfrak s^\ast$;

\vspace{0.1cm}

\item $\delta_{\phi_{ij}} (\mathfrak s \mathfrak s') = (\delta_{\phi_{ij}} \mathfrak s) \mathfrak s' + \mathfrak s (\delta_{\phi_{ij}} \mathfrak s') - \frac{\I}{2} (\partial_i \mathfrak s \, \partial_j \mathfrak s' - \partial_j \mathfrak s \, \partial_i \mathfrak s')$;

\vspace{0.1cm}

\item $\partial_{\phi_{ij}} \Tt \big (\mathfrak s(\phi) \big ) = \Tt \big ( (\delta_{\phi_{ij}} \mathfrak s)(\phi) \big )$.

\end{enumerate}

\end{proposition}

As it is well known, the ordinary perturbation theory w.r.t. the magnetic field cannot be used for magnetic Hamiltonians. This makes the computation of the response functions w.r.t. the magnetic field very difficult when approached by traditional methods. With the above basic rules of calculus, however, these calculations become straightforward. The examples below show this calculus in action.

\begin{exercise} Let $\mathfrak p \in C^1(\Gamma)$ be a projection. Then 
\begin{equation}
\mathfrak p(\partial_{\phi_{ij}} \mathfrak p) \mathfrak p = \tfrac{\I}{2}\mathfrak p [\partial_i \mathfrak p,\partial_j \mathfrak p] = \tfrac{\I}{2}[\partial_i \mathfrak p,\partial_j \mathfrak p] \mathfrak p\;.
\end{equation}
Observe that on the righthand side only the regular derivations appear. The identity follows from:
\begin{equation}
0 = \delta_{\phi_{ij}} \big (\mathfrak p (1-\mathfrak p) \big ) = (\delta_{\phi_{ij} \mathfrak p} ) (1- \mathfrak p) - \mathfrak p(\delta_{\phi_{ij}} \mathfrak p) - \tfrac{\I}{2} (-\partial_i \mathfrak p \partial_j \mathfrak p + \partial_j \mathfrak p \partial_i \mathfrak p) \; ,
\end{equation}
after multiplying from the right by $\mathfrak p$.
\hfill $\diamond$
\end{exercise}

\begin{exercise}\label{Ex-InvDer2} Let $\mathfrak s \in C^1(\Gamma)$ be invertible in $\Gamma$. Then $\mathfrak s^{-1} \in C^1(\Gamma)$ and:
\begin{equation}
\delta_{\phi_{ij}} \mathfrak s^{-1} = \mathfrak s^{-1} \big ( -\delta_{\phi_{ij}} \mathfrak s - \tfrac{\I}{2} (\partial_i \mathfrak s \, \mathfrak s^{-1} \, \partial_j \mathfrak s - \partial_j \mathfrak s \, \mathfrak s^{-1} \, \partial_i \mathfrak s) \big ) \mathfrak s^{-1} \; .
\end{equation}
The identity is important because it gives an effective way to compute the magnetic derivative for any holomorphic function $G(\mathfrak s)$, or smooth function if $\mathfrak s$ is self-adjoint. The identity follows from:
\begin{equation}
0=\delta_{\phi_{ij}} (\mathfrak s \mathfrak s^{-1}) = (\delta_{\phi_{ij}} \mathfrak s) \mathfrak s^{-1} + \mathfrak s (\delta_{\phi_{ij}} \mathfrak s^{-1}) - \tfrac{\I}{2} (\partial_i \mathfrak s \partial_j \mathfrak s^{-1} - \partial_j \mathfrak s \partial_i \mathfrak s^{-1}) \; ,
\end{equation}
after applying the rule of calculus from Exercise~\ref{Ex-InvDer1}.
\hfill $\diamond$
\end{exercise}

The rule (iv) of Proposition~\ref{Pro-MagCalculus} gives us a simple tool to diagnose how smooth a correlation function is.

\begin{proposition}\label{Pro-BSmooth} Assume that $h \in \Aa_d^\infty$ and that its Fourier coefficients are $k$-times differentiable w.r.t. the fluxes $\phi$. Let $G_1$, $G_2$, \ldots, be smooth functions on $\sigma(h)$. Then:
\begin{equation}
\Tt \big (\partial^\alpha G_\alpha(h) \, \partial^\beta G_\beta(h) \ldots \big )
\end{equation}
is $k$-times differentiable w.r.t. the magnetic field.
\end{proposition}

\proof Using the functional calculus from Proposition~\ref{Pro-SmoothFC}, it is clear that it is enough to consider just products of the form: 
\begin{equation}
\label{Eq-SimplProd}
\partial^\alpha (z_\alpha -h)^{-1} \, \partial^\beta (z_\beta - h)^{-1} \ldots \; .
\end{equation}
Using the rules of calculus from Proposition~\ref{Pro-MagCalculus}, one can expand the magnetic derivations $\delta^\gamma$, $|\gamma|=k$, of \eqref{Eq-SimplProd} and convince himself that the final result depends only on $\partial_\phi^\rho h_x$, $|\rho| \leq k$. \qed

\vspace{0.1cm}

Note that the generic Hamiltonians (see Example~\ref{Ex-HamRep}) have Fourier coefficients which are entirely independent of the magnetic field. In this case, the correlation functions considered above are actually smooth w.r.t. to the magnetic field.

\begin{remark}{\rm In the recent series of works \cite{CCL,CMF}, the authors investigate the electronic properties of multi-layered materials, with the layers twisted relative to each other. The issue of commensurate versus incommensurate twist angles considered in these works is very similar to what has been discussed in this section. It will be interesting to see if Ito-like derivations can be constructed for this context and if the smoothness of the electronic properties w.r.t. the twist angles can be confirmed using similar methods.
\hfill $\diamond$
}
\end{remark}

\section{Physics Formulas}
\label{Sec-PhysFormulas}

At this point we have the entire formal calculus in place and perhaps it will be interesting to see it in action right away. As such, we end this Chapter with a collection of closed-form expressions for various physical coefficients, taken from the published literature. Several of these formulas will be further discussed in the following Chapters, and, for the ones that are not, we provide here citations to related numerical works. Throughout, we will use physical units such that Boltzmann's constant, the electron charge, electron mass and Plank's constant $\hbar$ are all set to 1.

\begin{example}[Finite-temperature Kubo-formula, \cite{3BES,3SBB1,3SBB2}] Let us recall that the linear conductivity tensor $\hat \sigma$ relates a uniform electric field $E$ and the induced charge current-density $J$ in a solid, $J= \hat \sigma E$. Such linear relation between the electric field and the charge current-density is typically observed at small $E$. In the presence of disorder, magnetic fields and dissipation, and under suitable physical assumptions, the conductivity tensor at temperature $T$ accepts the following non-commutative Kubo-formula:
\begin{equation}\label{Eq-NCKubo}
\sigma_{ij}(\epsilon_F,T) = 2\pi N \, \Tt \Big ( (\partial_i h) \, (\Gamma + \Ll_h)^{-1} \big ( \partial_j \Phi_{\mathrm{FD}}(h) \big ) \Big) \; .
\end{equation}
Throughout, the conductivity tensor will be expressed in the natural unit of $e^2/h$. In \eqref{Eq-NCKubo}, $\Phi_{\rm FD}$ is the Fermi-Dirac distribution:
\begin{equation}
\Phi_{\rm FD}(h) = \frac{1}{1 + e^{(h-\epsilon_F)/T}} \; ,
\end{equation}
${\mathcal L}_h$ is the inner derivation $\mathcal L_h (a)=\I [a,h]$, and $\Gamma$ is a super-operator ({\it i.e.} a map on $\Aa_d$) called the dissipation super-operator. This Kubo-formula will be the subject of Chapter~\ref{Cha-ApplicationsI}.
\hfill $\diamond$
\end{example}

\begin{remark}{\rm One difference with {\it e.g.} \cite{3BES} can be readily spotted in \eqref{Eq-NCKubo}, namely the presence of $N$, the dimension of the fiber. It is there because we insisted on working with normalized traces. For this same reason, the factor $N$ appears in all the physics formulas mentioned below.
\hfill $\diamond$
}
\end{remark} 

\begin{example}[Even Chern numbers, \cite{3BES,3PLB,3ProdanSpringer2016ds}] The real-space formulas for the even Chern numbers from Section~\ref{Sec-TopInv} can be reformulated as ($d=$ even):
\begin{equation}
{\rm Ch}_d(p)=\tfrac{(2 \pi \I)^\frac{d}{2}}{\frac{d}{2}!} N \sum_{\rho \in \Ss_d}(-1)^\rho \, \Tt \Big ( p \prod_{j=1}^{d} \partial_{\rho_j}p\Big ) \;, \quad p^2=p^\ast=p \;.
\end{equation}
More generally, if $I \subseteq \{1,2, \ldots,d\}$ is an ordered subset such that its cardinality $|I|$ is even, then one can also define the lower even Chern numbers:
\begin{equation}
{\rm Ch}_I(p)=\tfrac{(2 \pi \I)^\frac{|I|}{2}}{\frac{|I|}{2}!} N \sum_{\rho \in \Ss_I}(-1)^\rho \, \Tt \Big ( p \prod_{j=1}^{|I|} \partial_{\rho_j}p\Big ) \;, \quad p^2=p^\ast = p \; . 
\end{equation}
It follows from \cite{3ProdanSpringer2016ds} that ${\rm Ch}_I$, $I \subseteq \{1,\ldots,d\}$, $|I|=$ even, define a complete set of topological invariants for class A, in the sense that knowing their $2^{d-1}$ numerical values is necessary and sufficient to locate the Fermi projection $p_F$ in the $K_0$-group of $\Aa_d$. As already mentioned, one of the main results of Ref.~\cite{3BES} is the following connection between the lowest even Chern numbers of the Fermi projection and the Kubo-formula for conductivity:
\begin{equation}
{\rm Ch}_{\{i,j\}}(p_F) = \lim_{T,\Gamma \rightarrow 0} \sigma_{ij}(\epsilon_F,T,\Gamma) \; .
\end{equation}
Then the physical interpretation of the higher Chern numbers follows from the following generalized Streda formula:
\begin{equation}\label{Eq-GenStreda}
\partial_{\phi_{ij}} {\rm Ch_I}(p_F) = \frac{1}{2 \pi} {\rm Ch}_{\{i,j\} \cup I}(p_F) \; , \quad I \cap \{i,j\} = \emptyset \;,
\end{equation}
which was proven in \cite{3ProdanSpringer2016ds} using the rules of calculus from Proposition~\ref{Pro-MagCalculus}. Starting from the lower Chern numbers and applying the generalized Streda multiple times, one can conclude as in \eqref{Eq-EvenChernPhys} that the even topological invariants are connected to the linear and non-linear magneto-electric transport coefficients at zero temperature.
\hfill $\diamond$
\end{example}

\begin{example}[Odd Chern Numbers, \cite{3PS,3ProdanSpringer2016ds}] Similarly, the real-space formulas for the odd Chern numbers from Section~\ref{Sec-TopInv} can be reformulated as ($d =$ odd):
\begin{equation}
{\rm Ch}_d(u)= \tfrac{\I (\I \pi)^\frac{d-1}{2}}{ d!!} N \sum_{\rho \in \Ss_d}(-1)^\rho \, \Tt \Big ( \prod_{j=1}^{d} u^\ast \partial_{\rho_j} u \Big ) \; , \quad uu^\ast = u^\ast u =1 \; ,
\end{equation}
and lower odd Chern numbers can be introduced as ($|I|=$ odd):
\begin{equation}
{\rm Ch}_I(u)= \tfrac{\I (\I \pi)^\frac{|I|-1}{2}}{ |I|!!} N \sum_{\rho \in \Ss_I}(-1)^\rho \, \Tt \Big ( \prod_{j=1}^{|I|} u^\ast \partial_{\rho_j} u \Big )  , \quad uu^\ast = u^\ast u =1 \; .
\end{equation}
It follows from \cite{3ProdanSpringer2016ds} that ${\rm Ch}_I(u_F)$, $I \subseteq \{1,\ldots,d\}$, $|I|=$ odd, define a complete set of invariants for class AIII, in the sense that knowing their $2^{d-1}$ numerical values is necessary and sufficient to locate the Fermi unitary operator $u_F$ in the $K_1$-group of $\Aa_d$. As for the physical interpretation of the invariants, the lowest odd Chern numebers ${\rm Ch}_{\{j\}}(u_F)$, $j=1,\ldots,d$, are identical with the components of the macroscopic vector of the so called chiral polarization \cite{3ProdanSpringer2016ds}. The generalized Streda formula \eqref{Eq-GenStreda} holds for the odd case too, and by iterating this formula one then concludes as in \eqref{Eq-OddChernPhys} that the odd topological invariants are connected with the chiral linear and non-linear magneto-electric response coefficients. 
\hfill $\diamond$
\end{example}

\begin{example}[Electric polarization, \cite{3ST}] The adiabatic change of the macroscopic vector of electric polarization $P$, {\it e.g.} under a mechanical deformation of the crystal, accepts the formula:
\begin{equation}\label{Eq-ElecPol}
\Delta P = \I N \, \int_0^T {\rm d} t \ {\mathcal T}\big ( p_F(t)[\partial_t p_F(t), \partial p_F(t)]\big ) \; .
\end{equation}
Numerical calculations of the macroscopic polarization based on \eqref{Eq-ElecPol} can be found in \cite{3SP0}.
\hfill $\diamond$
\end{example}

\begin{example}[Orbital magnetization,  \cite{3ST}] {\rm The macroscopic vector of magnetization accepts the formula:
\begin{equation}
M_j=\tfrac{\I}{2}N \, {\mathcal T}\big (|h-\epsilon_F|[\partial_{j+1}p_F,\partial_{j+2}p_F]\big ) \; .
\end{equation}
}
\end{example}

\begin{example}[Magneto-electric response, \cite{3LP,3ProdanSpringer2016ds}] The magneto-electric response tensor is defined as:
\begin{equation}
\alpha_{ij} = \frac{\partial P_i}{\partial B_j} = \frac{\partial M_i}{\partial E_j} \; ,
\end{equation}
where $P$ and $M$ are the macroscopic vectors of electrical polarization and magnetization, already introduced above, while $B$ and $E$ are the vectors of a uniform magnetic and electric field, respectively. The adiabatic variation of the isotropic part $\alpha = \frac{1}{d}\sum_{i=1}^d \alpha_{ii}$ accepts the formula:
\begin{equation}\label{Eq-MagElec}
\Delta \alpha =\tfrac{1}{2}N \, \int_0^T {\rm d} t  \ \sum_{\rho \in S_d}(-1)^\rho \, {\mathcal T} \big ( p_F \prod_{i=1}^{d+1} \partial_{\rho_i}p_F\big ) \; , 
\end{equation}
where the $d+1$ dimension is the time axis. As one can see, if the system is taken in an adiabatic cycle, then the right-hand side becomes the top even Chern number in dimension $d+1$ (provided $d$ is odd). In the presence of time-reversal invariance, this can be used to demonstrate that $\alpha$ can take only integer or half-integer values which gives the $\ZM_2$ classification of the AII class in odd dimensions \cite{3QHZ}. Numerical computations of the magneto-electric response function base on \eqref{Eq-MagElec} can be found in \cite{3Leu}.
\hfill $\diamond$
\end{example}

\chapter{Auxiliary $C^\ast$-Algebras} 

\abstract{This Chapter introduces and characterizes the approximating algebras used later to define the canonical finite-volume approximations and to analyze the numerical errors. In particular, the approximating finite algebra generates the ordinary finite super-cell models with periodic boundary conditions, which are commonly used in the numerical investigations. Formalizing them in an algebraic setting will enable us to bridge more naturally with the thermodynamic limit, to put forward a canonical finite-volume approximation and to ultimately resolve the error estimates with minimal effort. Several key ideas of the Chapter already appeared in \cite{4Pro4} but other are reported here for the first time.}

\section{Periodic Disorder Configurations}\label{Sec-PeriodicDisorder} 

We start by defining the space of periodic disorder configurations $\widetilde \Omega$ and by establishing connections between $\widetilde \Omega$ and $\Omega$, and between the algebras related with the two spaces. In the following, we denote by $V_L$ the cube in $\ZM^d$ centered at the origin and defined by $|x_j|\leq L$, $j = 1, \ldots,d$, $L \in \NM$. It will be often referred to as the super-cell, for obvious reasons. Note that $|V_L| = (2L + 1)^d$ and this somewhat odd choice is because we want the super-cell to be balanced, {\it i.e.} if $x \in V_L$ than also $-x \in V_L$.  

\begin{definition} {\rm The topological space $\widetilde \Omega$ of $(2L +1)$-periodic disorder configurations is defined as the closed subset of $\Omega$:
\begin{equation}
\widetilde \Omega = \big \{ \tilde \omega \in \Omega \ \big | \tau_j^{2L +1} \tilde \omega = \tilde \omega, \ j = 1,\ldots,d \big \} \; .
\end{equation}
This topological sub-space is invariant w.r.t. the $\tau$-action and the restriction of this action to $\widetilde \Omega$ will be denoted by $\tilde \tau$.
}
\end{definition}

When the size of the super-cell needs to be specified explicitly, we will use the more pointed notation $\widetilde \Omega_L$. Next, we endow $\widetilde \Omega$ with a natural $\tilde \tau$-invariant probability measure of full support.

\begin{definition}{\rm Let $\tilde{\mathfrak q} : \Omega \rightarrow \widetilde \Omega \subset \Omega$ be defined by $\tilde{\mathfrak q} \omega = \tilde \omega$, where $\tilde \omega$ is the unique point of $\widetilde \Omega$ such that $\tilde \omega_x = \omega_x$ for all $x \in V_L$. Then $\tilde{\mathfrak q}$ is continuous, because the elementary maps $\Omega \ni \omega \rightarrow \omega_x \in \Omega_0$ are continuous for the product topology on $\Omega$, and $\tilde{\mathfrak q}$ is also onto and idempotent. We endow the topological space $\widetilde \Omega$ with the push-forward measure $\widetilde \PM = \tilde{\mathfrak q}_\ast \PM$, or equivalently:
\begin{equation}
\int_{\widetilde \Omega} {\rm d} \widetilde \PM(\tilde \omega) \, \tilde f(\tilde \omega) = \int_\Omega {\rm d} \PM(\omega) \, \tilde f (\tilde{\mathfrak q}\omega) \; , \quad \forall \, \tilde f \in C(\widetilde \Omega) \; .
\end{equation}
}
Since $\tilde{\mathfrak q}$ is onto, the measure has full support and $\widetilde \PM(\widetilde \Omega) =1$.
\end{definition}

\begin{remark}\label{Re-PerTransInv} {\rm The invariance of the measure ${\rm d} \widetilde \PM$ against the $\tilde \tau$-action does not follow solely from the invariance of the parent ${\rm d} \PM$ against the $\tau$-action, for the simple reason that $\tilde \tau \circ \tilde{\mathfrak q} \neq \tilde{\mathfrak q} \circ \tau$. Nevertheless, the $\tilde \tau$-invariance does follow once the Assumption~\ref{Ass-PermInv} is in place.
\hfill $\diamond$
}
\end{remark}

For the case when ${\rm d} \PM$ is a product measure, the above construction can be characterized more directly.  

\begin{proposition} Let ${\rm d} \PM$ be the product measure from Proposition~\ref{Pro-ProdMeasure}. Then:
\begin{equation}
{\rm d} \widetilde \PM(\tilde \omega) = \prod_{x \in V_L} {\rm d} \PM_0(\tilde \omega_x) \; .
\end{equation}
\end{proposition}

\proof From definition, we have:
\begin{equation}\label{Eq-T15}
\int_{\widetilde \Omega} {\rm d} \widetilde \PM(\tilde \omega) \, \tilde f(\tilde \omega) = \prod_{x \in \ZM^d} \int_{\Omega_0} {\rm d}\PM_0(\omega_x) \, \tilde f (\tilde{\mathfrak q} \omega)  \; .
\end{equation}
Since $\tilde{\mathfrak q} \omega$ is constant as function of $\omega_x$ for $x \notin V_L$, we can integrate out all $\omega_x \notin V_L$. Furthermore, since $\omega_x = (\tilde{\mathfrak q} \omega)_x = \tilde \omega_x$ for $x \in V_L$, the righthand side of \eqref{Eq-T15} is nothing but: 
\begin{equation}
\prod_{x \in V_L} \int_{ \Omega_0}  {\rm d} \PM_0(\tilde \omega_x) \, \tilde f (\tilde \omega) \;,
\end{equation}
and the affirmation follows. \qed

\begin{definition} {\rm The periodic disorder configurations are defined and characterized by the measure-preserving dynamical system $(\widetilde \Omega, \tilde \tau, \ZM^d, {\rm d} \widetilde \PM)$. This system is definitely not ergodic.
}
\end{definition}

We now examine the dual picture. The following statements are obvious and are stated mostly to fix the notation.

\begin{proposition}\label{Pro-OmegaPer} Let $\tilde{\mathfrak q} : \Omega \rightarrow \widetilde \Omega \subset \Omega$ be the map defined above. Then the following hold:

\begin{enumerate}[\rm (i)]

\vspace{0.1cm}

\item $\tilde{\mathfrak q}$ induces the imbedding:
\begin{equation}
\mathfrak i : C_N(\widetilde \Omega) \hookrightarrow C_N(\Omega) , \quad  \mathfrak i (\tilde f) = \tilde f \circ \tilde{\mathfrak q} \; .
\end{equation}

\vspace{0.1cm}

\item There exists the epimorphism of $C^\ast$-algebras:
\begin{equation}
\tilde{\mathfrak p} : C_N(\Omega) \rightarrow C_N(\widetilde \Omega) \; , \quad \big (\tilde{\mathfrak p} (f)\big )(\tilde \omega) = f(\tilde \omega) \; ,
\end{equation} 
such that:
\begin{equation}
\tilde{\mathfrak p} \circ \mathfrak i = {\rm id}_{C_N(\widetilde \Omega)} \;, \quad (\mathfrak i \circ \tilde{\mathfrak p}) = \tilde{\mathfrak q}_\ast \; .
\end{equation}

\vspace{0.1cm}

\item The following identity holds:
\begin{equation}
\int_\Omega {\rm d} \PM(\omega) \, \tr \big ( \mathfrak i (\tilde f) (\omega) \big ) = \int_{\widetilde \Omega} {\rm d} \widetilde \PM(\tilde \omega) \, \tr \big ( \tilde f(\tilde \omega) \big ) \; .
\end{equation}

\end{enumerate}
\end{proposition} 

\begin{remark} {\rm It is important to note, and this will be used later, that $\tilde{\mathfrak p}$ is the dual map of the inclusion $\widetilde \Omega \hookrightarrow \Omega$. Likewise, $\mathfrak i$ is the dual of $\tilde{\mathfrak q}$. Another simple but useful remark is that $\tilde{\mathfrak p}$ commutes with the action of the $\ZM^d$, $\tilde \tau \circ \tilde{\mathfrak p} = \tilde{\mathfrak p} \circ \tau$.
\hfill $\diamond$
}
\end{remark}

The algebra $C_N(\widetilde \Omega)$ can be endowed with the trace:
\begin{equation}\label{Eq-TildeTrace0}
\widetilde \Tt_0(\tilde f) = \int_{\widetilde \Omega} d \widetilde \PM(\tilde \omega) \, \tr \big ( \tilde f(\tilde \omega) \big ) \; .
\end{equation}
Then point (iii) of Proposition~\ref{Pro-OmegaPer} assures us that:
\begin{equation}\label{Eq-TracePer}
\Tt_0 \big (\mathfrak i(\tilde f) \big ) = \widetilde \Tt_0 (\tilde f) \;.
\end{equation}
Furthermore, the trace is faithful, normalized and invariant w.r.t. the $\tilde \tau$-action. Together, $\big ( C(\widetilde \Omega), \tilde \tau, \ZM^d, \widetilde \Tt_0 \big )$ define the dual $C^\ast$-dynamical system of  $(\widetilde \Omega, \tilde \tau, \ZM^d, {\rm d} \widetilde \PM)$.

\section{Periodic Approximating Algebra}\label{Sec-PerAlg}

\begin{definition}
\label{Def-PeriodicBulkAlgebra}
{\rm The periodic algebra is defined as the universal $C^\ast$-algebra:
$$
\tilde \Aa_d=C^\ast \big ( C_N(\widetilde \Omega), u_1,\ldots, u_d \big ) \; ,
$$
with the generators satisfying the old commutation relations but with $\Omega$ and $\tau$ replaced by $\widetilde \Omega$ and $\tilde \tau$, respectively.
}
\end{definition}

\begin{remark}{\rm When the size of the super-cell needs to be specified explicitly, we will use more pointed notation $\tilde \Aa_d^{(L)}$. It is important to note that no additional condition has been imposed on $u_j$'s and that the flux matrix $\phi$ is still taking generic values in Definition~\ref{Def-PeriodicBulkAlgebra}.
\hfill $\diamond$
}
\end{remark}

\begin{proposition}\label{Pro-TildeAlgEpi} The epimorphism defined at point (iii) of Proposition~\ref{Pro-OmegaPer} can be lifted to a map:
\begin{equation}
\tilde{\mathfrak p} : \Aa_d \rightarrow \tilde \Aa_d \;, \quad  \tilde{\mathfrak p} (a)_x= \tilde{\mathfrak p}(a_x) \; ,
\end{equation}
which is a $\ast$-epimorphism of $C^\ast$-algebras. Note that we adopted the same notation because we can easily differentiate between the two maps by checking their argument.
\end{proposition}

\proof (i) We only need to verify the invariance of the commutation relation $f u_j = u_j (f \circ \tau_j )$. This follows from:
\begin{equation}
\tilde{\mathfrak p} (f u_j) = \tilde{\mathfrak p}(f) u_j = u_j \big ( \tilde{\mathfrak p}(f) \circ \tilde \tau_j \big ) = u_j  \, \tilde{\mathfrak p}(f \circ \tau_j) \;,
\end{equation}
where in the last equality we used the fact that $\widetilde \Omega$ and $\tilde{\mathfrak p}$ are invariant w.r.t. the $\tau$-action. Then, since $\tilde{\mathfrak p}$ was a $\ast$-epimorphism in the first place, its lift is indeed a $\ast$-epimorphism of algebras.\qed

\vspace{0.1cm}

The generic elements of the periodic algebra take the form:
 \begin{equation}
 \tilde a = \sum_{x \in \ZM^d} \tilde a_x \, u_x \; , \quad \tilde a_x \in C_N(\widetilde \Omega)\;,
 \end{equation} 
 and they accept similar representations on $\CM^N \otimes \ell^2(\ZM^d)$ as the elements from $\Aa_d$:
\begin{equation}
\tilde \pi_{\tilde \omega}(\tilde a) = \sum_{x,y \in \ZM^d} \tilde a_y(\tilde \tau_y \tilde \omega) \otimes |x \rangle \langle x | U_y \; , \quad \tilde \omega \in \widetilde \Omega\; .
\end{equation} 
Lastly, one can define the derivations via the same procedure:
\begin{equation}
\tilde \partial_j \tilde a = - \I \sum_{x \in \ZM^d} x_j \tilde a_x \, u_x \;,
\end{equation}
and the periodic algebra can be equipped with the trace:
\begin{equation}
\widetilde \Tt(\tilde a) = \widetilde \Tt_0(\tilde a_0) = \int_{\widetilde \Omega} {\rm d}\widetilde \PM(\tilde \omega) \, \tr \big ( \tilde a_0(\tilde \omega) \big ) \; .
\end{equation}
This trace is faithful, normalized and invariant w.r.t. the $\tilde \tau$-action.

\section{Finite-Volume Disorder Configurations} 

Let us start by introducing useful notation. Consider the following exact sequence of abelian groups:
\begin{equation}
\begin{diagram}
0 & \rTo & \big ( (2 L+1)\ZM \big)^d & \rTo{i}  & \ZM^d & \pile{\rTo^{\rm ev} \\  \lTo_{s}} & \widehat \ZM^d=\big ( \ZM \big / (2 L+1)\ZM \big)^d & \rTo & 0 \; .
\end{diagram}
\end{equation}
We will denote the class in $\widehat \ZM^d$ of $x \in \ZM^d$ by $\hat x$. Also, $s$ in the above diagram is the map $s(\hat x) = y$ with $y$ being the unique point in $V_L$ such that $\hat y = \hat x$. This map has all the trades of a splitting map except that it is not a homomorphism.

\begin{definition} {\rm The space  of disorder configurations at finite volume is defined as $\widehat \Omega = \prod_{\hat x \in \widehat \ZM^d} \Omega_0$.  On $\widehat \Omega$, we consider the following action of $\ZM^d$:
\begin{equation}
\hat \tau_y (\hat \omega)=\hat \tau_{y}\{\hat \omega_{\hat x}\}_{\hat x \in \widehat \ZM^d}  = \{\hat \omega_{\widehat{x-y}}\}_{\hat x \in \widehat \ZM^d} \; .
\end{equation}
Note that the action defined above is independent of which representative is used for the class $\hat x$.}
\end{definition}

\begin{proposition}\label{Pro-HatIso1} There exists an isomorphisms of topological dynamical systems: 
\begin{equation}
(\widetilde \Omega, \tilde \tau, \ZM^d) \simeq (\widehat \Omega, \hat \tau, \ZM^d) \; .
\end{equation}
 given by the map: 
\begin{equation} 
\widetilde \Omega \ni \tilde \omega = \{\tilde \omega_x\}_{x \in \ZM^d} \rightarrow \hat {\mathfrak q} \tilde \omega \in \widehat \Omega, \quad (\hat{\mathfrak q} \tilde \omega)_{\hat x}= \tilde \omega_{s(\hat x)}, \quad \hat x \in \widehat \ZM^d\;.  
\end{equation}
and its inverse:
\begin{equation} 
\widehat \Omega \ni \hat \omega = \{\hat \omega_{\hat x}\}_{\hat x \in \widehat \ZM^d} \rightarrow \hat {\mathfrak q}^{-1} \hat \omega \in \widetilde \Omega\; , \quad (\hat {\mathfrak q}^{-1} \hat \omega)_x = \hat \omega_{\hat x}, \quad x \in \ZM^d \; .
\end{equation}
\end{proposition}

\proof We need to show that $\hat q (\tilde \tau_y \tilde \omega) = \hat \tau_y (\hat q \tilde \omega)$ and we have: 
\begin{equation}
\big(\hat q (\tilde \tau_y \tilde \omega)\big)_{\hat x} \;=\; (\tilde \tau_y \tilde \omega)_{s(\hat x)} \;=\; \tilde \omega_{s(\hat x)-y} \;,
\end{equation}
while:
\begin{equation}
\big ( \hat \tau_y (\hat q \tilde \omega) \big)_{\hat x} \;=\; (\hat q \tilde \omega)_{\widehat{x-y}} \;=\; \tilde \omega_{s(\widehat{x-y})} \;.
\end{equation}
The two results coincide because $s(\hat x)-y$ and $s(\widehat{x-y})$ differ by multiples of $2L+1$ and $\tilde \omega$ is $(2L+1)$-periodic. Similar arguments can be used for the inverse map.\qed

\vspace{0.1cm}

Although the two dynamical systems are virtually identical, the two separate definition are still useful. For example, in a numerical calculation, one deals exclusively with  $(\widehat \Omega, \hat \tau, \ZM^d)$. On the theoretical side, the first round of error estimates presented in Chapter~\ref{Cha-FiniteVolApproxI} are established by comparing the full and periodic algebras, hence $(\widetilde \Omega, \tilde \tau, \ZM^d)$ is indeed useful and relevant for our analysis. Now, given the above isomorphism, we can endow $(\widetilde \Omega, \tilde \tau, \ZM^d)$ with the $\hat \tau$-invariant probability measure ${\rm d} \widehat \PM = \hat{\mathfrak q}_\ast {\rm d} \widetilde \PM$.

\begin{definition}{\rm The disorder configurations at finite volume are defined and characterized by the measure-preserving dynamical system $(\widehat \Omega, \hat \tau, \ZM^d,{\rm d} \widehat \PM)$.
}
\end{definition}

\begin{definition}{\rm The dual $C^\ast$-dynamical system associated with $(\widehat \Omega,\hat \tau,\ZM^d, {\rm d} \widehat \PM)$ is defined by $(C_N(\widehat \Omega), \hat \tau,\ZM^d, \widehat \Tt_0)$, where:

\begin{enumerate}[$\ \circ$]

\item $C_N(\widehat \Omega)$ is the $C^\ast$-algebra of continuous functions from $\widehat \Omega$ to $M_N(\CM)$, equipped with the $C^\ast$-norm:
\begin{equation}
\|\hat f\| = \sup_{\hat \omega \in \widehat \Omega} \|\hat f(\hat \omega)\| \; .
\end{equation}

\vspace{0.1cm}

\item The action of $\ZM^d$ on $C_N(\widehat \Omega)$ is given by $\hat \tau_y(\hat f) = \hat f \circ \hat \tau_y$, where on the right we have the old action of $\ZM^d$ on $\widehat \Omega$. As before, we will use the same notation for these two actions.

\vspace{0.1cm}

\item $\widehat \Tt_0$ is the trace 
\begin{equation}
\widehat \Tt_0(\hat f) = \int_{\widehat \Omega} {\rm d} \widehat \PM(\hat \omega)\ \tr\big ( \hat f(\hat \omega) \big ) \; , \quad \hat f \in C_N(\widehat \Omega) \; .
\end{equation} 
This trace is invariant w.r.t. the $\hat \tau$-action, faithful and normalized.

\end{enumerate}
}
\end{definition}

\begin{proposition}\label{Pro-HatIso2} There exists an isomorphisms of $C^\ast$-dynamical systems: 
\begin{equation}
(C_N(\widetilde \Omega),\tilde \tau,\ZM^d,\widetilde \Tt_0) \simeq (C_N(\widehat \Omega),\hat \tau,\ZM^d, \widehat \Tt_0) \; ,
\end{equation}
provided by $\hat{\mathfrak p}(\tilde f) = \tilde f \circ \hat{\mathfrak q}^{-1}$.
\end{proposition}

\proof The checks are straightforward. \qed

\section{Finite-Volume Approximating Algebra}
\label{Sec-FiniteAlg} 

It is here where we define the algebra which can be indeed operated by a computer. Since this algebra is used in the concrete applications, let us be as explicitly as possible.

\begin{definition}
\label{Def-FiniteBulkAlgebra}
{\rm The finite-volume approximating algebra is defined as the universal $C^\ast$-algebra:
$$
\widehat \Aa_d=C^\ast \big ( C_N(\widehat \Omega), \hat u_1,\ldots, \hat u_d \big )
$$
generated by the elements of $C_N(\widehat \Omega)$ and by $\hat u_j$, $j=1,\ldots,d$, satisfying the old commutation relations:
\begin{align*}
&  \hat u_j \hat u_j^\ast\; = \; \hat u_j^\ast \hat u_j\; = 1, \quad & j=1,\ldots,d \, , \\
& \hat u_i \hat u_j\;=\;e^{\I \, \phi_{ij}} \hat u_j \hat u_i \;,  \qquad&  i,j=1,\ldots,d \; ,\\
& \hat f \, \hat u_j \; =\; \hat u_j  \hat \tau_{e_j}(\hat f) \; =\; \hat u_j (\hat f \circ \hat \tau_{e_j})\, , \qquad & \forall \ \hat f \in C_N(\widehat \Omega),  \ \ j=1,\ldots,d\; ,
\end{align*}
together with the additional constraints: 
\begin{align*}
\hat u_j^{2L +1} = 1 \; , \quad j=1, \ldots,d \; .
\end{align*}
}
\end{definition}

As opposed to the approximating periodic algebra, the above commutation relations and the constraints are consistent with each other only if $\phi$ takes quantized values. Indeed, if we raise the commutation relation $\hat u_j^\ast \hat u_i \hat u_j = e^{\I \phi_{ij}} \hat u_i$ to the $(2 L+1)$-th power:
\begin{equation}
(\hat u_j^\ast \hat u_i \hat u_j)^{2 L+1} = e^{\I (2L+1)\phi_{ij}} \hat u_i^{2 L +1} \Leftrightarrow 1= e^{\I (2L+1)\phi_{ij}} \; ,
\end{equation}
or: 
\begin{equation}
\label{Eq-PhiQuant}
\phi_{ij} = \frac{2n_{ij}\pi}{2L +1} \; , \quad n_{ij} \in \ZM \; .
\end{equation}
This quantization condition will be automatically assumed and implied whenever we work with the finite-volume approximating algebra. With this constraint in place, all the commutation relations and the constraints are consistent with each other. For example, if we iterate the commutation relation $\hat u_j^\ast \hat f \, \hat u_j= \hat f \circ \hat \tau_j$ for $2 L +1$ times:
\begin{equation}\label{Eq-QRelation}
(\hat u_j^\ast)^{2 L +1} \hat f \, \hat u_j^{2L +1} \; =\; \hat f \circ \hat \tau_j^{2 L + 1} \; .
\end{equation}
The lefthand side of \eqref{Eq-QRelation} equals $\hat f$, because of our constraints on $\hat u_j$, and the righthand side of \eqref{Eq-QRelation} also equals $\hat f$ because $\hat \tau_j^{2L +1}$ is the identity map on $\widehat \Omega$.

\vspace{0.1cm}

The algebra $\hat \Aa_d$ is generated by $C(\widehat \Omega)$ and by a finite number of monomials: 
\begin{equation}
\hat u_x = e^{\frac{\I}{2} \sum_{i<j} \Phi_{ij}x_i x_j} \, \hat u_1^{x_1} \ldots \hat u_d^{x_d}, \quad x \in V_L \; .
\end{equation}
The monomials $\hat u_x$ obey the following commutations relations:
\begin{equation}
\hat u_x \hat u_y = e^{2\I \, x \wedge y} \hat u_y \hat u_x = e^{\I \, x \wedge y} \hat u_{s(\widehat{x+y})}\; , \quad \forall \ x,y \in V_L \; ,
\end{equation}
and
\begin{equation}
\hat u_x^\ast = \hat u_{-x} \; , \quad \forall \ x \in V_L \; .
\end{equation}
We should note that it is at this point where one appreciates that $V_L$ was chosen to be a balanced set. A generic element of $\hat \Aa_d$ then takes the form:
\begin{equation}\label{Eq-TildeGenericElem}
\hat a = \sum_{x \in V_L} \hat a_x \, \hat u_x \;, \quad \hat a_x \in C_N(\widehat \Omega) \; ,
\end{equation} 
and explicit rules for multiplication and conjugation can be obtained as in \ref{Eq-GenericProd} and \ref{Eq-GenericStar}:
\begin{align}\label{Eq-FiniteAlgProd}
\hat a\,  \hat a' & = \sum_{x,y \in V_L} \hat a_y \, \hat u_y \, \hat a'_{x} \, \hat u_x = \sum_{x,y \in V_L} \hat a_y \, (\hat a'_{x}\circ \hat \tau_{-y}) \, \hat u_y \, \hat u_x\\
\nonumber & = \sum_{x,y \in V_L} e^{i y \wedge x} \hat a_y \, (\hat a'_{x}\circ \hat \tau_{-y}) \, \hat u_{s(\widehat{x+y})} \\
\nonumber & =\sum_{x \in V_L} \Big ( \sum_{y \in V_L} e^{\I \, y \wedge x} \, \hat a_y \, (\hat a'_{s(\widehat{x-y})} \circ \hat \tau_{-y}) \Big ) \hat u_x \; ,
\end{align}
and
\begin{equation}
\hat a^\ast = \sum_{x \in V_L} \hat u_x^\ast \, \hat a_x^\ast = \sum_{x \in V_L} (\hat a_x^\ast \circ \hat \tau_x) \, \hat u_x^\ast = \sum_{x \in V_L} (\hat a_{-x}^\ast \circ \hat \tau_{-x}) \hat u_x \; .  
\end{equation}

\begin{remark}\label{Re-FiniteAlgIssue}{\rm Note that the algebra $\hat A_d$ is not finite because of the $C_N(\widehat \Omega)$ component. Nevertheless, according to \eqref{Eq-FiniteAlgProd}, to evaluate the Fourier coefficient of a product of elements $\hat a \, \hat a' \, \ldots$ at some $\hat \omega \in \widehat \Omega$, the only required input is the value of the Fourier coefficients $\hat a_x$, $\hat a_x'$, $\ldots$, at the translates $\hat \tau_y \hat \omega$, $y \in V_L$, of $\hat \omega$, which are finite in number. This is also reflected in the canonical representation introduced below, which factors over $\widehat \Omega$ and become finite dimensional. As such, the algebra $\hat \Aa_d$ can indeed be manipulated on a computer.
\hfill $\diamond$
}
\end{remark}

\begin{proposition}[Canonical representation]\label{Pro-FiniteVolRep} The following relations define a field $\{\hat \pi_{\hat \omega} \}_{\hat \omega \in \widehat \Omega}$ of $\ast$-representations of $\hat \Aa_d$ on the finite dimensional Hilbert space $\CM^N \otimes \ell^2(V_L)$: 
\begin{align*}
& \hat \pi_{\hat \omega}(\hat u_j)
\;=\; I \otimes \hat U_j
\;, 
& \quad j=1,\ldots,d\;, \\
& \hat \pi_{\hat \omega}(\hat f) 
\;= \;
\sum_{x \in V_L} \hat f(\hat \tau_x \hat \omega) \otimes |x \rangle \langle x|
\;, 
\quad & \forall \ \hat f \in C_N(\widehat \Omega)
\;,
\end{align*}
where $\hat U_j \;=\; e^{\I \, e_j \wedge \hat X} \hat S_j$, $\hat X$ is the position operator on $\ell^2(V_L)$, $\hat X|x\rangle = x |x\rangle$, $x \in V_L$,  and $\hat S_j$, $j = \overline{1,d}$, are the periodic shifts of $\ell^2(V_L)$.
\end{proposition}

For a generic element as in \eqref{Eq-TildeGenericElem}:
\begin{equation}\label{Eq-Rep3}
\hat \pi_{\hat \omega}(\hat a) = \sum_{x,y \in V_L} \hat a_y(\hat \tau_x \hat \omega) \otimes |x\rangle \langle x| \hat U_y \; .
\end{equation}
These expressions generate all the approximating models typically used in the super-cell computer simulations of homogeneous condensed matter systems. Note that the representation still has a covariant property:
\begin{equation}
\hat V_y \hat \pi_{\hat \omega}(\hat a) \hat V_y^\ast = \hat \pi_{\hat \tau_y \hat \omega}(\hat a) \; , \quad \hat V_y \;=\; e^{-\I \, y \wedge \hat X} \hat S_y \; .
\end{equation}

\begin{exercise} The representations $\hat \pi_{\hat \omega}$ are no longer $\widehat \PM$-almost surely faithful. Indeed, the argument used for the representations $\pi_\omega$ of $\Aa_d$ does not apply anymore since the dynamical system $(\widehat \Omega, \hat \tau,\ZM^d,d\widehat \PM)$ is not ergodic. In fact, we can see directly that $\hat \pi_{\hat \omega}(\hat f) =0 $ if $\hat f$ happens to be zero at the finite set of points $\hat \tau_x \hat \omega$, $x \in V_L$.
\hfill $\diamond$
\end{exercise}

\vspace{0.1cm}

Let us conclude this section by stating the essential connection between the periodic and finite approximating algebras. Recall that we automatically assume the quantization \eqref{Eq-PhiQuant} of the flux matrix.

\begin{lemma}\label{Lem-PerFinite} The isomorphism defined in Propositions~\ref{Pro-HatIso1} and \ref{Pro-HatIso2} can be lifted to a map:
\begin{equation}
\hat {\mathfrak p} : \tilde \Aa_d \rightarrow \hat \Aa_d \;, \quad  \hat{\mathfrak p} (u_j)= \hat u_j \; , \quad  \hat{\mathfrak p} (\tilde f) = \tilde f \circ \hat {\mathfrak q}^{-1}\; ,
\end{equation}
which is a $\ast$-epimorphism of $C^\ast$-algebras.
\end{lemma}

\proof (i) We only need to verify the invariance of the commutation relation $\tilde f u_j = u_j (\tilde f \circ \tilde \tau_j )$. This follows from:
\begin{equation}
\hat{\mathfrak p} (\tilde f u_j) = \hat{\mathfrak p}(\tilde f) \hat u_j = \hat u_j \big ( \hat{\mathfrak p}(\tilde f) \circ \hat \tau_j \big ) = \hat u_j  \, \hat{\mathfrak p}(\tilde f \circ \tilde \tau_j) \;,
\end{equation}
where in the last equality we used the fact that $\hat{\mathfrak q}^{-1} \circ \hat \tau = \tilde \tau \circ \hat{\mathfrak q}^{-1}$. The map clearly commutes with the $\ast$-operation and is surjective.\qed

\begin{exercise} There is a simple way to see that $\tilde \Aa_d$ is a much larger algebra then $\hat \Aa_d$. Indeed, the map $\hat{\mathfrak p}$ is not injective for the identity of $\tilde \Aa_d$ and $u_j^{2L+1}$ are both mapped into identity of $\hat \Aa_d$.  As we shall see below, $\tilde \Aa_d$ is actually isomorphic to a field of ``almost" finite-algebras, which we call the Bloch algebras.
\hfill $\diamond$
\end{exercise}

\begin{proposition}\label{Pro-PerFiniteF} For generic elements, we have:
\begin{equation}\label{Eq-HatForierMap}
\tilde \Aa_d \ni \tilde a = \sum_{x \in \ZM^d} \tilde a_x u_x \mapsto \hat{\mathfrak p}(\tilde a) = \sum_{x \in V_L} \hat a_x \hat u_x, \quad \hat a_x = \sum_{y\in \ZM^d,\; \hat y = \hat x} \tilde a_y \circ \hat{\mathfrak q}^{-1} \; .
\end{equation}
\end{proposition}
The above relation will become very useful later as it gives an explicit mapping between the Fourier coefficients of the two algebras.

\section{Approximate Differential Calculus}
\label{Sec-ApproxDiffCalculus}

We now turn our attention to the non-commutative differential calculus. If we examine the real space expressions of the topological invariants in \eqref{Eq-EvenChernNR} and \eqref{Eq-OddChernNR}, we see the commutators with the position operator entering in an essential way. The position operator, however, is incompatible with the periodic boundary conditions and it was not clear how to generate a canonical finite-volume representation of these commutators. It is perhaps at this place where the $C^\ast$-algebraic approach contributed in an essential way to the present computational program. Indeed, recall \eqref{Eq-BulkDerivationRep}, which says that the commutators with the position operator are just the physical representation of the derivations on the infinite volume algebra $\Aa_d$. Then our task is to find the optimal approximation of these derivations and this can be achieved in a surprisingly straightforward manner.

\vspace{0.1cm}

The commutation relations in Definition~\ref{Def-FiniteBulkAlgebra} are still invariant w.r.t. phase twists of Section~\ref{Sec-Fourier}, but the constraints $\hat u_j^{2L+1}=1$ force these phases to take only discrete values, more precisely, $\lambda_j^{2 L +1} = 1$, $j=1,\ldots,d$. Hence, there is no way to define the generators and one cannot define derivations using the phase shifts. The inescapable conclusion is that we have to settle for approximate derivations. An important point to keep in mind is that these approximate derivations can be defined in a natural canonical way.

\begin{definition}[Approximate differential calculus, \cite{4Pro4}] {\rm The approximate differential calculus on $\hat \Aa_d$ is define by the approximate derivations:
\begin{equation}\label{Eq-HatDer}
\hat \partial_j \sum_{x \in V_L} \hat a_x \, \hat u_x = -\I \sum_{x \in V_L} x_j \hat a_x \, \hat u_x \; ,
\end{equation}
and by the trace:
\begin{equation}\label{Eq-HatTr}
\widehat \Tt(\hat a)  = \widehat \Tt_0 (\hat a_0) = \int_{\widehat \Omega} {\rm d} \widehat \PM(\hat \omega) \; \tr \big ( \hat a_0(\hat \omega) \big )\; .
\end{equation}
}
\end{definition}

\begin{remark}{\rm One can check directly that the property of the trace $\widehat \Tt(\hat a \hat a') = \widehat \Tt(\hat a' \hat a)$ is indeed satisfied. Furthermore, as in Example~\ref{Ex-TraceFaith}, one can check explicitly that the trace $\widehat \Tt$ is faithful. The derivations $\hat \partial$, however, do not conform with the Leibniz rule.
\hfill $\diamond$
}
\end{remark}

\begin{remark} {\rm The approximate derivations and the trace are uniquely defined by the natural relations: 
\begin{equation}
(\hat{\mathfrak p} \circ \tilde{\mathfrak p})(\partial^\alpha a) = \hat \partial^\alpha \big ((\hat{\mathfrak p} \circ \tilde{\mathfrak p})(a)\big ) \; , \quad \widehat \Tt\big((\hat{\mathfrak p} \circ \tilde{\mathfrak p})(a)\big)=\Tt(a)\; ,
\end{equation}
whenever the Fourier coefficients $\Phi_x(a)$ vanish for $x \notin V_L$. Above, $\hat{\mathfrak p}$ and $\tilde{\mathfrak p}$ are the epimorphisms  from Propositions~\ref{Lem-PerFinite} and \ref{Pro-TildeAlgEpi}, respectively.
\hfill $\diamond$
}
\label{Re-CanDiffCalc}
\end{remark}

 The expression of the approximate derivations are natural and very simple. Their physical representation, however, has a surprising form, which would have been hard to guess if one insisted on working on the physical Hilbert space.

\begin{proposition}\label{Pro-FVCommutator} Under the canonical representations:
\begin{equation}
\hat \pi_{\hat \omega}(\hat \partial_j \hat a) = \sum_{\lambda^{2L+1}=1} c_\lambda \, \lambda^{\hat X} \hat \pi_{\hat \omega}(\hat a) \lambda^{-\hat X} \; , \quad c_\lambda = \left \{ 
\begin{array}{ll}
\frac{\lambda^{L}}{1-\lambda} , \ \  & \lambda \neq 1\;, \\
0, & \lambda = 1\;.
\end{array}
\right .
\end{equation}
Furthermore:
\begin{equation}\label{Eq-FiniteVolTrace}
\widehat \Tt(\hat a) = \int_{\widehat \Omega} {\rm d} \widehat \PM(\hat \omega) \, \tr \big ( \langle 0 | \hat \pi_{\hat \omega} (\hat a) |0 \rangle \big ) = \frac{1}{|V_L|} \sum_{x \in V_L} \int_{\widehat \Omega} {\rm d} \widehat \PM(\hat \omega) \, \tr \big ( \langle x | \hat \pi_{\hat \omega} (\hat a) |x \rangle \big ) \; .
\end{equation}
\end{proposition}

\proof Recall the identity: 
\begin{equation}
n = \sum_{\lambda^{2 L +1}=1} c_\lambda \lambda^n \; , \quad n \in \{ -L, \ldots, L\}\; ,
\end{equation} 
with the coefficients $c_\lambda$ as given in the statement. Then: 
\begin{align}
\hat \pi_{\hat \omega}(\hat \partial_j \hat a)&  = -\I \sum_{x,y \in V_L} y_j \hat a_y(\hat \tau_x \hat \omega) \otimes |x\rangle \langle x| \hat U_y \\
\nonumber & = -\I \sum_{\lambda^{2 L +1}=1} c_\lambda \sum_{x,y \in V_L}  \hat a_y(\hat \tau_x \hat \omega) \otimes \lambda^{\hat X_j} |x\rangle \langle x| \hat U_y \lambda^{-\hat X_j} \; ,
\end{align}
and the first affirmation follows. The second affirmation follows from the very definition of $\hat \pi_{\hat \omega}$ in Proposition~\ref{Pro-FiniteVolRep} and from the invariance of the measure w.r.t. the $\hat \tau$-action. \qed

\vspace{0.2cm}

Coming back to the discussion from the beginning of the Section, let us point out that Proposition~\ref{Pro-FVCommutator} delivers the long sought canonical approximation of the commutator $\I [\pi_\omega(a),X]$ on finite volumes.

\section{Bloch Algebras}
\label{Sec-BlochAlg} 

If the quantization \eqref{Eq-PhiQuant} is in place, then the periodic algebra $\widetilde \Aa_d$ has a center and it is isomorphic to a continuous field of finite algebras over the torus $\TM^d$. We call the latter the Bloch algebras because of their close relation with the ordinary Bloch-Floquet transformation. We instruct the reader, however, that the Bloch algebras will not play any essential role in our numerical program. The reason is because one wants to avoid the use of both a super-cell and a continuous fibration in the numerical computations. Nevertheless, the Bloch algebras will play an important theoretical role (see Chapter~\ref{Cha-CanAlg}).

\begin{definition}
\label{Def-BlochAlgebras}
{\rm The Bloch algebras are defined as the following universal $C^\ast$-algebras indexed by $k \in \TM^d$:
$$
\hat \Aa_d(k)=C^\ast \big ( C_N(\widehat \Omega), \hat u_1(k),\ldots, \hat u_d(k) \big ) \;,
$$
generated by the elements of $C_N(\widehat \Omega)$ and by $\hat u_j(k)$, $j=1,\ldots,d$, satisfying the old commutation relations:
\begin{align*}
&  \hat u_j(k) \hat u_j(k)^\ast\; = \; \hat u_j(k)^\ast \hat u_j(k)\; = 1\;, \quad & j=1,\ldots,d \; , \\
& \hat u_i(k) \hat u_j(k)\;=\;e^{\I \, \phi_{ij}} \hat u_j(k) \hat u_i(k) \;,  \qquad&  i,j=1,\ldots,d \; ,\\
& \hat f \, \hat u_j(k) \; =\; \hat u_j(k) \hat \tau_{e_j}(\hat f) \; =\; \hat u_j(k) (\hat f \circ \hat \tau_{e_j})\; , \qquad & \forall \ \hat f \in C_N(\widehat \Omega),  \ \ j=1,\ldots,d\; ,
\end{align*}
together with the additional constraints 
\begin{align}\label{Eq-ConstrBloch}
\hat u_j(k)^{2L +1} = e^{ik_j} \; , \quad j=1, \ldots,d \; .
\end{align}
The Bloch algebras are well defined only if the quantization condition \eqref{Eq-PhiQuant} holds.
}
\end{definition}

\begin{lemma}\label{Lem-PerBloch} The isomorphism defined in Proposition~\ref{Pro-HatIso2} can be lifted to a map:
\begin{equation}
\hat {\mathfrak p}_k : \tilde \Aa_d \rightarrow \hat \Aa_d(k) \;, \quad  \hat{\mathfrak p}_k (u_j)= \hat u_j(k) \; , \quad  \hat{\mathfrak p}_k (\tilde f) = \tilde f \circ \hat {\mathfrak q}^{-1}\; ,
\end{equation}
which is a $\ast$-epimorphism of $C^\ast$-algebras.
\end{lemma}

\proof The checks are similar to the proof of Lemma~\ref{Lem-PerFinite}. \qed

\begin{proposition}\label{Pro-BlochFloquet} Let $\tilde a \in \tilde A_d$ and $\hat a_k = \hat{\mathfrak p}_k (\tilde a) \in \hat \Aa_d(k)$, as well as $\hat \omega = \hat{\mathfrak q} \tilde \omega$. Then the following relations between the Fourier coefficients hold:
\begin{equation}\label{Eq-Bloch1}
\hat a_{k,x}(\hat \omega) = \sum_{y \in \ZM^d} e^{\I k \cdot y} \, \tilde a_{x+(2L+1)y}(\tilde \omega) \; , \quad x \in V_L \; ,
\end{equation}
and
\begin{equation}\label{Eq-Bloch2}
\tilde a_{x+(2L+1)y}(\tilde \omega) = \tfrac{1}{(2\pi)^d}\int_{\TM^d} {\rm d} k \, e^{- \I k\cdot y} \hat a_{k,x}(\hat \omega) \; , \quad x \in V_L, \quad y \in \ZM^d \; .
\end{equation}
\end{proposition}

\proof The identity \eqref{Eq-Bloch1} follows from the very definition of $\hat{\mathfrak p}_k$ and the constraints \eqref{Eq-ConstrBloch}, while \eqref{Eq-Bloch2} is just the Fourier inverse of \eqref{Eq-Bloch1}. Note that the above identities are compatible with those in Proposition~\ref{Pro-PerFiniteF}, which deals with the case $k=0$.\qed 

\begin{theorem}[Bloch decomposition]\label{Th-BlochDeco} Let:
\begin{equation}
\Gamma = \Big \{ \mathfrak s \in \prod_{k \in \TM^d } \hat \Aa_d(k), \  \mathfrak s (k)=\sum_{x \in V_L} \hat a_x(k) \hat u_x(k) \ \big | \ \hat a_x \in C \big (\TM^d, C_N(\widehat \Omega) \big ) \Big \} \; .
\end{equation} 
Then $\big (\{\hat \Aa_d(k)\}_{k \in \TM^d }, \Gamma \big )$ defines a field of $C^\ast$-algebras, as defined in \cite{4Dix}, which is isomorphic with $\tilde A_d$. The isomorphism is defined by the relations \eqref{Eq-Bloch1} and \eqref{Eq-Bloch2}.
\end{theorem}

\proof We only need to make sure that the Fourier coefficients in \eqref{Eq-Bloch1} are continuous of $k$. The continuity property is evident for non-commutative polynomials and, since they are dense in $\tilde \Aa_d$, the continuity follows in general. \qed

\begin{remark} {\rm The Bloch decomposition is related to the fact that the non-commutative torus has a large center whenever $\phi$ has only rational entries and that the center is isomorphic to $C(\TM^d)$ \cite{4LLS}.
\hfill $\diamond$
}
\end{remark}

\chapter{Canonical Finite-Volume Algorithms}
\label{Cha-CanAlg} 

\abstract{As we have seen in Section~\ref{Sec-PhysFormulas}, the response functions and the various thermodynamic coefficients are expressed as correlation functions of the type:
\begin{equation}\label{Eq-CorrFunc}
\Tt \big ( \partial^{\alpha_1} G_1(h) \, \partial^{\alpha_2} G_2(h) \ldots \big ) \; , \quad h \in \Aa_d \; ,
\end{equation}
and variations of it, such as the Kubo formula treated in Section~\ref{Sec-KuboFormula}. In this Chapter, we use the previously introduced auxiliary algebras and propose a canonical finite-volume algorithm for computing the above correlation functions. Our main goals for the Chapter are to integrate the algorithm in a broader context and then to summarize for the reader the concrete steps of the algorithm, as applied to homogeneous disordered crystals. The error estimates are the subjects of the following Chapters.}

\section{General Picture}

The role of finite algebra approximations in computer assisted calculations has been recognized for some time \cite{5Ave,5Lan,5LLS,5Bro,5BO}. Their use for the rotation algebras has been essentially made possible by the works \cite{5PV,5LLS} (see also \cite{5Dav} for an extremely pedagogical exposition). Relevant for our discussion are the almost-finite (AF) algebras, generated by limits of inductive towers of finite $C^\ast$-algebras:
\begin{equation}
\begin{diagram}
\Aa = \varinjlim \, \Aa_n \; , \quad \Aa_0 & \rInto{\mathfrak i_0} & \Aa_1 & \rInto{\mathfrak i_1}  & \ldots & \rInto{\mathfrak i_{n-1}} & \Aa_n & \rInto{\mathfrak i_n} & \ldots \; .
\end{diagram}
\end{equation}
Since we will refer to them later, let us also introduce the class of $C^\ast$-algebras obtained as limits of projective towers of $C^\ast$-algebras (finite or not):
\begin{equation}
\begin{diagram}
\Aa = \varprojlim \, \Aa_n \; , \quad \Aa_0 & \lOnto{\ \ \ \mathfrak p_0} & \Aa_1 & \lOnto{\ \ \ \mathfrak p_1}  & \ldots & \lOnto{\ \ \ \mathfrak p_{n-1}} & \Aa_n & \lOnto{\ \ \ \mathfrak p_n} & \ldots \; .
\end{diagram}
\end{equation}
Note that the category of $C^\ast$-algebras is complete, hence both the inductive and projective limits exist and are essentially unique \cite{5War}. Still, it is sometimes more natural to consider the projective limits of $C^\ast$-algebras in the category of topological $\ast$-algebras \cite{5Phi1,5Phi2}. This will be avoided here but let us note that the works \cite{5Phi1,5Phi2} on the latter are also useful for understanding the projective limits in the category of $C^\ast$-algebras.

\vspace{0.1cm}

If we were dealing with AF-algebras, then the task of defining the canonical finite-volume approximations will be reduced to finding explicit representations of the algebras $\Aa_n$ and of the connecting homomorphisms. However, regardless of the rational/irrational character of $\phi$, the non-commutative tori are not AF-algebras, a fact that can be seen by examining the $K_1$-groups, which for the former algebras are isomorphic to $\ZM^{2^{d-1}}$, independently of $\phi$ \cite{5ProdanSpringer2016ds}, while for the latter they are always trivial. This topological obstruction, however, can be circumvented by seeking AF-algebras which, instead of being Morita equivalent, they only imbed the non-commutative tori. Indeed, for the purpose of generating finite approximations, this is equally fine. In $d=2$ and for irrational values of $\phi$, such program was accomplished by Pimsner and Voiculescu in \cite{5PV}. These authors were not concerned that much about the computer assisted calculations but Landi, Lizzi and Szabo, the authors of \cite{5LLS}, were. One of their main goal was to put the irrational and rational non-commutative tori on the same footing, by completing the program from \cite{5PV} for rational values of $\phi$, and then to give a uniform algorithm for computing the correlation functions using finite approximates. We should point out that, since there are qualitative differences between the rational and irrational rotational algebras, {\it e.g.} the former has a large center while the latter is always central simple, the results of \cite{5LLS} cleared a longstanding puzzle by showing how and why no discontinuities occur when switching from rational to irrational $\phi$'s. 

\vspace{0.1cm}

Our approach differs from the finite approximations of \cite{5PV,5LLS}, mainly because we are dealing with the disordered non-commutative tori, hence we need to devise approximations for $C(\Omega)$ and for the non-commutative tori, which at the end need to be compatible with each other. The two-step approach hinted by the previous Chapter will enable us to deal with these two problems separately. Indeed, note that the non-commutative torus was left untouched when we introduced the approximating periodic algebras $\tilde \Aa_d$. As for $C(\Omega)$, recall that it was approximated by $C(\widetilde \Omega)$ and we can easily build on Sections~\ref{Sec-PeriodicDisorder} and \ref{Sec-PerAlg} and integrate the $C(\widetilde \Omega)$'s in a projective tower of $C^\ast$-algebras. For example, by considering an increasing sequence: 
\begin{equation}\label{Eq-Lk}
L_{n+1} = c L_n + \frac{c}{2}-\frac{1}{2} \; ,
\end{equation} 
with $c \in \NM$ odd, so that the sizes of the super-cells obey the relation $2L_{n+1} + 1 = c(2 L_n +1)$, we obtain an inductive tower of compact topological spaces:
\begin{equation}
\widetilde \Omega_{L_0} \subset \widetilde \Omega_{L_1} \ldots \subset \widetilde \Omega_{L_n} \ldots \; ,
\end{equation}
whose dual gives a projective tower of $C^\ast$-algebras:
\begin{equation}
\begin{diagram}
C(\widetilde \Omega_{L_0}) & \lOnto{\ \ \ \tilde{\mathfrak p}_0} & \widetilde C(\widetilde \Omega_{L_1}) & \lOnto{\ \ \ \tilde{\mathfrak p}_1}  & \ldots & \lOnto{\ \ \ \tilde{\mathfrak p}_{n-1}} & C(\widetilde \Omega_{L_n}) & \ldots \; ,
\end{diagram}
\end{equation}
whose limit is known to be $C(\Omega)$. The construction carries over to the periodic algebras:
\begin{equation}
\begin{diagram}
\tilde \Aa_d^{(L_0)} & \lOnto{\ \ \ \tilde{\mathfrak p}_0} & \tilde \Aa_d^{(L_1)} & \lOnto{\ \ \ \tilde{\mathfrak p}_1}  & \ldots & \lOnto{\ \ \ \tilde{\mathfrak p}_{n-1}} & \tilde \Aa_d^{(L_n)} & \ldots \; ,
\end{diagram}
\end{equation}
hence we can recover the algebra of the physical observables as the projective limit $\Aa_d = \varprojlim \, \tilde \Aa_d$, and this summarizes the strategy of the first step of our approximation.

\vspace{0.1cm}

In the second step, we propose to work with the finite-volume approximating algebra introduced in Section~\ref{Sec-FiniteAlg}, which leaves $C(\widetilde \Omega)$ unchanged (see Proposition~\ref{Pro-HatIso2}) but modifies the non-commutative torus. To understand this step, let us assume that we are trying to compute correlation functions for a magnetic field with $\phi_{ij} = \frac{2n_{ij}\pi}{M}$, for some fixed integers $M$ and $n_{ij}$. If $\hat \Uu_d^{(L)}$ denotes the $C^\ast$-algebra generated by the $\hat u_j$'s from Section~\ref{Sec-FiniteAlg}, then, by taking $2L_0+1$ a multiple of $M$ and with $L_n$'s as in \eqref{Eq-Lk}, we can generate the projective tower:
\begin{equation} 
\begin{diagram}
\hat \Uu_d^{(L_0)} & \lOnto{\ \ \ \hat{\mathfrak p}_0} & \hat \Uu_d^{(L_1)} & \lOnto{\ \ \ \hat{\mathfrak p}_1}  & \ldots & \lOnto{\ \ \ \hat{\mathfrak p}_{n-1}} & \hat \Uu_d^{(L_n)} & \ldots \; .
\end{diagram}
\end{equation}
Its limit is generated by the coherent sequences 
\begin{equation}
\{\hat u^{(L_0)}_j,\hat u^{(L_1)}_j, \ldots \}, \quad j = 1,\ldots,d \; ,
\end{equation}
which are unitary operators obeying the commutation relations of the non-commutative torus. In other words, the projective limit coincides with $C^\ast(u_1, \ldots, u_d)$.

\vspace{0.1cm}

If we now combine both steps, for same fixed quantized $\phi$, we can generate the projective tower:
\begin{equation}\label{Eq-FinalTower}
\begin{diagram}
\hat \Aa_d^{(L_0)} & \lOnto{\ \ \ \tilde{\mathfrak p}_0} & \hat \Aa_d^{(L_1)} & \lOnto{\ \ \ \tilde{\mathfrak p}_1}  & \ldots & \lOnto{\ \ \ \tilde{\mathfrak p}_{n-1}} & \hat \Aa_d^{(L_n)} & \ldots \; ,
\end{diagram}
\end{equation}
and recover the algebra of physical observables as its limit, $\Aa_d = \varprojlim \,  \hat \Aa_d$. We recall that the algebras in the tower are not finite but they accept canonical finite dimensional representations. This summarizes the first part of our general strategy for computing the correlations functions for quantized flux matrices $\phi$.  

\begin{remark}{\rm We recall that the strategy in \cite{5LLS} was to imbed the non-commutative torus in an AF-algebra, hence the approximation scheme would be captured by an inductive tower. Instead, our strategy is based on a projective tower of approximations. The latter has the extremely important property that the resolvent set of an element from $\Aa_d$ is not tainted by spurious spectrum while moving down the approximating tower. This is particularly important for topological insulators, which are prone to such effects if (unwanted) defects are introduced. 
\hfill $\diamond$
}
\end{remark}

\begin{exercise}{\rm When the flux is quantized, we can generate an exact finite-volume representation of the non-commutative torus using the Bloch algebras. However, the presence of disorder forces on us to take the infinite-volume limit anyway. It is then instructive to look at the behavior of the Bloch algebras in this limit.  Recall Theorem~\ref{Th-BlochDeco}, which assures us that, once the flux is quantized, the non-commutative torus $C^\ast(u_1, \ldots, u_d)$ is isomorphic with the field of Bloch algebras $\{\hat \Uu_d^{(L_n)}(k)\}_{k \in \TM^d}$ generated by the $\hat u_j(k)$'s from Section~\ref{Sec-BlochAlg} and with $L$ as in \eqref{Eq-Lk}. This isomorphism holds for each $n$ and there exists an isomorphism between the fields of Bloch algebras given by:
\begin{equation}
\Big \{\sum_{x \in V_{L_n}} \hat a_{k,x} \hat u^{(L_n)}_x(k) \Big \}_{k \in \TM^d} \mapsto 
\Big \{\sum_{x' \in V_{L_{n+1}}} \hat a'_{k,x'} \hat u^{(L_{n+1})}_{x'}(k) \Big \}_{k \in \TM^d} \; ,
\end{equation}
where the connection between the Fourier coefficients can be easily found from Proposition~\ref{Pro-BlochFloquet}:
\begin{equation}
\hat a'_{k,x'} = \tfrac{1}{c}\sum_{s \in \{0,\ldots,c-1\}^d} \, e^{- \I (k/c-2\pi s/c)\cdot r} \hat a_{k/c-2\pi s/c,x} \; ,
\end{equation}
where $r \in \{0,\ldots,c-1\}^d$ and $x \in V_{L_n}$ are the unique integers such that $x' = x + (2L_n +1) r$. It is convenient to redefine the quantization condition as $\hat u_j(k)^{2L+1} = e^{\I k \cdot (2L+1)}$, so that $k$ becomes a point of the torus $L^{-1} \, \TM^d$. In this case, the relation between the Fourier coefficients becomes:
\begin{equation}
\hat a'_{k,x'} = \tfrac{1}{c}\sum_{s\in \{0,\ldots,c-1\}^d} \, e^{- \I (k-2\pi s/c)\cdot r} \hat a_{k-2\pi s/c,x} \; , \quad k = L_{n+1}^{-1} \, \TM^d \;, 
\end{equation}
and now we can see explicitly that the dependence on $k$ of the Fourier coefficients becomes weaker and weaker as the infinite-volume limit is approached. It is precisely this effect which enables us to actually achieve fast convergent algorithms using just the Bloch algebra at $k=0$.  
\hfill $\diamond$
}
\end{exercise}

Let us recall that, in order to compute the correlations \eqref{Eq-CorrFunc}, we actually need a finite-volume approximation of the entire non-commutative Brillouin torus $(\Aa_d,\partial,\Tt)$. The projective tower of approximations seems to hold the key of this aspect too, because, according to Remark~\ref{Re-CanDiffCalc}, the optimal approximations $\hat \partial$ and $\widehat \Tt$ are uniquely determined by the epimorphism $\hat{\mathfrak p}$. For us, this is an indication that the canonical approximation of the differential calculus found in Section~\ref{Sec-ApproxDiffCalculus} is not bound to our particular context. 

\vspace{0.1cm}

We now open the issue of irrational values of $\phi$. Of course, every irrational number can be approximated by the rationals and a concrete algorithm based on continued fraction expansions can be found {\it e.g.} in \cite{5LLS}. In practice we have never made appeal to it because, according to Proposition~\ref{Pro-BSmooth}, the correlation functions involving smooth $G$'s are smooth functions of $\phi$. Using techniques similar with the ones employed in Section~\ref{Sec-Sobolev}, it seems that a similar conclusion can be reached when the $G$'s are discontinuous but the discontinuities occur in the Anderson localized spectrum. It is important to mention here that disorder has a smoothening effect. This effect can be witness, for example, in the plots of the density of states shown in Fig.~\ref{HallMap}. As a result, we can use various extrapolations schemes to fill the entire range of $\phi$ once the correlation functions were mapped for a dense enough rational values. 

\vspace{0.1cm}

While \eqref{Eq-FinalTower} may appear as a merely reformulation of the previous Sections, we believe that it incapsulates the essence of a generic program in non-commutative geometry as applied to homogeneous condensed matter, and it is our hope that concrete computer implementations can be achieved for the already existing theoretical works mentioned at the beginning of Section~\ref{Sec-DiffCalculus}. 

\section{Explicit Computer Implementation}
\label{Sec-ExplImple}

We now return to our particular context and discuss the steps of the algorithm in detail. The input for such calculation consists of:
\begin{enumerate}

\item A uniform magnetic field characterized by the flux matrix $\phi$.

\vspace{0.1cm}

\item The disorder configuration space $\Omega$ equipped with a probability measure.

\vspace{0.1cm}

\item A covariant family of Hamiltonians $H_\omega$ as in \eqref{Eq-GenericDisModel}.

\end{enumerate}
Then the goal is to compute correlation functions of the type:
\begin{equation}
\lim_{V \rightarrow \ZM^d} \tr_V \big( [[[G_1(H_\omega),X_{\alpha_{1,1}}],X_{\alpha_{1,2}},\ldots] \,    [[[G_2(H_\omega),X_{\alpha_{2,1}}],X_{\alpha_{2,2}}],\ldots] \ldots \big ) \; ,
\end{equation}
where $G$'s satisfy certain regularity conditions, to be specified precisely in the following Chapters. It is also important to mention that the number of terms inside the trace is finite. In broad general lines, the proposed algorithm proceeds as follows:
\begin{enumerate}[1.]

\item Choose $L$ and quantize the flux matrix as in \eqref{Eq-PhiQuant}.

\vspace{0.1cm}

\item Identify the element $h \in \Aa_d$ which generates the family of Hamiltonians.

\vspace{0.1cm}

\item Define $\hat h = (\hat{\mathfrak p} \circ \tilde{\mathfrak p})(h) \in \hat \Aa_d$ using the epimorphisms defined in the previous section.

\vspace{0.1cm}

\item Compute $G_i(\hat h)$ using various methods of functional calculus.

\vspace{0.1cm}

\item Evaluate the approximate derivations $\hat \partial^{\alpha_i} G_i(\hat h)$.

\vspace{0.1cm}

\item Evaluate the trace:
\begin{equation}
\widehat \Tt \big (\hat \partial^{\alpha_1}G_1(\hat h) \ldots \hat \partial^{\alpha_n} G_n (\hat h) \big ) \; .
\end{equation}

\vspace{0.1cm}

\item Repeat steps 2-6 for a sequence of quantized flux matrices centered around $\phi$ and extrapolate at $\phi$.

\vspace{0.1cm}

\item Repeat steps 1-7 for an increasing sequence of $L$'s in order to probe the convergence towards the thermodynamic limit.

\end{enumerate}

Some of the above steps seems obvious, from a pure computational point of view, but some are not. For example, the steps 2 and 3 are quite obvious due to the explicit formulas \eqref{Eq-HGenerator1} and \eqref{Eq-HatForierMap},  but step 4, for example, can be somewhat involved. The smooth functions can be approximated in norm by polynomials and this provides an obvious route for evaluating a given smooth function $G$ on an element $\hat h \in \hat \Aa_d$ (see Remark~\ref{Re-FiniteAlgIssue}). The functional calculus with piecewise smooth functions can be reduced to previous case since, in the topology of $\bar \Aa_d^\infty$, $G(\hat h)$ can be approximated by the functional calculus with smooth functions, whenever the discontinuities of the functions occur in the mobility gaps of the Hamiltonians (see Section~\ref{Sec-Sobolev}). Step 5 is again straightforward due to the explicit formula in \eqref{Eq-HatDer}  but step 6, while formally very simple, see \eqref{Eq-HatTr}, it is involved because one needs to approximate the integration over $\hat \omega$.  

\vspace{0.1cm}

 For the reasons expressed above, we provide below the exact computational method used in all the applications presented here, which is actually based on the physical representations:
 \begin{enumerate}[1.]

\item Choose $L$ and quantize the flux matrix as in \eqref{Eq-PhiQuant}.

\vspace{0.1cm}

\item Identify the element $h \in \Aa_d$ which generates the family of Hamiltonians.

\vspace{0.1cm}

\item Define $\hat h = (\hat{\mathfrak p} \circ \tilde{\mathfrak p})(h) \in \hat \Aa_d$ as above.

\vspace{0.1cm}

\item Represent $\hat h$ on the finite dimensional Hilbert space $\CM^N \otimes \ell^2(V_L)$:
\begin{equation}
\hat \pi_{\hat \omega}(\hat h) = \sum_{x,y \in V_L} \hat h_y(\hat \tau_x \hat \omega) \otimes |x\rangle \langle x| \hat U_y \; .
\end{equation}

\item Diagonalize $\hat H_{\hat \omega}= \hat \pi_{\hat \omega} (\hat h)$, using {\it e.g.} Intel's Math Kernel Library.

\vspace{0.1cm}

\item Compute $G_i(\hat H_{\hat \omega})$ using the spectral theorem.

\vspace{0.1cm}

\item Evaluate the approximate derivations $\hat \partial^{\alpha_i} G_i(\hat H_{\hat \omega})$ using Proposition~\ref{Pro-FVCommutator}.

\vspace{0.1cm}

\item Evaluate the trace using \eqref{Eq-FiniteVolTrace}:
\begin{equation}\label{Eq-FTrace}
\widehat \Tt (\hat a) = \lim_{M\rightarrow \infty} \tfrac{1}{M}\sum_{i = 1}^M \tr_{V_L} \big ( \hat \pi_{\hat \omega_i}(\hat a) \big ) \; ,
\end{equation}
where $\{\hat \omega_i\}_{i=\overline{1,M}}$ is an independently generated sequence of disorder configurations.

\vspace{0.1cm}

\item Repeat steps 2-8 for a sequence of quantized flux matrices centered around $\phi$ and extrapolate at $\phi$.

\vspace{0.1cm}

\item Repeat steps 1-9 for an increasing sequence of $L$'s in order to probe the convergence towards the thermodynamic limit.

\end{enumerate}

As we shall see in the concrete applications, if the size of the simulation super-cell is larger than Anderson's localization length, the fluctuations of the summand in \eqref{Eq-FTrace} are very small from one disorder configuration to another, due to the self-averaging property of the covariant observables. As such, in practice, a small $M$ is sufficient to converge the calculations. 

\vspace{0.1cm}

Let us conclude by mentioning that, besides the algorithms based on the physical representation, we have investigated algorithms which implemented directly the algebraic operations and proceeded with the functional calculus as outlined in Remark~\ref{Re-FiniteAlgIssue}. These algorithms were found to be efficient and precise when computing the transport and response coefficients at high temperatures. In the zero temperature regime, their performance was rather poor. 

\bibliographystyle{plain}

\chapter{Error Bounds for Smooth Correlations}
\label{Cha-FiniteVolApproxI}

\abstract{In this Chapter we derive upper bounds on the error introduced by restricting to a finite-volume, and show that they decay to zero faster than any inverse power of the volume. The error bounds derived here are relevant for the finite-temperature correlation functions, which in general involve only the smooth functional calculus with the Hamiltonian. In these cases, the fast convergence to the thermodynamic limit of the numerical algorithms is not conditioned by the localized or delocalized character of the energy spectrum. This is important, for example, when studying the metal-insulator transition, where inherently the character of the spectrum changes.}

\section{Assumptions}\label{Sec-Assumptions} 

Here we state the precise assumptions about the infinite volume model. They are:
\begin{enumerate}[{\rm a}1.]

\item The Hamiltonian $h$ belongs to the smooth algebra $\Aa_d^\infty$. 

\vspace{0.1cm}

\item The Fourier coefficients of the Hamiltonian are localized w.r.t. $\omega$, in the sense that, for any $K \in \NM$, there exists a positive and finite $A_K$ such that:
\begin{equation}
\| h_x(\omega) - h_x(\omega') \| \leq \frac{A_K}{(1 + |V_M|)^K} \; , \quad \forall x \in \ZM^d \;, 
\end{equation}
whenever $\omega_y = \omega'_y$ for $y \in V_M$, $M \in \NM$. The norm on the left is the matrix norm.
\end{enumerate}

Let us point out that most if not all the Hamiltonians used in concrete applications are actually of finite-range, hence assumption a1 is not restrictive at all. Also, condition a2 says that the dependence of the Fourier coefficients on $\omega_x$ become weaker and weaker as $|x|$ becomes larger and larger. Examining the canonical representation from Proposition~\ref{Pro-CanRep}, this means that the fluctuations in the hopping matrices connecting to the site $x$ of the lattice depend primarily on the local environment, that is, on $\omega_y$'s with $y$ in a neighborhood of $x$. The condition is definitely satisfied by the linearized disordered Hamiltonians from \eqref{Eq-LinDisHam} and is also expected to hold for the Hamiltonian derived from first principles.

\begin{proposition} \label{Pro-CondA12} The assumptions a1 and a2 are stable under multiplication. Furthermore, if $a \in \Aa_d$ is invertible and satisfies a1 and a2, the $a^{-1}$ also satisfies a1 and a2. In other words, a1 and a2 are stable under the inversion.
\end{proposition}

\proof Assume $a,a' \in \Aa_d$ satisfy a1 and a2. Since $\Aa_d^\infty$ is known to be a Frech\'et sub-algebra which is stable under the holomorphic calculus \cite{6Con}, condition a1 is automatically stable under multiplication and inversion. Now, consider $\omega,\omega' \in \Omega$ such that $\omega_x = \omega'_x$ for all $x \in V_M$. Then, using \eqref{Eq-GenericProd}:
\begin{align}
(aa')_x(\omega) - (aa')_x(\omega')  & = \sum_{y \in \ZM^d} e^{\I \, y \wedge x} \, \big (a_y(\omega)-a_y(\omega') \big ) \, a'_{x-y}(\tau_{-y} \omega) \\
\nonumber & + \sum_{y \in \ZM^d} e^{\I \, y \wedge x} \, a_y(\omega') \, \big (a'_{x-y}(\tau_{-y}\omega) - a'_{x-y}(\tau_{-y} \omega') \big ) \; .
\end{align}
Note that $(\tau_{-y}\omega)_x$ and $(\tau_{-y} \omega')_x$ coincide for $x \in V_{M-\bar y}$, with $\bar y = \max |y_i|$. In the following, we will use the convention that $V_{M-\bar y} = \emptyset$ if $M-\bar y < 0$. Then:
\begin{align}\label{Eq-Step1}
\big \| (aa')_x(\omega) - (aa')_x(\omega') \big \| &\leq A_K\sum_{y \in \ZM^d} \Big ( \frac{\|a'_{x-y}\|}{(1+|V_M|)^K} + \frac{\|a_y \|}{(1+|V_{M-\bar y}|)^K} \Big ) \; .
\end{align}
At this point, we use $|V_{M}| \leq |V_{M-\bar y}|+|V_M|\, |V_{\bar y}|$ together with the rapid decay of $\|a_y\|$ in the following way:
\begin{equation}
\frac{\|a_y \|}{(1+|V_{M-\bar y}|)^K} \leq \frac{{\rm ct.}}{(1+|V_{\bar y}|)^{K+2}(1+|V_{M-\bar y}|)^K} \leq \frac{{\rm ct.}}{(1+|V_{\bar y}|)^{2} (1+|V_{M}|)^K} \; .
\end{equation}
Plugging this back into \eqref{Eq-Step1} and using the fast decay of $\|a'_{x-y}\|$, we finally obtain:
\begin{equation}
\big \| (aa')_x(\omega) - (aa')_x(\omega') \big \| \leq \frac{\rm ct.}{(1+|V_{M}|)^K} \;,
\end{equation}
and the first affirmation follows.

\vspace{0.1cm}

For the second affirmations, we write:
\begin{equation}
a^{-1}_x(\omega')-a^{-1}_x(\omega) = \langle 0 | \pi_\omega(a^{-1})\big (\pi_\omega(a) - \pi_{\omega'}(a) \big ) \pi_{\omega'}(a^{-1})|-x \rangle \; .
\end{equation}
Writing out explicitly:
\begin{equation}
a^{-1}_x(\omega')-a^{-1}_x(\omega) = \sum_{y,y' \in \ZM^d} e^{\I \varphi}a_y^{-1}(\omega)\big (a_{y'}(\tau_{-y} \omega) - a_{y'}(\tau_{-y} \omega') \big ) a_{x-y-y'}^{-1}(\tau_{y + y'}\omega')\; ,
\end{equation}
where $e^{\I \varphi}$ is just a phase factor. Then:
\begin{equation}
\|a^{-1}_x(\omega')-a^{-1}_x(\omega)\| \leq \sum_{y,y' \in \ZM^d} \|a_y^{-1}\|\big \| a_{y'}(\tau_{-y} \omega) - a_{y'}(\tau_{-y} \omega') \big \| \|a_{x-y-y'}^{-1}\| \; .
\end{equation}
Note again that $(\tau_{-y}\omega)_x$ and $(\tau_{-y} \omega')_x$ coincide for $x \in V_{M-\bar y}$ and that $a^{-1} \in \Aa_d^\infty$, hence its Fourier coefficients have the rapid decay property. Then:
\begin{equation}
\|a^{-1}_x(\omega')-a^{-1}_x(\omega) \| \leq {\rm ct.}\sum_{y,y' \in \ZM^d} \frac{\|a_{x-y-y'}^{-1}\|}{(1+|V_{\bar y}|)^{K+2} (1+|V_{M-\bar y}|)^K}  \; .
\end{equation}
Then the affirmation follows by using the same manipulations as above, together with the rapid decay of $\|a_{x-y-y'}^{-1}\|$.\qed

\vspace{0.1cm}

The above statement assures us that the elements which satisfies a1 and a2 form a sub-algebra which is stable under the holomorphic functional calculus and also under the $C^\infty$-functional calculus with self-adjoint elements.

\section{First Round of Approximations} 

In this Section, we derive an upper bound on the error induced by replacing the algebra $\Aa_d$ with the periodic algebra $\tilde \Aa_d$.
 
\begin{theorem}\label{Th-Bounds1} Let $h \in \Aa_d$ be a Hamiltonian in the infinite-volume setting and assume a1-a2 from above. Recall the epimorphism $\tilde{\mathfrak p}$ from Proposition~\ref{Pro-TildeAlgEpi} and define $\tilde h \in \tilde A_d$ as $\tilde h = \tilde{\mathfrak p}(h)$. Then, for any $K \in \NM$, there exists the finite constant $A_K$ such that:
\begin{equation}
\Big | \Tt \big (\partial^{\alpha_1} G_1(h) \ldots \partial^{\alpha_n}G_n(h) \big )- \widetilde \Tt \big (\tilde \partial^{\alpha_1} G_1(\tilde h) \ldots \tilde \partial^{\alpha_n}G_n(\tilde h) \big ) \Big | \leq \frac{A_K}{(1+ |V_L|)^K} \; ,
\end{equation}
where $G_i:\RM \rightarrow \CM$ are are smooth functions over $\sigma(h)$ and $\alpha_i$'s are multi-indices. 
\end{theorem}

\begin{remark} {\rm The statement tells that the convergence to the thermodynamic limit of the periodic approximation is faster than any inverse power of the super-cell's size. Note that the constants $A_K$ depend on all the inputs, that is, $h$, $\alpha_i$'s and $G_i$'s. For the particular case of disordered Hofstadter model, these constants were evaluated in \cite{6Pro4}. No such attempt will be pursued here.
\hfill $\diamond$
}
\end{remark}

\proof  The proof of the statement is surprisingly short, in great part due to the algebraic connections established so far. Recall that $\sigma(\hat h) \subset \sigma(h)$, hence $G$'s are smooth over $\sigma(\hat h)$. Note that, since $\tilde{\mathfrak p}$ is a homomorphisms, it commutes with the functional calculus and, as a consequence:
\begin{equation}
\tilde \partial^{\alpha_i} G_i \big (\tilde{\mathfrak p}(h) \big ) = \tilde{\mathfrak p}\big ( \partial^{\alpha_i} G_i(h) \big )\; .
\end{equation}
Furthermore:
\begin{equation}
\tilde \partial^{\alpha_1} G_1 \big (\tilde{\mathfrak p}(h) \big ) \ldots \tilde \partial^{\alpha_n} G_n \big (\tilde{\mathfrak p}(h) \big ) = \tilde{\mathfrak p}\Big ( \partial^{\alpha_1} G_1(h) \ldots \partial^{\alpha_n} G_n(h) \Big )\; .
\end{equation}
From Proposition~\ref{Pro-CondA12}, we know that the argument of $\tilde{\mathfrak p}$ in the above equation satisfies conditions a1 and a2. Thus, the statement follows if we can show that:
\begin{equation}
\big | \Tt (a )- \widetilde \Tt \big (\tilde{\mathfrak p}(a) \big ) \big | \leq \frac{A_K}{(1+ |V_L|)^K} \; ,
\end{equation} 
for any $a$ satisfying conditions a1 and a2 and $K \in \NM$, where $A_K$ is finite parameter which can dependend on $a$. We have: 
\begin{equation}
\Tt (a )- \widetilde \Tt \big (\tilde{\mathfrak p}(a) \big ) = \Tt_0 (a_0 )- \widetilde \Tt_0 \big (\tilde{\mathfrak p}(a)_0 \big ) = \Tt_0 (a_0 )- \widetilde \Tt_0 \big (\tilde{\mathfrak p}(a_0) \big ) \; ,
\end{equation}
and from \eqref{Eq-TracePer}:
\begin{equation}
\Tt (a )- \widetilde \Tt \big (\tilde{\mathfrak p}(a) \big ) = \Tt_0 (a_0 - (\mathfrak i \circ \tilde{\mathfrak p})(a_0) \big ) = \int_\Omega {\rm d}\PM(\omega) \; \tr \big (a_0(\omega) - a_0(\tilde{\mathfrak q} \omega) \big ) \;.
\end{equation}
Note that $\omega_x$ and $(\tilde{\mathfrak q} \omega)_x$ coincide on $V_L$, hence, for any $K \in \NM$:
\begin{equation}
\big \|a_0(\omega) - a_0(\tilde{\mathfrak q} \omega) \big \| \leq \frac{A_K}{(1+ |V_L|)^K} \;,
\end{equation} 
and the affirmation follows. \qed

\section{Second Round of Approximations} 

In this Section, we derive an upper bound on the errors induced by replacing the periodic algebra $\tilde \Aa_d$ with the finite algebra $\hat \Aa_d$. The quantization \eqref{Eq-PhiQuant} is assumed.

\begin{theorem}\label{Th-Bounds2} Let $\tilde h \in \tilde \Aa_d^\infty$ be a Hamiltonian from the periodic algebra with rapid decay Fourier coefficients. Recall the epimorphism $\hat{\mathfrak p}$ from Proposition~\ref{Lem-PerFinite} and define $\hat h \in \hat \Aa_d$ as $\hat h = \hat{\mathfrak p}(\tilde h)$. Then, for any $K \in \NM$, there exists the finite constant $A_K$ such that:
\begin{equation}\label{Eq-Diff2}
\Big | \widetilde \Tt \big (\tilde \partial^{\alpha_1} G_1(\tilde h) \ldots \tilde \partial^{\alpha_n}G_n(\tilde h) \big )- \widehat \Tt \big (\hat \partial^{\alpha_1} G_1(\hat h) \ldots \hat \partial^{\alpha_n}G_n(\hat h) \big ) \Big | \leq \frac{A_K}{(1+ |V_L|)^K} \; ,
\end{equation}
where $G_i:\RM \rightarrow \CM$ are are smooth functions over $\sigma(\tilde h)$ and $\alpha_i$'s are multi-indices. 
\end{theorem} 

\proof Since $\hat{\mathfrak p}$ is again a homomorphisms, it commutes with the functional calculus and, as a consequence $G_j (\hat h) = \hat{\mathfrak p}\big ( G_j(\tilde h) \big )$. Unfortunately, $\hat{\mathfrak p}\big (\tilde \partial^{\alpha_j} G_j (\tilde h) \big ) \neq \hat \partial^{\alpha_j} G_j (\hat{\mathfrak p}(\tilde h) \big )$ but they are close. In fact:
\begin{equation}
\hat \partial^{\alpha_j} G_j (\hat h) = \hat{\mathfrak p}\big (\hat \partial^{\alpha_j} G_j(\tilde h) \big ) \;,
\end{equation}
where the abusive notation $\hat \partial^\alpha \tilde a$ stands for the element of $\tilde \Aa_d$ having Fourier coefficients:
\begin{equation}\label{Eq-FakeDeriv}
(\hat \partial^\alpha \tilde a)_x = (-\I)^{|\alpha|} s(\hat x)^\alpha \tilde a_x \; , \quad x \in \ZM^d \; .
\end{equation} 
Clearly $\hat \partial$ maps $\tilde \Aa_d^\infty$ into itself, hence $\hat \partial^{\alpha_j} G_j(\tilde h) \in \tilde \Aa_d^\infty$. At this point, the difference in \eqref{Eq-Diff2} has been reduced to:
\begin{equation}\label{Eq-Diff3}
\widetilde \Tt \big (\tilde \partial^{\alpha_1} G_1(\tilde h) \ldots \tilde \partial^{\alpha_n}G_n(\tilde h) \big ) - \widehat \Tt \big (\hat{\mathfrak p}\big (\hat \partial^{\alpha_1} G_1(\tilde h) \ldots \hat \partial^{\alpha_n}G_n(\tilde h) \big )\big ) \; .
\end{equation}
Note that the whole argument of $\tilde{\mathfrak p}$ belongs to $\tilde \Aa_d^\infty$. Now, if $\tilde a \in \tilde \Aa_d^\infty$, then, from Proposition~\ref{Pro-PerFiniteF}, we have:
\begin{equation}
\hat{\mathfrak p}(\tilde a)_0 = \sum_{\hat y = \hat 0} \tilde a_y \circ \hat{\mathfrak q}^{-1} \; ,
\end{equation}
hence:
\begin{equation}
\widetilde \Tt(\tilde a) - \widehat \Tt \big ( \hat{\mathfrak p}(\tilde a) \big ) = \widetilde \Tt_0 (\tilde a_0) - \widehat \Tt_0 \big (\hat{\mathfrak p}(\tilde a)_0 \big ) = \sum_{\hat y = \hat 0, y \neq 0} \widetilde \Tt_0(\tilde a_y) \; .
\end{equation}
Using the rapid decay of the Fourier coefficients of $\tilde a$:
\begin{equation}
\big | \widetilde \Tt(\tilde a) - \widehat \Tt \big ( \hat{\mathfrak p}(\tilde a) \big ) \big | \leq \sum_{\hat y = \hat 0, y \neq 0} \frac{A_K}{(1+ |V_y|)^{2K}} = \sum_{y \in \ZM^d \setminus \{0\}} \frac{A_K}{(1+ |V_L| \, |V_y|)^{2K}} \; ,
\end{equation}
and from:
\begin{equation}
\frac{1}{(1+ |V_L| \, |V_y|)^{2K}} \leq \frac{1}{(1+ |V_L|)^K(1+ |V_y|)^K}\; , \quad y \neq 0\;,
\end{equation}
we obtain:
\begin{equation}
\big | \widetilde \Tt(\tilde a) - \widehat \Tt \big ( \hat{\mathfrak p}(\tilde a) \big ) \big | \leq \frac{A_K}{(1+ |V_L|)^K},
\end{equation}
if we take $K\geq 2$ so that the sum over $y$ is convergent. The conclusion is that, up to such corrections, the difference in \eqref{Eq-Diff3} reduces to:
\begin{equation}\label{Eq-Diff4}
\widetilde \Tt \big (\tilde \partial^{\alpha_1} G_1(\tilde h) \ldots \tilde \partial^{\alpha_n}G_n(\tilde h) \big ) - \widetilde \Tt \big (\hat \partial^{\alpha_1} G_1(\tilde h) \ldots \hat \partial^{\alpha_n}G_n(\tilde h)\big ) \;,
\end{equation} 
which can be casted in the following form:
\begin{equation}
\sum_{j=1}^n \widetilde \Tt \big (\tilde \partial^{\alpha_1} G_1(\tilde h) \ldots  \big (\tilde \partial^{\alpha_j} G_j(\tilde h)\big ) - \hat \partial^{\alpha_j} G_j(\tilde h) \big ) \ldots \hat \partial^{\alpha_n}G_n(\tilde h) \big ) \; .
\end{equation}
We can use the cyclic property of the trace and cast the arguments of $\widetilde \Tt$ in the form:
\begin{equation}
\big (\tilde \partial^{\alpha_j} G_j(\tilde h)\big ) - \hat \partial^{\alpha_j} G_j(\tilde h) \big ) \tilde b_j \; ,
\end{equation}
with $\tilde b_j \in \tilde \Aa_d^\infty$. Note that:
\begin{equation}
\tilde \partial^{\alpha} G (\tilde h)_x - \hat \partial^{\alpha} G (\tilde h)_x  = 
\left \{ 
\begin{array}{ll}
0 \;, \ & x \in V_L \; , \\
 (-\I)^{|\alpha|} (x^{\alpha} - s(\hat x)^{\alpha}) G (\tilde h)_x\; , \ & x \in \ZM^d \setminus V_L \; .
 \end{array}
 \right.
\end{equation}
As such, the problem has been reduced to estimating $\widetilde \Tt( \tilde a \, \tilde b)$, with both elements from $\tilde \Aa_d^\infty$ and $\tilde a$ such that $\tilde a_x =0$ for $x \in V_L$. Writing out explicitly:
\begin{align}
\widetilde \Tt(\tilde a \, \tilde b) = \widetilde \Tt_0\big ( (\tilde a \, \tilde b)_0 \big ) = \sum_{y \in \ZM^d\setminus V_L} \int_{\widetilde \Omega} {\rm d}\widetilde \PM(\tilde \omega) \, \tilde a_y(\tilde \omega) \, \tilde b_{-y}(\tilde \tau_{-y} \tilde \omega) \; ,
\end{align}
and using the fast decay of the Fourier coefficients:
\begin{align}
\big | \widetilde \Tt(\tilde a \, \tilde b) \big | \leq \sum_{y \in \ZM^d\setminus V_L} \frac{A_J}{(1+|y|)^{2J}} =  \sum_{x \in V_L} \sum_{y \in \ZM^d\setminus \{0\}} \frac{A_J}{(1+|x + (2L+1)y|)^{2J}}\; .
\end{align}
Note that $|x_i + (2L+1)y_i| \geq L|y_i|$, hence:
\begin{align}
\big | \widetilde \Tt(\tilde a \, \tilde b) \big | \leq   |V_L| \sum_{y \in \ZM^d\setminus \{0\}} \frac{A_J}{(1+L|y|)^{2J}} \leq |V_L| \sum_{y \in \ZM^d\setminus \{0\}} \frac{A_J}{(1+L)^J (1+|y|)^J}\; .
\end{align}
Finally, by taking $J = d(K+1)$, and $K\geq 1$ so that the remaining sum over $y$ is convergent, we obtain:
\begin{align}
\big | \widetilde \Tt(\tilde a \, \tilde b) \big | \leq \frac{A_K}{(1+|V_L|)^K}\; ,
\end{align}
and the statement follows. \qed

\section{Overall Error Bounds} 

In this Section, we derive an upper bound on the errors induced by replacing the algebra $\Aa_d$ of physical observables with the finite algebra $\hat \Aa_d$. They follow straightforwardly from the previous two Sections but we still want to state them explicitly.

\begin{theorem}\label{Th-OverallBounds} Let $h \in \Aa_d$ be a Hamiltonian in the infinite-volume setting and assume a1-a2 from above. Recall the epimorphism $\tilde{\mathfrak p}$ from Proposition~\ref{Pro-TildeAlgEpi} and the epimorphism $\hat{\mathfrak p}$ from Proposition~\ref{Lem-PerFinite} and define $\hat h \in \hat A_d$ as $\hat h = (\hat{\mathfrak p} \circ \tilde{\mathfrak p})(h)$. Then, for any $K \in \NM$, there exists the finite positive constant $A_K$ such that:
\begin{equation}
\Big | \Tt \big (\partial^{\alpha_1} G_1(h) \ldots \partial^{\alpha_n}G_n(h) \big )- \widehat \Tt \big (\hat \partial^{\alpha_1} G_1(\hat h) \ldots \hat \partial^{\alpha_n}G_n(\hat h) \big ) \Big | \leq \frac{A_K}{(1+ |V_L|)^K} \; ,
\end{equation}
where $G_i:\RM \rightarrow \CM$ are are smooth functions over $\sigma(h)$ and $\alpha_i$'s are multi-indices.
\end{theorem}

\proof The statement follows from Theorems~\ref{Th-Bounds1} and \ref{Th-Bounds2} once we verify that the assumptions match. Clearly the assumptions are matched for Theorem~\ref{Th-Bounds1}. Taking $\tilde h = \tilde{\mathfrak p}(h)$, then $\tilde h$ is indeed an element of $\tilde \Aa_d^\infty$, hence the assumptions in Theorem~\ref{Th-Bounds2} are also matched. \qed

\chapter{Applications: Transport Coefficients at Finite Temperature}
\label{Cha-ApplicationsI}

\abstract{In the first part of the Chapter, we show that the non-commutative Kubo-formula for the transport coefficients is covered by the theory developed so far. In the second part, we present computer assisted calculations of the finite-temperature transport coefficients, for model Hamiltonians relevant for the integer quantum Hall effect and Chern insulators. The focus will be on the physical signature of the Anderson transition occurring at the boundary between the topological phases.}

\section{The Non-Commutative Kubo Formula}\label{Sec-KuboFormula} 

The non-commutative Kubo-formula was already formulated in \eqref{Eq-NCKubo}. Various models for the dissipation super-operator $\Gamma$ together with the physical regimes where they are expected to apply are discussed in \cite{7SBB2,7SpB,7Bel1,7ABS}. Here we will restrict ourselves to the case where the dissipation super operator is simply proportional with the identity map, $\Gamma \times {\rm id}$, $\Gamma \in \RM_+$.  In this case, the finite-volume approximation of \eqref{Eq-NCKubo} takes the form:
\begin{equation}\label{Eq-FVKubo}
\hat{\sigma}_{i j}(\epsilon_F,T,\Gamma)= 2\pi N \, \widehat \Tt \Big ( (\hat \partial_i \hat h) \,   (\Gamma + \Ll_{\hat h})^{-1} \big (\hat \partial_j \Phi_{\mathrm{FD}}(\hat h) \big ) \Big ) \;,
\end{equation}
in units of $e^2/h$. At the first sight, this expression looks more complicated than what we have covered so far, due to the presence of the derivation $\Ll_{\hat h}$. As we shall see, however, the error bounds for this approximation follow quite directly from the results established in the previous section \cite{7Pro4}.

\begin{corollary} For any $K \in \NM$, there exists a positive and finite constant $A_K$ such that:
\begin{equation}
\big | \sigma_{ij}-\hat \sigma_{ij} \big | \leq \frac{A_K}{(1 + |V_L|)^K} \; , \quad i,j = 1, \ldots d \;.
\end{equation}
The parameter $A_K$ depends on the dissipation and temperature. 
\end{corollary}

\proof The non-commutative Kubo formula \eqref{Eq-NCKubo} can be also represented as:
\begin{equation}
\sigma_{ij}= 2\pi N \, \int_0^\infty {\rm d}t \, e^{-t\Gamma} \, \Tt \Big ( (\partial_j h) \, e^{-\I t h}
\, \big ( \partial_j \Phi_{\mathrm{FD}}(h) \big ) \, e^{\I t h}\Big ) \; .
\end{equation}
As in \cite{7Pro4}, we perform one more step and write the righthand side as:
\begin{equation}\label{Eq-X10}
\frac{-N}{2 \pi}\int\limits_0^\infty {\rm d} t \, e^{-t\Gamma}  \int\limits_\Cc {\rm  d}z e^{-\I t z}  \int\limits_\Cc {\rm  d}z' e^{\I t z'}
 \Tt \Big ( (\partial_j h) \, (z - h)^{-1}
\, \big ( \partial_j \Phi_{\mathrm{FD}}(h) \big ) \, (z'-h)^{-1}\Big ) \; ,
\end{equation}
where $\Cc$ is the curve (curves) in the complex plane defined by the equation $|z -\sigma(h)|=\Gamma/4$. Recall that $\sigma(\hat h) \subseteq \sigma(h)$. The argument of the trace in \eqref{Eq-X10} fulfills all the conditions in Theorem~\ref{Th-OverallBounds} and $|e^{\pm\I t z}|\leq e^{t\Gamma/4}$ when $z \in \Cc$. Hence:
\begin{equation}
\big | \sigma_{ij}-\hat \sigma_{ij} \big | \leq \int_0^\infty {\rm d} t \, e^{-t\Gamma/2} \frac{A_K(\Gamma)}{(1 + |V_L|)^K} = \frac{2}{\Gamma}\frac{A_K(\Gamma)}{(1 + |V_L|)^K} \; .
\end{equation}
As one can see, the finite temperature as well as the dissipation are essential for this rapid convergence to the thermodynamic limit. This conclusion is not an artifact of the proof as it was witnessed in the numerical computations. \qed

\vspace{0.1cm}

In the applications reported in the following Sections, the finite-volume Kubo-formula \eqref{Eq-FVKubo} was evaluated using its physical representation:
\begin{equation}\label{Eq-PracticalKubo}
\hat \sigma_{jk} = \int_{\widehat \Omega} {\rm d} \widehat \PM(\hat \omega) \  \tfrac{2\pi}{|V_L|}\sum_{a,b=1}^{N|V_L|}\frac{\langle \phi_b|\hat \pi_{\hat \omega}(\hat \partial_j \hat h)|\phi_a \rangle \langle \phi_a | \hat \pi_{\hat \omega} (\hat \partial_k \Phi_{\mathrm{FD}}(\hat h))|\phi_b \rangle}{\Gamma + \I (\epsilon_a-\epsilon_b)} \; .
\end{equation}
Here $\{\epsilon_a, \phi_a\}_{a = 1,\ldots,N|V_L|}$ are the eigenvalues and the corresponding eigenvectors of the finite-volume Hamiltonian $\hat h$. As already explained in Section~\ref{Sec-ExplImple}, the integration over $\hat \omega$ was computed by averaging over a finite number of disorder configurations.

\begin{figure}
\center
  \includegraphics[width=1\textwidth]{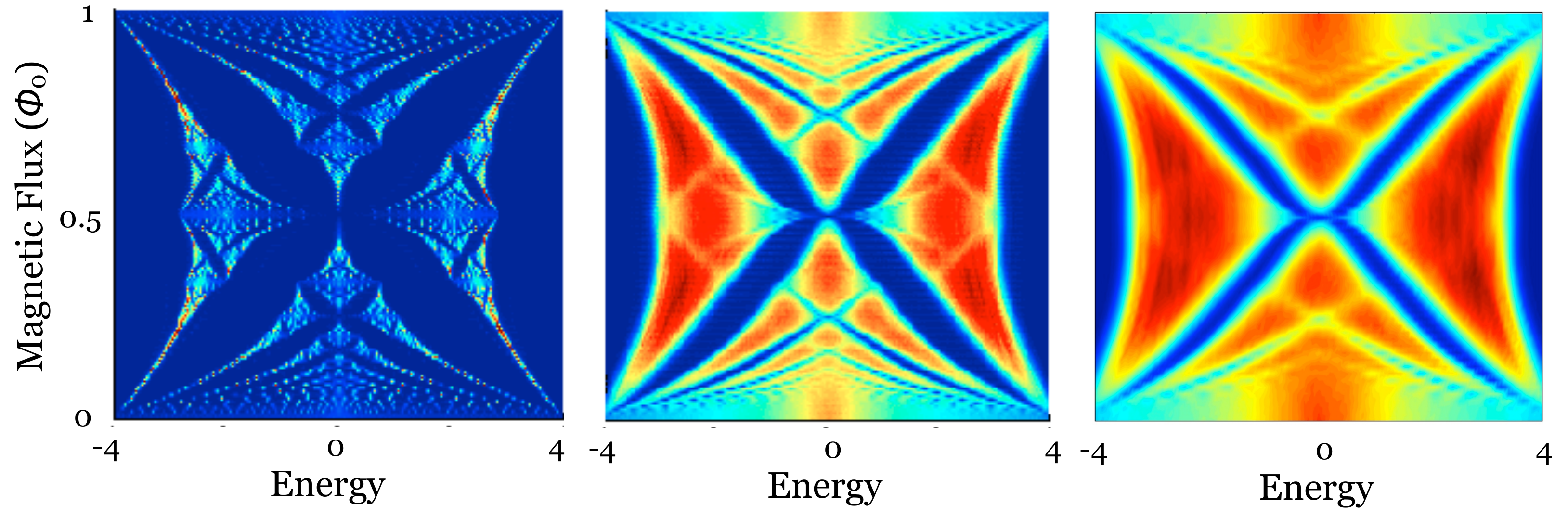}\\
  \caption{\small (Adapted with permission from \cite{7Pro4}) The density of states of the model Hamiltonian \eqref{Eq-Hofstadter} as function of energy and magnetic flux through the unit cell, for three degrees of disorder: $\lambda=0$ (left), $\lambda=2$ (middle) and $\lambda=3$ (right). At $\lambda = 0$, one can recognize the fractal nature of the famous Hofstadter butterfly \cite{7Hof}. For finite $\lambda$'s, however, this fractal structure is smoothen out by the disorder.}
 \label{Spectrum}
\end{figure}

\section{The Integer Quantum Hall Effect}\label{Sec-IQHE} 

In this Section, we report applications to the integer quantum Hall effect (IQHE), which is modeled here by the disordered Hofstadter Hamiltonian \cite{7Hof}:
\begin{equation}\label{Eq-Hofstadter}
h \in \Aa_2 \; , \quad h = u_1 + u_1^\ast + u_2 + u_2^\ast + \lambda w_0 \; , 
\end{equation}
where:
\begin{equation}
w_0 \in C(\Omega) \; , \quad \Omega = \prod_{x \in \ZM^2} \big [ -\tfrac{1}{2},\tfrac{1}{2} \big ] \; , \quad w_0(\omega) = \omega_0 \; .
\end{equation}
The probability measure on $\Omega$ is just the product of the Lebesque measures. Most of the results are reproduced from \cite{7Pro4}, \cite{7SP1} and \cite{7PB}. The parameters of the model are the disorder strength $\lambda$ and the magnetic flux $\phi$ through one unit cell, also equal to  $\phi_{12}$ in the notation of the previous Chapters. The density of states, computed on a finite $120\times 120$-lattice as $D(\epsilon) = \Tt \Big ( \frac{2 \pi \delta }{(h-\epsilon)^2 + \delta^2} \Big )$ with $\delta = 0.01$, is reported in Fig.~\ref{Spectrum} in the form of an intensity plot in the plane of energy and magnetic flux. The density of states provides a reasonable representation of the spectrum of $h$. In the clean case, 
the spectrum has a fractal structure known as the Hofstadter butterfly \cite{7AO,7AOS,7AKY,7AEG}. Note, however, the presence of several prominent spectral gaps. As the disorder is turned on, most of the fractal structure is washed away and for $\lambda =3$ only the most prominent spectral gaps survive in the spectrum. 

\begin{table}
\caption{\small (Adapted with permission from \cite{7Pro4}) The numerical values of $\sigma_{11}$ at $T=\Gamma=0.1$, $\lambda=\phi =0$ and various Fermi energies spread uniformly over the positive part of the spectrum. They were obtained with the proposed algorithm for increasing lattice sizes, as indicated. The last column displays the value of $\sigma_{11}$ computed with machine precision using the Bloch decomposition.}
\small
\begin{center}
\begin{tabular}{|c|c|c|c|c|c|}
\hline
$E_F$ & $80\times 80$ & $100 \times 100$ & $120 \times 120$ &  $140 \times 140$ & Exact \\
\hline
  0.0   &4.0339628247 & 4.0339630615  & 4.0339630708  &   4.0339630712  & 4.0339630712 \\   
 -0.4   & 3.9394154619    & 3.9394154735   & 3.9394154621 &  3.9394154624   & 3.9394154624     \\
 -0.8   & 3.7040304262   & 3.7040301193   & 3.7040301310 &  3.7040301307   & 3.7040301307     \\
  -1.3   & 3.3684805414     & 3.3684801617   & 3.3684801517 &  3.3684801516  & 3.3684801516     \\
  -1.7    &  2.9522720814   & 2.9522713926 & 2.9522714007 & 2.9522714009   & 2.9522714009     \\
  -2.2   &  2.4678006935     & 2.4678005269   & 2.4678005093 &  2.4678005104  & 2.4678005104     \\
  -2.6   & 1.9239335953    & 1.9239338070  & 1.9239338090 & 1.9239338089   & 1.9239338089     \\
  -3.1   & 1.3274333126   & 1.3274333067  & 1.3274333084 &  1.3274333085   & 1.3274333085     \\
  -3.5    & 0.6854442914    & 0.6854442923 & 0.6854442923 & 0.6854442923   & 0.6854442923     \\
  -4.0   & 0.1086465150  & 0.1086465150  & 0.1086465150 &  0.1086465150 & 0.1086465150 \\
  \hline
\end{tabular}
\end{center}
\label{Ta-Table1}
\end{table}

We start our analysis with the simplest check possible by turing off the disorder and the magnetic field, in which case the transport coefficients can be computed virtually exact using the Bloch decomposition. The performance of the numerical algorithm in such conditions is reported in Table~\ref{Ta-Table1}, which is reproduced from \cite{7Pro4}. Here, one can see that the first five exact digits are already reproduced on the $80\times 80$-lattice. Furthermore, the number of the exact digits reproduced by the algorithm increases with the lattice-size and, for example, the first ten exact digits are all reproduced when the lattice-size reaches $140 \times 140$, for all Fermi levels. This is indeed the fast convergence to the thermodynamic limit that is expected. Further tests of the convergence in the presence of strong disorder will be reported in Chapter~\ref{Cha-TopInv}, which analyzes the topological invariants.
 
\begin{figure}
\center
  \includegraphics[width=0.8\textwidth]{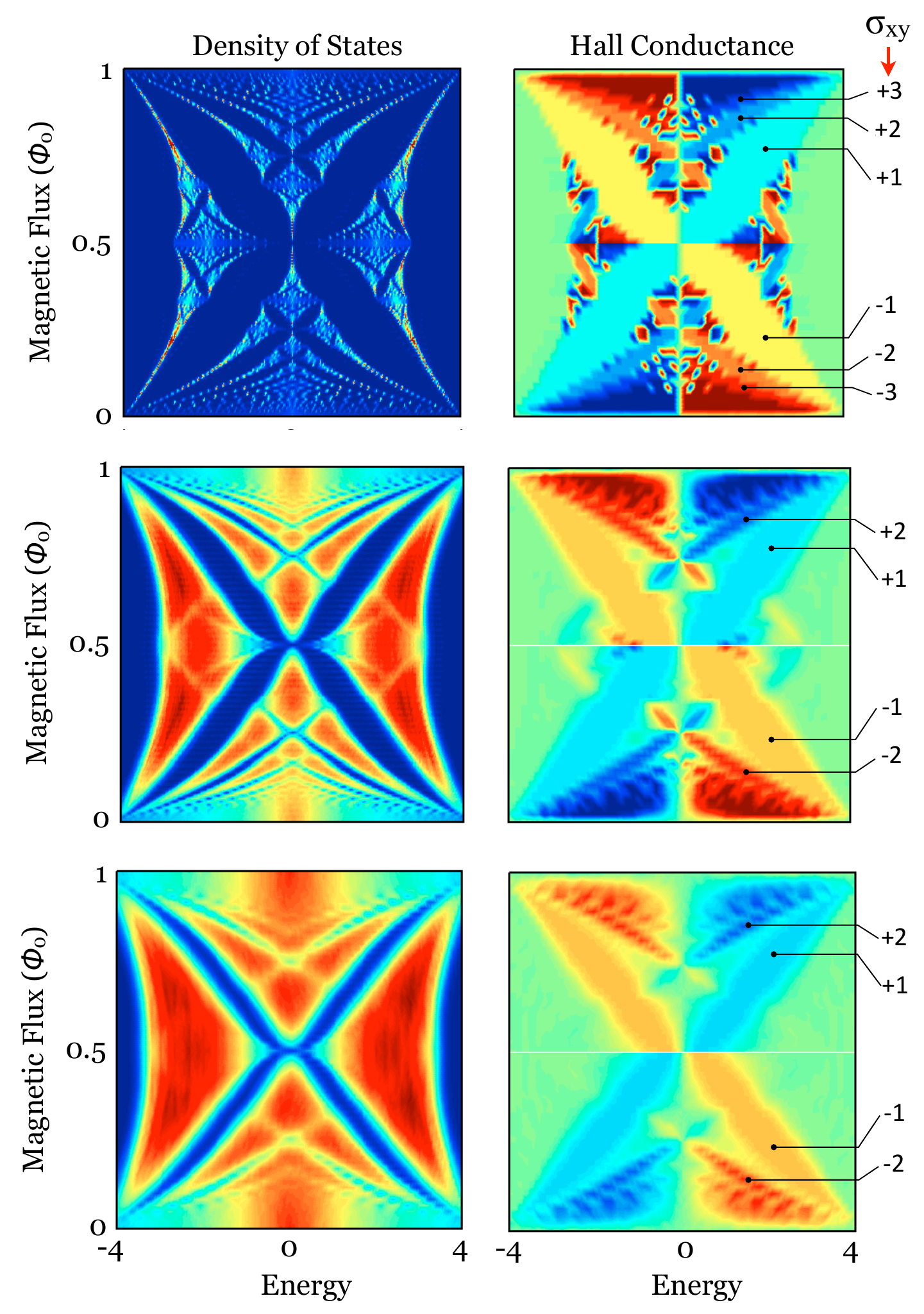}\\
  \caption{\small (Adapted with permission from \cite{7Pro4}) Intensity maps of density of states (left) and the Hall conductivity (right) for $\lambda=0$ (first row), $\lambda=2$ (second row) and $\lambda = 3$ (third row). The regions of quantized Hall conductivity, which appear as well defined patches of same color, are indicated at the right.
}
 \label{HallMap}
\end{figure}

\vspace{0.1cm}

Now, recall the limit \eqref{Eq-ChernLimit} from \cite{7BES}, which says that the Hall conductance converges as $T \searrow 0$ to a topological invariant whose quantized value can change only if the Fermi level crosses a spectral region were the direct conductivity is strictly positive. Hence, if the temperature is small enough, one expects to see regions of quantized Hall conductance. Maps of $\sigma_{12}$ are reported in Fig.~\ref{HallMap} for $\lambda=0$, 2, and 3. These calculations, which are reproduced from \cite{7Pro4}, were performed using a single disorder configuration on a $100 \times 100$-lattice, with the parameters fixed at $T=\Gamma=0.01$. The Fermi energy was varied over the entire energy spectrum, which was sampled at 60 equally spaced points. All allowed values of the magnetic flux were considered. In all three cases, the maps display regions where $\sigma_{12}$ indeed takes the expected quantized values. They correspond to the plateaus in the Hall conductance seen in the experiments. For the cases with disorder, it is important to notice that this quantization occurs beyond the regions empty of spectrum, that is, in the regions where $\epsilon_F$ is immersed in the essential, though, localized spectrum. The Hall conductivity maps in Fig.~\ref{HallMap} are fairly rough since the grid was coarse, the lattice size was relatively small and only one disorder configuration was considered. Still, they give a useful panoramic view of the Hall conductivity, which can serve as a good starting point for our investigation. In fact, the most interesting aspect of the IQHE physics is not the quantization of the Hall conductance per se but rather the transition between such quantized values. This is studied next. 

\begin{figure}
\center
\includegraphics[height=4.5cm]{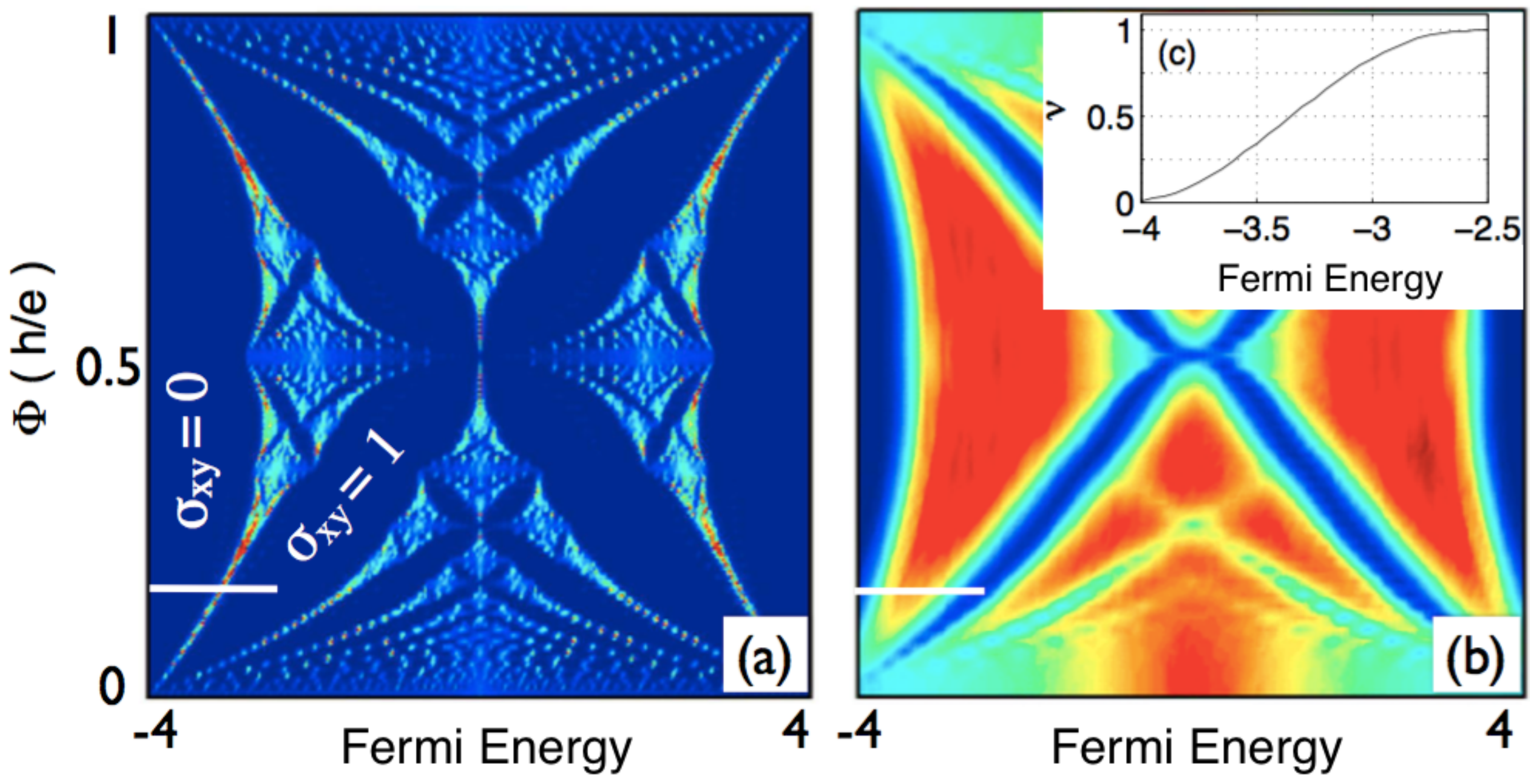}
\caption{\small (Adapted with permission from \cite{7SP1}) The density of states of the model Eq.~\ref{Eq-Hofstadter} as function of the magnetic flux $\phi$, for (a) $\lambda=0$ and (b) $\lambda=3$, together with the values of the Hall conductance for two prominent spectral gaps. The horizontal white line shows the range of energies and the magnetic flux used in the simulations. The inset (c) shows the relationship between the filling factor and the Fermi energy along the simulation range, which is linear for most of the part.}
\label{DOS}
\end{figure}

\vspace{0.1cm}
The plateau-plateau transition \cite{7WTP,7KW,7Pru1,7KHKP1,7KHKP2,7DSMM,7KHKP3,7Huc,7LVX} and at the plateau-insulator transition \cite{7SVPW,7DGPP,7PLVM,7PLPV,7VPGL,7LPVP} in the phase diagram of IQHE has been the subject of intense experimental and theoretical scrutiny. The plateau-insulator transition refers to the jump of $\sigma_{12}$ from 1 to 0 (hence to a normal insulator), while the plateau-plateau transition refers to all the other jumps. There are qualitative differences between the two transitions, which is the reason why they are mentioned and studied separately. When the temperatures are low enough, the following scaling law has been observed in most if not all the experiments:
\begin{equation}\label{Eq-TScaling}
\sigma(\epsilon_F,T) =F\Big ( \big (\epsilon_F-\epsilon_c \big )\left(\tfrac{T}{T_0} \right )^{-\kappa}\Big) \; .
\end{equation}
A similar scaling law applies to the resistivity tensor $\rho(\epsilon_F,T)$, which is just the inverse of $\sigma$. In \eqref{Eq-TScaling}, $\kappa$ is a constant called the finite-temperature scaling exponent, $F$ is a system-specific ({\it i.e.} non-universal) function, $T_0$ is a reference temperature and $\epsilon_c$ is the critical Fermi energy where the transition happens. The finite-temperature scaling exponent $\kappa$ can be related to the finite-size scaling exponent $\nu$ \cite{7Pru1}, the latter being set by the diverging asymptotic behavior of Anderson's localization length \cite{7AALR} near the critical energy:
\begin{equation}\label{Eq-EScaling}
\Lambda(\epsilon_F) \sim \frac{\alpha_0}{| \epsilon_F-\epsilon_c |^\nu } \; .
\end{equation}
The precise definition of Anderson's localization length is given in \eqref{Eq-AndersonLL}. 

\begin{figure}
\center
\includegraphics[height=6.5cm]{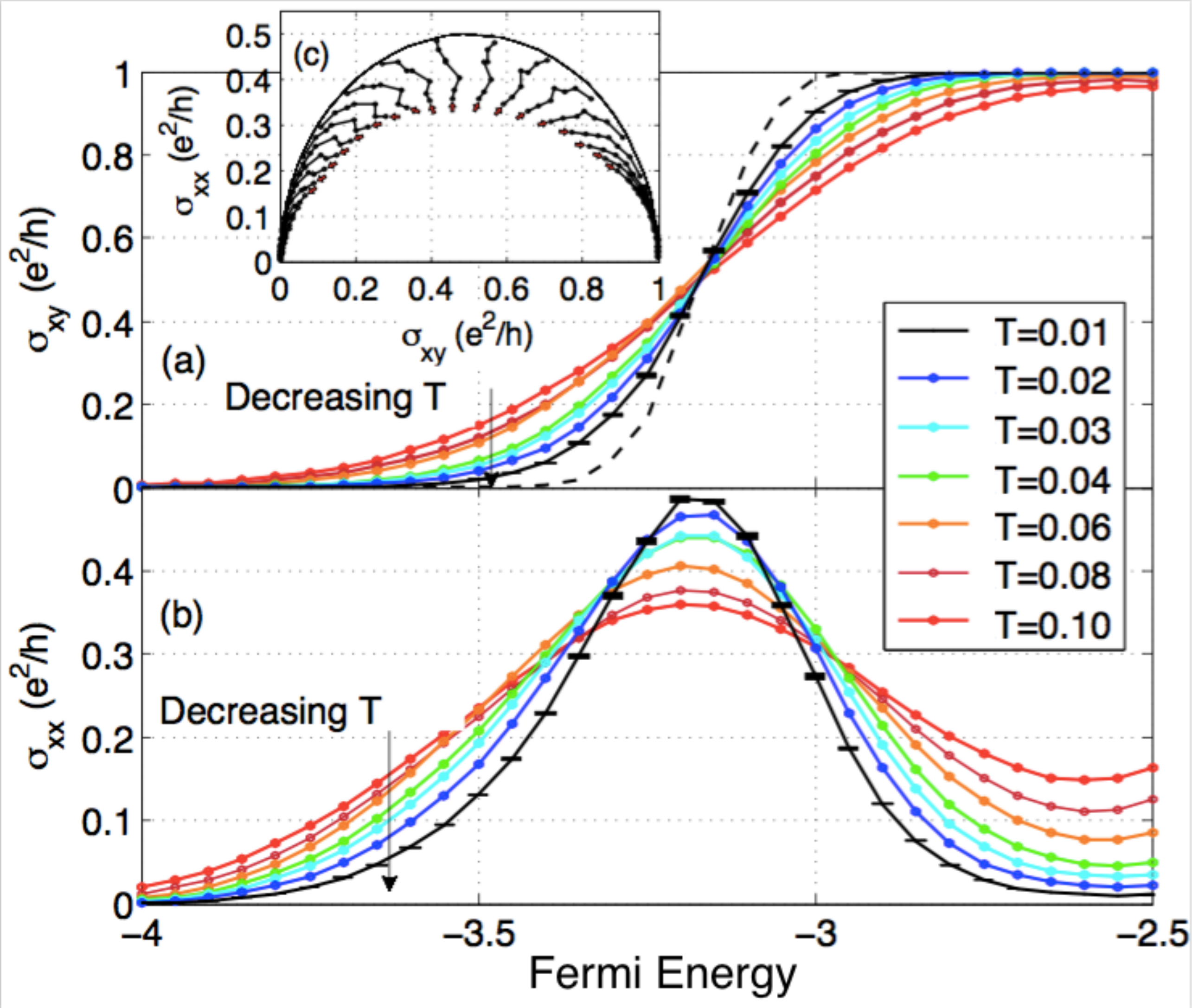}
\caption{\small (Adapted with permission from \cite{7SP1}) The simulated (a) Hall conductance and (b) direct conductance, as functions of Fermi energy, at different temperatures. The dashed line in panel (a) represents the Chern number. The marker-sizes (black rectangle) corresponding to $T = 0.01$ reflect the actual statistical errors. The inset (c) shows the flow of the conductivity tensor with temperature.}
\label{SigmaVsEf}
\end{figure}

\vspace{0.1cm}
The relation between the critical exponents $\kappa$ and $\nu$ can be heuristically understood from the so called single-parameter scaling paradigm \cite{7Pru1}, which consists of the empirical observation that, in the critical regime where \eqref{Eq-EScaling} applies, the physical properties measured in the limit $T\searrow 0$ on samples of different finite sizes $L \lesssim \Lambda$ depend on $\epsilon_F$ and $L$ only through the ratio $L/\Lambda$. We will see this empirical law manifested in Chapter~\ref{Cha-TopInv}. Now, the transport experiments are performed at finite temperatures and on virtually infinite samples. But in this case, there exists an effective sample-size, set by the coherence length $L_{\rm eff}(T)$ introduced by Thouless \cite{7Tho}. The latter is determined by the dissipation processes $L_{\rm eff}(T) \sim 1/\sqrt{\Gamma}$ and, since $\Gamma$ typically scales with the temperature as a power law $\Gamma \sim T^p$ in the regime $T \searrow 0$, one has $L_{\rm eff}(T) \sim T^{-p/2}$ at low temperatures. The exponent $p$ is often referred to as the dynamical exponent for dissipation. Then, according to the single-parameter scaling paradigm, $\sigma(\epsilon_F,T) = F\big ( L_{\rm eff}(T)/\Lambda(\epsilon_F)\big)$, and this leads automatically to the scaling law \eqref{Eq-TScaling} observed experimentally, and to the following precise relation between the critical exponents:
\begin{equation}\label{Eq-Kappa}
\kappa=\frac{p}{2\nu} \; .
\end{equation}
It is important to keep in mind that both critical exponents, $\kappa$ and $\nu$ (hence $p$ too), are accessible from experiments \cite{7LVX}. We also want to point out that \eqref{Eq-Kappa} also follows from the critical behavior of the current-current correlation function \cite{7PB}.

\begin{figure}
\center
\includegraphics[height=5.5cm]{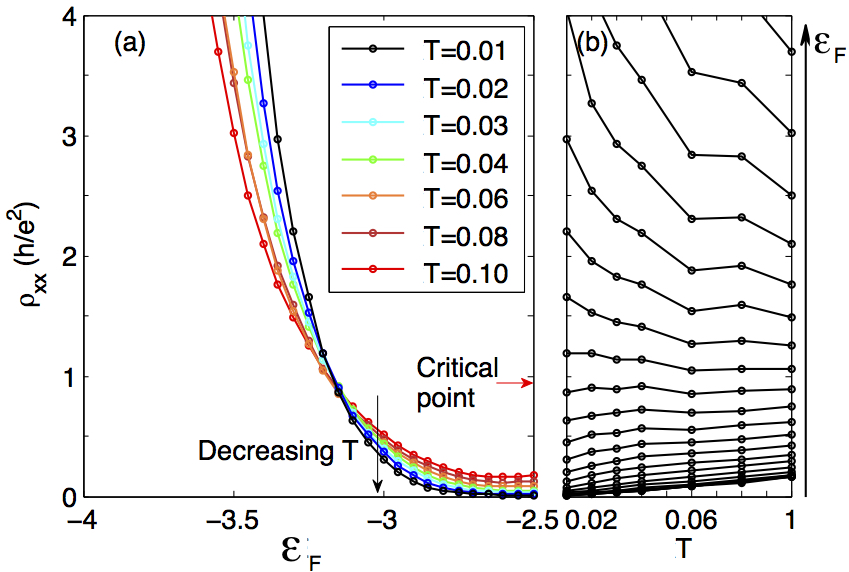}
\caption{\small (Adapted with permission from \cite{7SP1}) The direct resistivity (a) as function of Fermi energy, at different temperatures, and (b)  as function of temperature for various Fermi energy values. The red arrow indicates the plateau-insulator transition.}
\label{RhoXX}
\end{figure}

\vspace{0.1cm}
Below, we reproduce from \cite{7SP1} the simulations of the plateau-insulator transition for the disordered Hofstadter model based on the non-commutative Kubo-formula for transport. This formalism enabled us to converge the simulations at $T$'s low enough to enter the quantum critical regime, where we were able to determine the scaling functions and extrapolate to $T=0$. The results reproduced all qualitative and almost all quantitative experimental signatures of the plateau-insulator transition. Particularly, we were able to detect and characterize the so called Quantized Hall Insulator phase \cite{7LPVP}. The simulations also pointed to possible new effects, such as the anti-levitation of the critical points which was later observed experimentally \cite{PBW}. Before diving into the numerical results, let us comment on why simulating the critical behavior is a very difficult task and on how the new algorithms helped. First, let us point out that the transport simulations simplify tremendously at $T=0$ or at high $T$'s. At $T=0$ and with no dissipation, the Hall conductance reduces to the Chern number, whose computation does not require the inversion of $\Gamma + \Ll_h$. Furthermore, the diagonal conductivity can be obtained from the Landauer formula, which involves only the quantum states at the Fermi level \cite{7SW}. In fact, the whole conductivity tensor can be obtained from a 4-terminal Landauer approach \cite{7PA}. In the critical regime we are dealing with here, the entire energy spectrum has to be taken into account. Truncating the spectrum to a small region around Fermi level will affect even the first digit of the output. The temperatures are low so the details and the fine character of the energy spectrum are not washed away, as it happens at high temperature, but instead they are the main factors determining the values of the transport coefficients. As such, the super-operator $\Gamma + \Ll_h$ must be accurately inverted. But the main challenge is the convergence of the numerical algorithms to the thermodynamic limit. This is essential because one needs to eliminate any artificial finite-size effect in order to correctly assess the critical behavior. The traditional implementations \cite{7HH,7MA,7Roc,7SG} are known \cite{7XP} to converge only as the inverse of $L$ to the thermodynamic limit, hence the finite size effects persist even after the super-cell exceeds the effective system size (set by the temperature and Anderson's localization length, see \cite{7SP1}). With the present algorithms, the simulations start to converge faster then any inverse power law of $L$, once the supercell exceeds this effective size. This enabled us to eliminate any spurious finite-size effects and to converge the calculations at temperatures small enough to enter the critical regime. To our knowledge, this has been achieved for the first time in \cite{7SP1}.

\vspace{0.1cm}
We now present the results of the simulations, which are reproduced from \cite{7SP1}. In Fig.~\ref{DOS} we display again the density of states together with the line which indicates the transition is being simulated. Note that we are dealing with the plateau-insulator transition ({\it cf.} Fig.~\ref{HallMap}). The data is reported as function of Fermi energy, but note that in the experiments one does not have control on $\epsilon_F$ but rather on the electron density or on the filling factor. Hence, it is important to notice the almost linear relation, for most of the part, between $\epsilon_F$ and the filling factor in the inset of Fig.~\ref{DOS}. Fig.~\ref{SigmaVsEf} reports the simulated $\sigma$ as function of $\epsilon_F$ at various temperatures. The dissipation was chosen as $\Gamma = T$, hence we fixed $p=1$. The simulations were performed on a $140 \times 140$-lattice and the output was averaged over 50 independent disorder configurations. Due to the self-averaging property, this was more than enough to reduce the statistical errors below the marker sizes used in Fig.~\ref{SigmaVsEf}. The other parameters were fixed as $\lambda=3$ and the magnetic flux $\phi=22/140\approx 0.157$. In panel (a) we see the Hall conductance $\sigma_{xy}$ (same as $\sigma_{12}$ in the previous notation) transitioning from 0 to 1, and the transition becoming sharper as $T \searrow 0$. To a high degree, all $\sigma_{xy}$-curves intersect each other at one point, the critical point $\epsilon_c$, and the slopes of the curves at $\epsilon_c$ obey the expected scaling with temperature \cite{7Pru1}. In panel (b), $\sigma_{xx}$ (same as $\sigma_{11}$ in the previous notation) displays an insulating behavior for most energies, manifested in the decrease of $\sigma_{xx}$ as $T \searrow 0$, except for a small region around $\epsilon_c$ where  $\sigma_{xx}$ increases as $T \searrow 0$, signaling the presence of extended quantum states. The data suggests that this region subsides to one point and that the maximum value of $\sigma_{xx}$ saturates at $\frac{1}{2}$ as $T \searrow 0$, in line with the accepted theoretical picture of the plateau-insulator transition \cite{7Pru1}. The inset in Fig.~\ref{SigmaVsEf}, which is just a re-plotting of the data from the main panels, shows the flow of $\sigma$ with $T$, where different traces corresponds to different Fermi energies and the direction of the arrows indicate the decreasing of $T$. There is quite a substantial dependence on $T$, but $\sigma$ ultimately flows to a separatrix which is shaped as a perfect semi-circle. An unstable fixed-point at $\sigma_{xx}=\sigma_{xy}=\frac{1}{2}$ can be clearly seen, in agreement with the renormalization theory of the IQHE \cite{7Pru1}. Experimentally, the $T$-flow of $\sigma$ was reported in Ref.~\cite{7SVPW}, and it looks very similar to the flow of $\sigma$ in our simulations.

\begin{figure}
\center
\includegraphics[height=6cm]{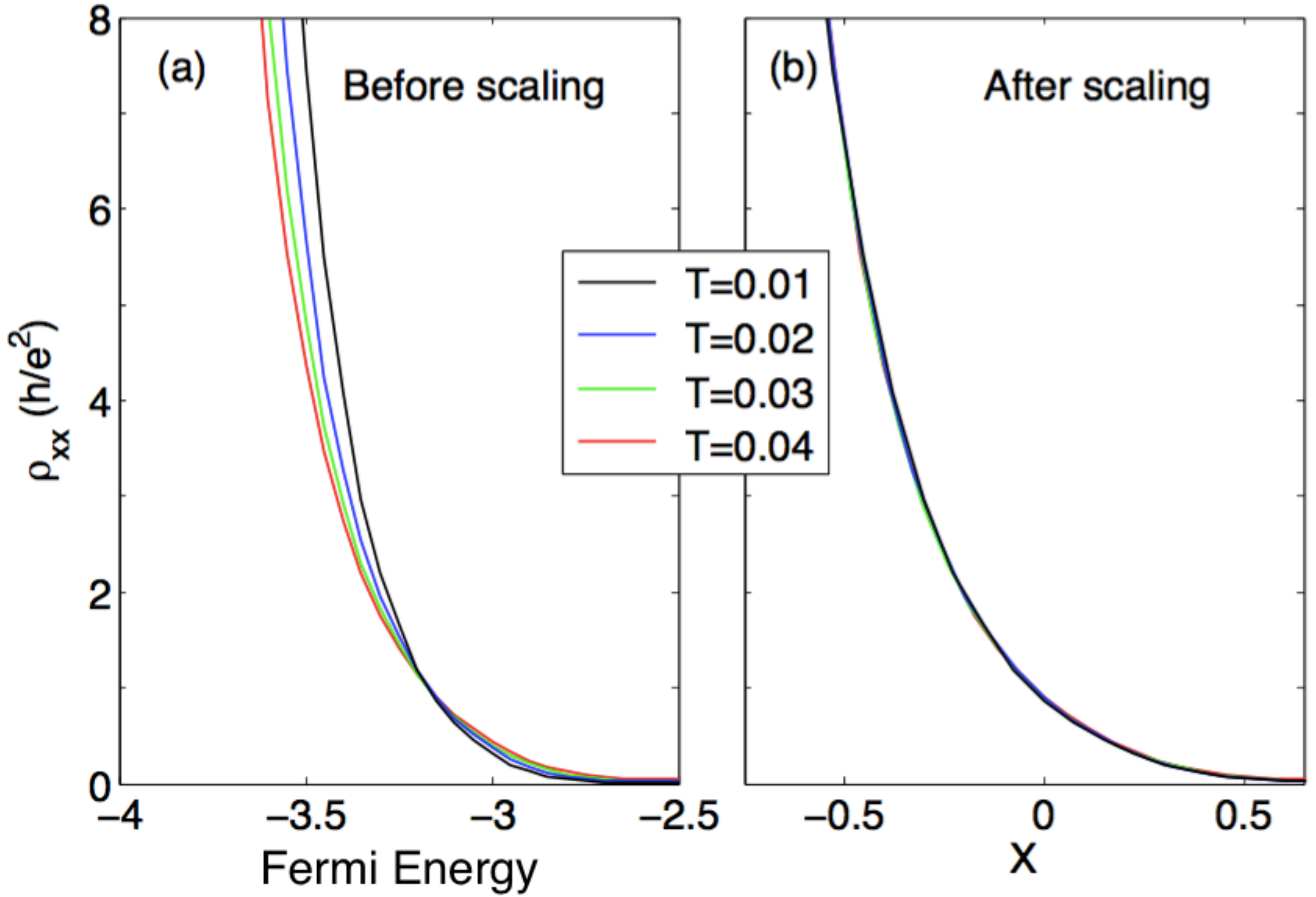}
\caption{\small (Adapted with permission from \cite{7SP1}) The direct resistivity data (a) before and (b) after the single-parameter rescaling \eqref{Eq-Rescale} of the horizontal axis. The parameters for the rescaling were $\epsilon_c=-3.15$, $T_0=0.08$ and $\kappa=0.197$.}
\label{Scaling}
\end{figure}

\vspace{0.1cm}
The existence of a critical point is best revealed in Fig.~\ref{RhoXX}, which reports $\rho_{xx}=\sigma_{xx}/(\sigma_{xx}^2+\sigma_{xy}^2$) as function of $\epsilon_F$ (panel a) and as function of $T$ (panel b). Here, one can see $\rho_{xx} \rightarrow \infty$ in the normal insulator side, and $\rho_{xx} \rightarrow 0$ in the quantum Hall insulator side. While the latter behavior is usually attributed to a metallic phase, in the present context it is due to the fact that $\sigma_{xy}\neq 0$. These opposite behaviors make the $\rho_{xx}$-curves intersect each other, very much like $\sigma_{xy}$-curves do in Fig.~\ref{SigmaVsEf}(a). To a high degree, all the curves (even those at higher $T$'s) intersect at a single critical point, exactly as the experimental data showed (see Fig.~1 in \cite{7HS1}, \cite{7SVPW} and \cite{7VPGL}). As already demonstrated experimentally \cite{7HS2}, the coordinates of the critical point can be extracted with great precision if $\rho_{xx}$ is plotted as function of $T$, as in Fig.~\ref{RhoXX}(b). Here one can see that, as $T \searrow 0$, the $\rho_{xx}$-lines curve  downwards/upwards below/above a sharp critical point $\epsilon_c \approx -3.15$. The critical value $\rho_{xx}^c$ at the transition is virtually equal to 1 in the natural units, and from Fig.~\ref{SigmaVsEf}(a) we estimate $\sigma_{xy}^c$ to be virtually equal to $\frac{1}{2}$, exactly as seen in experiments  \cite{7HS1,7SVPW,7VPGL}. Furthermore, in Fig.~\ref{Scaling} we demonstrate that we entered the quantum critical regime with the four lowest $T$'s. Indeed, upon a single-parameter rescaling of the energy axis: 
\begin{equation}\label{Eq-Rescale}
\epsilon_F \rightarrow X=(\epsilon_F-\epsilon_c)\left ( \frac{T}{T_0}\right ) ^{-\kappa} \; ,
\end{equation} 
all $\rho_{xx}$-curves collapse almost perfectly on top of each other. The best overlap was obtained for $\kappa=0.197 \pm 0.004$, which compares well with the value $\kappa = 0.194 \pm 0.002$ obtained from the relation $\kappa = p/2\nu$, the presently accepted (average) theoretical value $\nu=2.58 \pm 0.03$ \cite{7SO1,7KMO,7OSF,7FHA,7DET,7AMS,7SO2,7OGE} and $p=1$ like in our simulations. 

\vspace{0.1cm}
Further insight into the Anderson transition is provided by the current-current correlation measure, introduced next. A rigorous result \cite{7KP} assures the existence of a Radon measures ${\rm d} m_{ij}(\epsilon,\epsilon')$ on $\mathbb R \times \mathbb R$ such that:
\begin{equation}\label{CCCMeasure}
  \mathcal T \big ( \partial_i h \, G(h) \, \partial_j h \, G'(h) \big ) =  \int_{\RM^2}  G(\epsilon) \, G'(\epsilon') \, {\rm d} m_{ij} (\epsilon,\epsilon') \; .
 \end{equation}
This equality defines the current-current correlation measure, consisting of the matrix of measures ${\rm d} m_{ij}$. The support of the measures is $\sigma(h) \times \sigma(h)$. We will denote its isotropic part by ${\rm d} m=\frac{1}{d}\sum_{j=1}^d {\rm d} m_{jj}$. The isotropic direct conductivity $\sigma = \frac{1}{d}\sum_{j=1}^d \sigma_{jj}$ of a homogenous systems accepts the following formula in terms of the current-current correlation measure \cite{7SBB1,7SBB2}:
\begin{equation}\label{Kubo2}
\sigma(\beta,\epsilon_F,\Gamma) =  \int_{\RM^2} \frac{\Phi_{\rm FD}(\epsilon') - \Phi_{\rm FD}(\epsilon)}{\epsilon-\epsilon'} \frac{4\pi \Gamma}{\Gamma^2 + (\epsilon-\epsilon')^2} {\rm d} m (\epsilon,\epsilon') \; .
\end{equation}
The Anderson localization length also accepts a formula in term of the current-current correlation measure. Let us recall first the so called $\Delta$-localization length $\Lambda(\Delta)$ \cite{7BES}, with $\Delta$ an interval from $\mathbb R$ centered at the Fermi level:
\begin{equation}
\Lambda^2(\Delta) = \mathcal T \big ( | \partial p_\Delta |^2 \big ) \; .
\end{equation}
Here, $p_\Delta$ the spectral projector of $h$ onto $\Delta$: $p_\Delta = \chi(h \in \Delta)$. Then \cite{7BES}:
\begin{equation}
\Lambda^2(\Delta) = \int_{\Delta \times \mathbb R} \frac{{\rm d} m (\epsilon,\epsilon')}{(\epsilon-\epsilon')^2} \;.
\end{equation}
It is a well established fact \cite{7BES,7SBB2} that, whenever there exists a finite interval $\Delta$ centered at $E_F$ and such that $\Lambda(\Delta) < \infty$, the direct conductivities $\sigma_{ii}$ vanish in the limit $T \searrow 0$. The Anderson localization length, defined strictly at the Fermi level, can be expressed as:
\begin{equation}\label{Eq-AndersonLL}
\Lambda^2(\epsilon_F) = \lim_{|\Delta| \rightarrow 0} \frac{\Lambda^2(\Delta)}{|\Delta|} = \lim_{|\Delta| \rightarrow 0} \int_{\Delta \times \mathbb R} \frac{{\rm d} m (\epsilon,\epsilon')}{(\epsilon-\epsilon')^2} \; ,
\end{equation}
where by $|\Delta|$ is meant to be the length of the interval. 

\begin{figure}
{\includegraphics[width=1.00\textwidth]{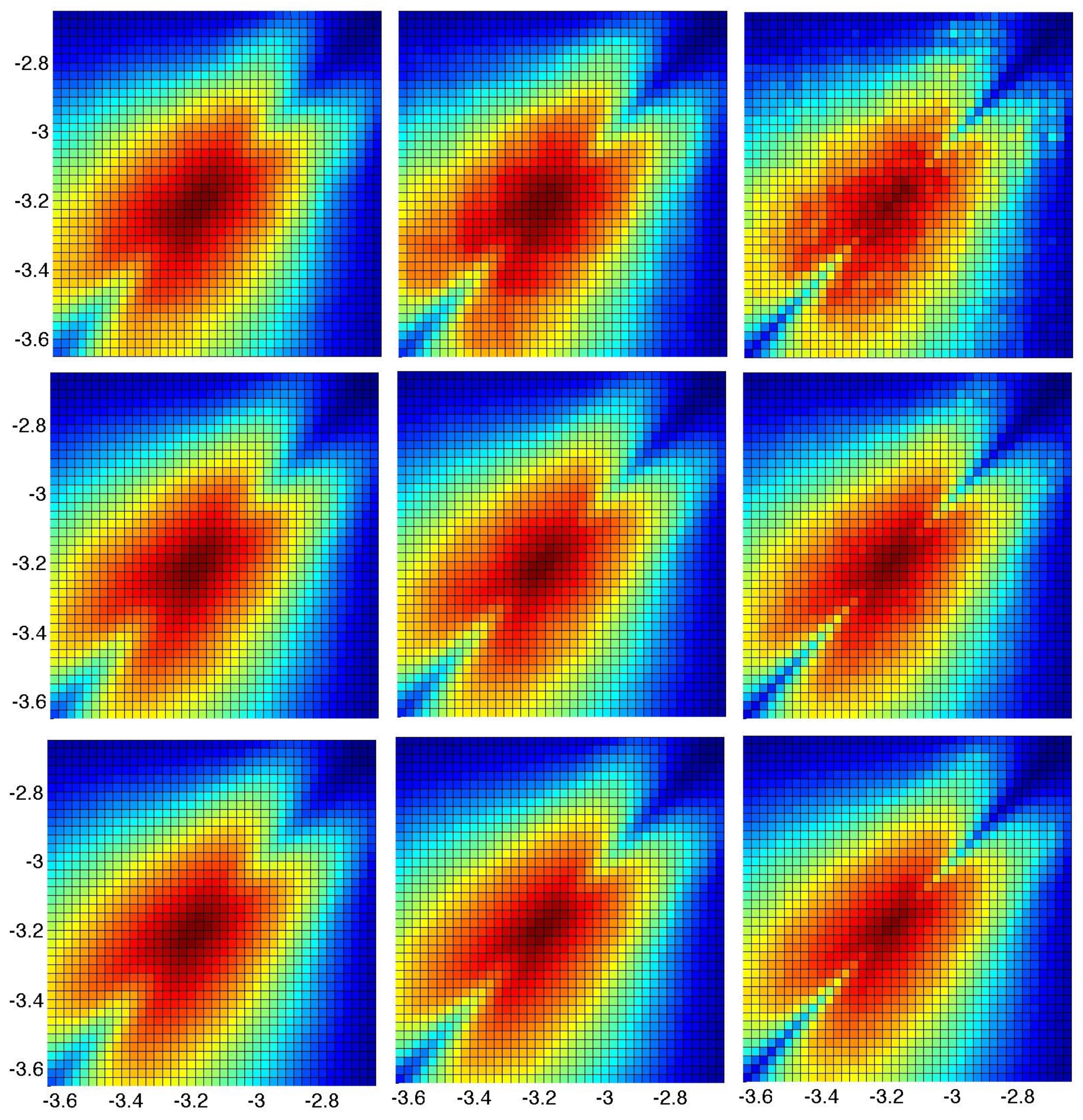}}
\caption{\small (Adapted with permission from \cite{7PB}) Convergence tests on the finite-volume $r$-approximation of the current-current correlation function. The first, second and third rows correspond to different lattice-sizes: $40 \times 40$ and $80 \times 80$ and $120 \times 120$, respectively. The first, second and third columns correspond to different $r$ values (see Eq.~\ref{DiracDelta}): $r=0.03$, $0.02$ and $0.01$, respectively.}
\label{ConvCheck1}
\end{figure}

\vspace{0.1cm}
As we have already discussed, both $\sigma$ and $\Lambda$ display critical behaviors at the Anderson transition. In \cite{7PB}, we found that they can be explained simultaneously by the following critical behavior of the current-current correlation measure:
\begin{equation}\label{Eq-CriticalCCCF}
{\rm d} m(\epsilon,\epsilon') = f(\epsilon,\epsilon') {\rm d} \epsilon {\rm d} \epsilon' \; , \quad f(\epsilon,\epsilon')=g\left( \frac{\epsilon+\epsilon'-2 \epsilon_c}{(\epsilon-\epsilon')^{\kappa/p}} \right ) \;,
\end{equation}
provided $\kappa$, $\nu$ and $p$ obey the relation \eqref{Eq-Kappa}. The density $f$ of the measure is often referred to as the current-current correlation function, which will be assumed to exists and be continuous. The following predictions can be made solely based on the above formula and the asymptotic characteristics of function $g$ (see \cite{7PB}):
\begin{enumerate}[{\rm (p1)}]

\item $f(\epsilon,\epsilon')$ is exactly zero along the diagonal $\epsilon=\epsilon'$, except at the critical point where the exact value is indeterminate.

\vspace{0.1cm}

\item If the critical point $(\epsilon_c,\epsilon_c)$ is approached along the diagonal $\epsilon=\epsilon'$, then the limit value of $f(\epsilon,\epsilon')$ is zero. But if the critical point is approached from any other direction, the limit value of $f(\epsilon,\epsilon')$ is $\frac{\sigma_c}{2\pi^2}$, where $\sigma_c$ is the value of the isotropic conductivity at the critical point (known not be $\frac{1}{2}$ for IQHE).

\vspace{0.1cm}

\item The level sets of $f(\epsilon,\epsilon')=g(t)$ near the critical point are well described by the equation
\begin{equation}\label{Match}
\epsilon+\epsilon' = t (\epsilon-\epsilon')^\frac{1}{2\nu}.
\end{equation}
This observation can also be used to map the function $g$ of \eqref{Eq-CriticalCCCF}.
\end{enumerate}

We now lay down the basic principles of a numerical algorithm for computing the current-current correlation function. Consider the approximation of the Dirac-delta distribution:
\begin{equation}\label{DiracDelta}
\delta_r(t) = \frac{1}{r} \varphi\left(\frac{t}{r}\right ) \overset{r \rightarrow 0}{\longrightarrow}\delta(t), \quad \varphi(t) = \tfrac{1}{\sqrt{2\pi}}e^{-t^2/2} \;.
\end{equation}
Then:
\begin{equation}\label{FEpsilon}
f_r(\epsilon,\epsilon')= \int\limits_{\mathbb R \times \mathbb R}  \delta_r(t-\epsilon) \delta_r(t-\epsilon') \, {\rm d} m (t,t')
\end{equation}
is a point-wise approximation of $f(\epsilon,\epsilon')$: 
\begin{equation}\label{Limit}
\lim_{r \rightarrow 0} f_r(\epsilon,\epsilon') = f(\epsilon,\epsilon'), \quad \epsilon, \epsilon' \in \mathbb R \;.
\end{equation}
The function $f_r$ was referred to in \cite{7PB} as the $r$-approximation of the current-current correlation function $f$. The point here is that the distribution $f_r$ can also be computed from:
\begin{equation}\label{Eq-Principle}
f_r(\epsilon,\epsilon') = \frac{1}{d}\sum_{j=1}^d \mathcal T \big ( (\partial_j h) \, \delta_r (h-\epsilon) \, (\partial_j h) \, \delta_r (h-\epsilon') \big ) \; .
 \end{equation}
The conditions of Theorem~\ref{Th-OverallBounds} are met, hence \eqref{Eq-Principle} is amenable on a computer via the canonical finite-volume algorithm. The convergence to the thermodynamic limit is expected to slow down as $r \searrow 0$, but away from possible singular points we can achieve a good representation of $f(\epsilon,\epsilon')$ even with a finite $r$. How small we need to take $r$ depends on the profile of $f(\epsilon,\epsilon')$ and on desired accuracy. 

\begin{table}\label{Table}
\begin{center}
\begin{tabular}{|c|c|c|c|c|c|c|c|c|c|}
\hline
\ & $r = r' = 0.03$ & $r = r' = 0.02$ & $r = r' = 0.01$ 
\\\hline
$40\times40 \rightarrow 80\times 80$  & $1.1\times 10^{-2}$ & $1.5 \times 10^{-2}$ & $1.9 \times 10^{-2}$  
\\\hline 
$80 \times 80 \rightarrow 120 \times 120$ & $6.5 \times 10^{-3}$ & $6.8\times 10^{-3}$ & $1.1 \times 10^{-2}$
\\
\hline
\end{tabular}
\end{center}
\caption{\small (Adapted with permission from \cite{7PB}) The estimator $D$ defined in \eqref{Eq-Estimator} as evaluated on the data from Fig.~\ref{ConvCheck1}.}
\label{tab-class}
\end{table}

\vspace{0.1cm}
We now discuss the numerical results, which are all reproduced from \cite{7PB}. The flux and the range of energies were chosen exactly as in Fig.~\ref{DOS} and the strength of the random potential was fixed at $\lambda=3$. Hence, we are dealing with exactly the same plateau-insulator transition as before. Fig.~\ref{ConvCheck1} reports a convergence test. In these simulations, we computed the finite-volume $r$-approximation $\hat f_r(\epsilon,\epsilon')$ for $(\epsilon,\epsilon')$ on the grid $\mathcal G$ that is clearly visible in the plots, and the value of $r$ was reduced from $0.03$ to $0.01$ and the lattice-size was increased from $40 \times 40$ to $120 \times 120$. Let us make some qualitative remarks first. Although not a perfect representation of $f(\epsilon,\epsilon')$, we can already detect the position of the critical point and the numerical values at this point agree well with the prediction (p2) from above. For example, the numerical value at the critical point, for $r = 0.01$ and lattice-size $120 \times 120$, is $0.9981$ in units of $\frac{1}{4\pi^2}$. In fact, the overall shape is consistent qualitatively with the theoretical predictions. We can already see that the current-current correlation function is smooth away from the critical point, and even featureless away from the diagonal $\epsilon=\epsilon'$, but it displays abrupt features near critical point, especially near the diagonal where $\hat f_r$ drops to lower values. This feature is consistent with the prediction (p1) and that of Ref.~\cite{7CGH}), which says that $f(\epsilon,\epsilon)=0$ except at $\epsilon_c$, but to fully resolve the behavior near the diagonal it will be a very difficult numerical task. In Fig.~\ref{ConvCheck1}, one can clearly see how the structure near the diagonal becomes sharper as $r$ is decreased but the graph itself becomes more rugged, as expected. Increasing the system-size makes the graph smooth again. 

\begin{figure}
{\includegraphics[width=1\columnwidth]{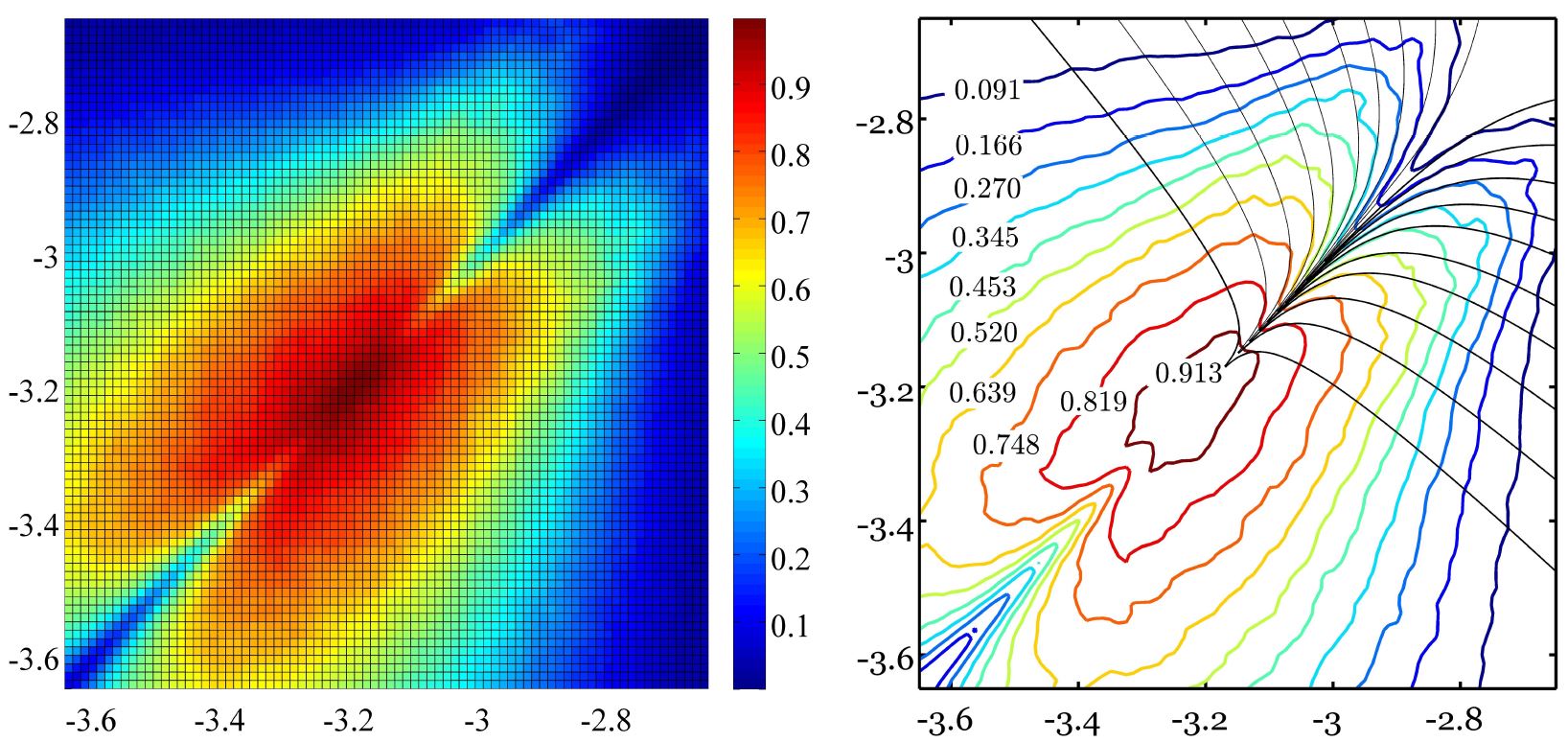}}
\caption{\small (Adapted with permission from \cite{7PB}) Left: A refined intensity plot of the current-current correlation distribution $f(\epsilon,\epsilon')$ in units of $\frac{1}{4 \pi^2}$. Right: The level sets of $f(\epsilon,\epsilon')$ together with  matching curves described in \eqref{Match}. The computation was performed on a 120$\times$120 lattice and the data was averaged over 100 random configurations.}
\label{CCM}
\end{figure}

\vspace{0.1cm}
To quantify the convergence tests, we evaluated the following quantity:
\begin{equation}\label{Eq-Estimator}
D = \frac{4\pi^2}{|\mathcal G|}\sum_{(\epsilon,\epsilon') \in \mathcal G} \Big |\hat f_r^L(\epsilon,\epsilon') - \hat f_{r'}^{L'}(\epsilon,\epsilon') \Big |\;,
\end{equation}
which is an estimator of the (absolute) variations from one simulation to another, and we report the findings in Table~\ref{Table}. Based on these numbers, we expect that, for these tested values of $r$, the $r$-approximation $f_r(\epsilon,\epsilon')$ be converged w.r.t. the system-size to at least two digits of precision (in units of $\frac{1}{4\pi^2}$). To quantify how far is $f_r(\epsilon,\epsilon')$ from the true current-current correlation function, we evaluated the estimator on the $120 \times 120$ lattice and found $D=9.5 \times 10^{-3}$ when decreasing $r$ from $0.03$ to $0.02$,  and $D= 7.2 \times 10^{-3}$ when decreasing $r$ from $0.02$ to $0.01$. We could infer from these numbers that, on average, the true current-current correlation function $f(\epsilon,\epsilon')$ was also resolved up to two digits of precision (in units of $\frac{1}{4\pi^2}$). 

\begin{figure}
\center
{\includegraphics[width=1.0\textwidth]{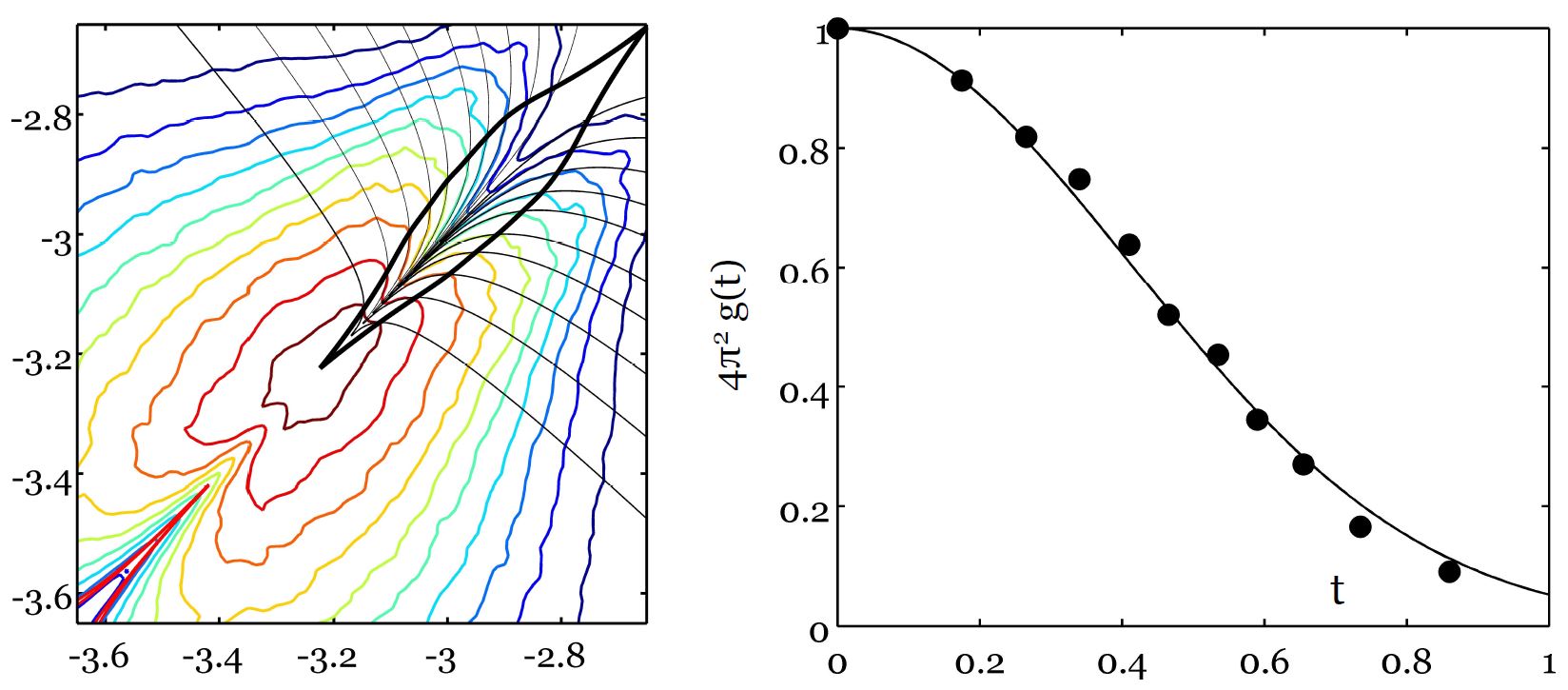}}
\caption{\small (Adapted with permission from \cite{7PB}) Left: The trace of the asymptotic region where the scaling invariance of the current-current correlation function occurs. Right: Plot of the 10 values of the function $g(t)$, as derived from the 10 contours from Fig.~\ref{CCM}, together with a Gaussian fit.}
\label{GFunction}
\end{figure}

\vspace{0.1cm}
We now analyze in more details the current-current correlation function. In Fig.~\ref{CCM}(a) we report the numerically computed  distribution $\hat f_r(\epsilon,\epsilon')$ on a $120 \times 120$ lattice, for $\epsilon=0.01$ but on a more refined grid than in Fig.~\ref{ConvCheck1} (which is again clearly visible from the graph). Based on our convergence tests, we believe that the data is an accurate representation of the exact $f(\epsilon,\epsilon')$. Panel (b) of Fig.~\ref{CCM} displays 10 level sets of $f(\epsilon,\epsilon')$, which will be used to test the prediction (p3) from above. For this we overlap in Fig.~\ref{CCM}(b) the matching contours generated with Eq.~\ref{Match}, $\epsilon =\epsilon' + t (\epsilon-\epsilon')^\frac{1}{2\nu}$. Here, we used $\nu = 2.58$ from the previous simulations \cite{7SP1} and only optimized the value of $t$ for each level set. Although the quality of the contours is somewhat low, the agreement between the numerical level sets and Eq.~\ref{Match} is surprisingly good in a region near the diagonal. Beyond this region, the two curves rapidly diverge from one another. This give us an estimate of the asymptotic region where the scaling invariance applies and this region is traced for the eye in the left panel of Fig.~\ref{GFunction}. Furthermore, by pairing the values of $t$ used to generate the matching contours in Fig.~\ref{CCM} with the values of the level sets, we can generate the profile of the function $g$ and this is shown in the right panel of Fig.~\ref{GFunction}. This profile is quite different from a Lorentzian or a Poisson profile but it is represented quite well by a Gaussian, as the fit shows.

\begin{figure}
\center
  \includegraphics[height=7cm]{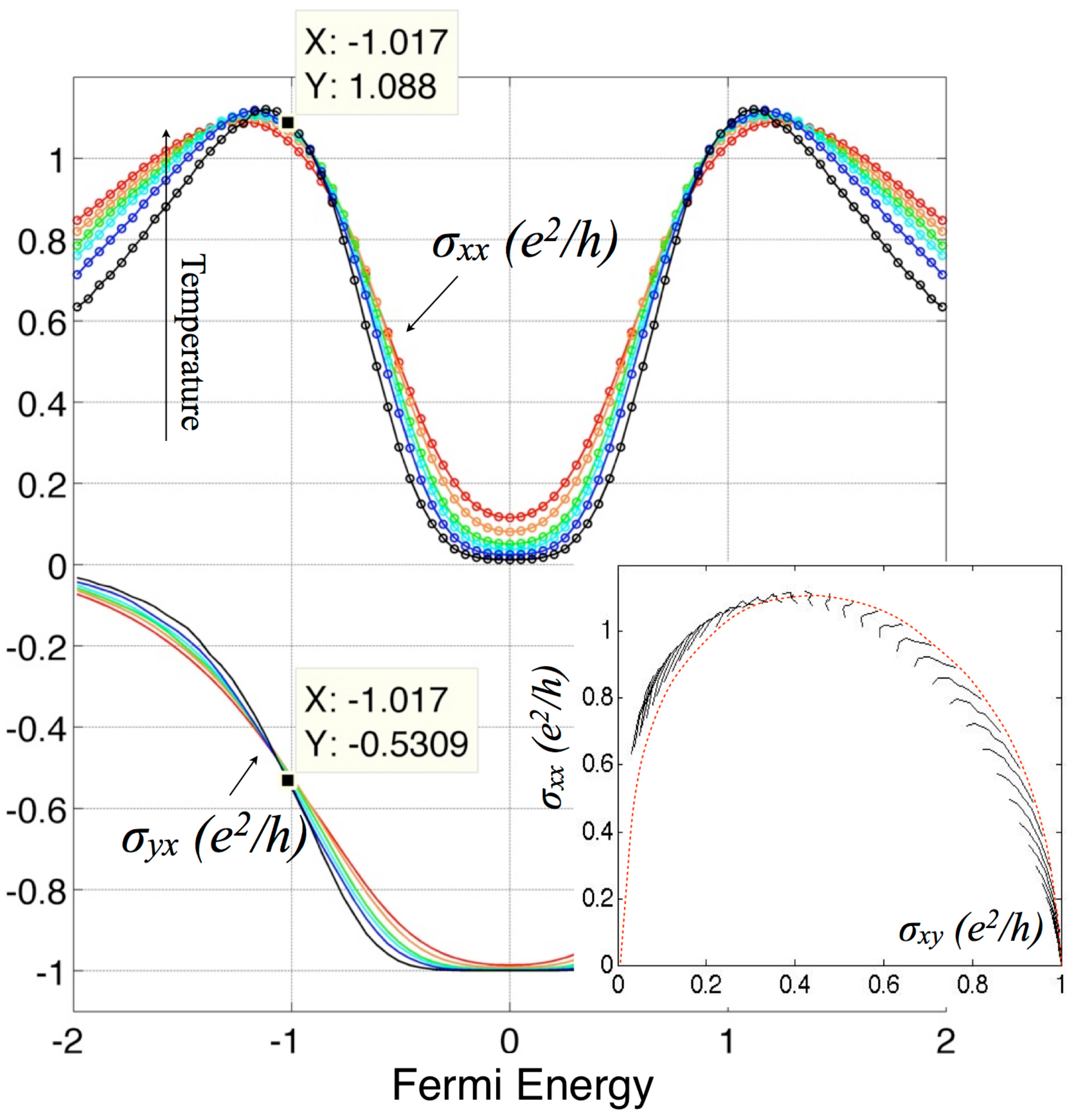}\\
  \caption{\small (Adapted with permission from \cite{7XP1}) The conductivity tensor of the model \eqref{Eq-KM} as function of Fermi ennegy at various temperatures, as simulated on a $80 \times 80$ lattice. An average over many disorder configurations was considered (as many as 67 for $T=0.01$ and 24 for $T=0.08$). The marks posted on the graph give $\sigma$ at the critical point. The inset shows the flow of our data in the $(\sigma_{xy},\sigma_{xx})$ plane as $T \searrow 0$. The dotted line suggests the separatrix for the flow.}
 \label{KMSigmaVsEf}
\end{figure} 

\section{Chern Insulators}
\label{Sec-ChernTI} 

IQHE studied in the previous section is the prototypical example of a topological condensed matter system from class A in the classification Table~\ref{Table1}. A topological insulator, though, is a material in which the physics of IQHE occurs without the need of any externally maintained magnetic field. In other words, it is intrinsic to the materials. There are many experimental realizations and fine characterizations of a Chern insulating phase \cite{7Cha1,7Cha2,7Kou1,7BWM,7KRR,7Col,7Kou2,7JJ}, and the quantum critical regime at the transition between the topological and trivial phases was resolved in \cite{7CYT,7CZL}. In this Section, we investigate this regime numerically on a model of a Chern insulator. The numerical results are reproduced from \cite{7XP1}, which unfortunately were performed before the experiments in \cite{7CYT,7CZL} and, as such, the models are not tuned to those materials.

 \begin{figure}
\center
  \includegraphics[height=6.5cm]{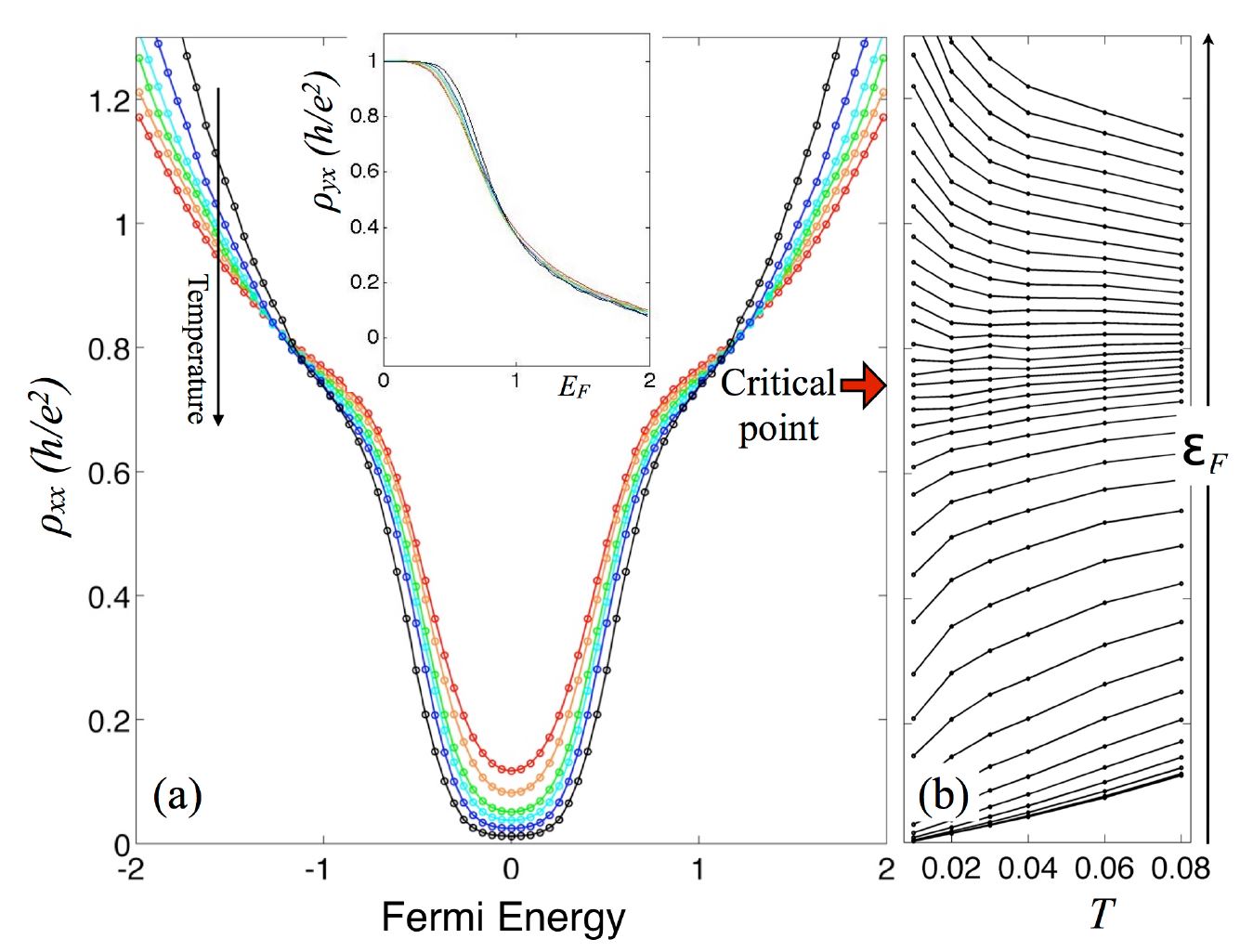}
  \caption{\small (Adapted with permission from \cite{7XP1}) (a) The direct resistivity (and the Hall resistivity in the inset) as function of $\epsilon_F$ at $T=0.01$, 0.02, 0.03, 0.04, 0.06 and 0.08. (b) The direct resistivity as function of temperature for various Fermi energies. The arrow indicates the transition from the Chern to the normal insulator.}
 \label{KMRhoVsEf}
\end{figure}

One may wonder if there are differences between a Chern insulator and an IQHE system. The answer is yes.  First, let us point that the clean limit of the two systems is markedly different: For IQHE systems, the energy spectrum is concentrated in sharp Landau levels while Chern insulators can have wide energy bands separated by spectral gaps. When disorder is introduced, the energy spectrum becomes entirely localized except at the critical points \cite{7Pro0}, but the localized spectra in the two systems can be of very different nature. In IQHE, the whole localized spectrum is generated by impurity states pulled out from the Landau levels into a region of empty spectrum. This type of spectrum is often called a Lifshitz tail. In a Chern insulator, part of the localized spectrum originates from the localization of the already existing energy bands; The rest of the localized spectrum is a  Lifshitz tail composed of impurity states pulled by the disorder into the clean spectral gaps. The density of states for the former is much higher then for the latter. As a result, the diagonal transport coefficients at finite temperature take appreciable values when the Fermi energy is not located in a Lifshitz tail, and this will qualitatively change the flow of $\sigma$ with the temperature, as discussed below.

\vspace{0.1cm}

In \cite{7XP1}, we worked with the spin-up sector of Kane-Mele model \cite{7KM1}, which is relevant to the itinerant electrons in graphene. Unfortunately, the spin-orbit interaction responsible for the emergence of the topological phase is extremely small in these systems, in fact, below the resolution of the present experiments. Nevertheless, when tuned in the middle of the topological phase, the model with disorder takes the form: 
\begin{equation}\label{Eq-KM}
H_\omega=\sum\limits_{\langle {x,y} \rangle} |x \rangle \langle y| +0.6 \I \sum\limits_{\langle \langle x, y \rangle \rangle}  \eta_x \big ( |x \rangle \langle y|- |y\rangle \langle x |\big ) + \lambda \sum_{x} \omega_{x} |x\rangle \langle x| \; .
\end{equation}
Here, $x$ and $y$ label sites of the honeycomb lattice, which contains two sites per unit cell indexed by the isospin number $\eta_x \in \{-1,1\}$ (see \cite{7Pro0} for more details). Furthermore, $\langle  \rangle$ and $\langle \langle \rangle \rangle$ symbolize first and second nearest neighbor relation on the honeycomb lattice, respectively, and $\omega_x$'s are drawn randomly and independently from the interval $[-\frac{1}{2},\frac{1}{2}]$. The strength of the disorder was fixed at $\lambda=4$, precisely twice the size of the clean spectral gap, in order to achieve the strong disorder regime where the spectral gap is closed and only a mobility gap remains. 

\vspace{0.1cm}

The simulated conductivity tensor is reported in Fig.~\ref{KMSigmaVsEf}, where the super-cell's size was $80 \times 80$ primitive cells.  Comparing with the results for the disordered Hofstadter model in Fig.~\ref{SigmaVsEf}, some obvious qualitative differences can be observed. First, the parametrix in the inset of Fig.~\ref{KMSigmaVsEf} is no longer a semi-circle. Secondly, the conductivity tensor flows with the temperature from the outside of the parametrix in the trivial side of the critical point and from inside of the parametrix in the topological side. This is a consequence of the qualitative difference between IQHE systems and Chern insulators mentioned above. Lastly, let us note that the value of $\sigma_{xx}$ saturates now at a value $\sigma_{xx}^c \approx 1$.

\vspace{0.1cm}

\begin{figure}
\center
  \includegraphics[height=6.5cm]{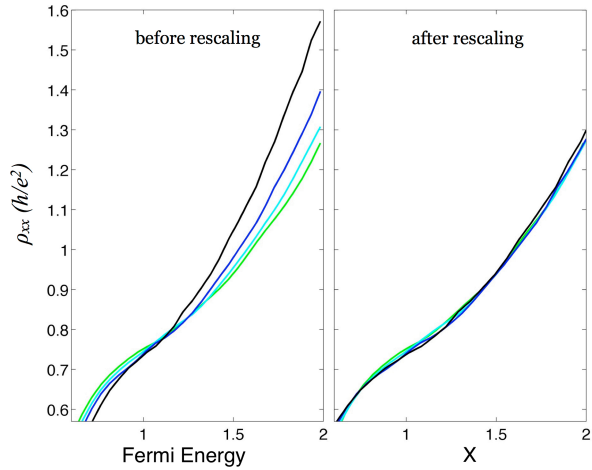}
 \caption{\small (Adapted with permission from \cite{7XP1}) The direct resistivity at different temperatures collapses onto a single curve after the single-parameter rescaling \eqref{Eq-Rescale}, with $\epsilon_F^c=1.017$, $T_0=0.04$ and $\kappa=0.21$.}
 \label{KMRescaling}
\end{figure}

 Fig.~\ref{KMRhoVsEf} reports the direct resistivity $\rho_{xx}$ as function of temperature, for different Fermi levels. The results reveal again the existence of critical points, which can be accurately located from the data, $\epsilon_F^c = \pm 1.017$. The precise value of the conductivity tensor at the critical point is: $\sigma_{xy}^c=-0.53$ and $\sigma_{xx}^c= 1.09$. Lastly, Fig.~\ref{KMRescaling} shows the scaling analysis, which reveals again that the $\rho_{xx}$-curves collapse almost perfectly on top of each other after the energy axis is rescaled as in \eqref{Eq-Rescale} ($T_0=0.04$). The best overlap of the rescaled curves is obtained for $\kappa=0.21 \pm 0.01$, a value that is in good agreement with $k=0.194 \pm 0.002$ obtained from the expression $\kappa=p/(2\nu)$ with the presently accepted value $\nu=2.58 \pm 0.03$, and $p=1$ like in our simulations.

\vspace{0.1cm}

Let us conclude the Section by mentioning that the experiments reported in \cite{7CYT} found a critical behavior quite similar to the IQHE systems, hence different from what we reported here. This indicates that the Chern insulators studied experimentally in \cite{7CYT} have very narrow energy bands, in which case is has been shown explicitly \cite{7PLS} that the critical behavior is similar to the one seen in IQHE systems. In \cite{7CZL}, however, the reported critical behavior is more similar to our simulations, indicating that those experimental samples have wider energy bands.

\chapter{Error Bounds for Non-Smooth Correlations}
\label{Cha-NSCorrelations}

\abstract{In this Chapter we analyze correlation functions of the type:
\begin{equation}\label{Eq-Correlation2}
\Tt( \partial^{\alpha_1} G_1(h) \, \partial^{\alpha_2} G_2(h) \ldots) \; , \quad h \in \Aa_d \; ,
\end{equation} 
where $G_i$'s are only piecewise continuous and the discontinuities occur inside the essential spectrum of $h$. We establish that, if the discontinuities occur inside the mobility gaps of $h$, then the canonical finite-volume algorithm presented in Chapter~\ref{Cha-CanAlg} continues to display a rapid convergence to the thermodynamic limit. Key to the error estimates is the Aizenman-Molchanov bound \cite{8AM,8AW}, which is discussed and adapted to the present context.}

\section{The Aizenman-Molchanov Bound} 

Here we adapt the analysis from \cite{8Aiz} to the case of periodic boundary conditions. We recall that \cite{8Aiz} dealt only with Dirichlet boundary conditions. Our goal here is two-fold: 1) To exemplify for the reader a simple derivation of an Aizenman-Molchanov bound and 2) to verify that indeed the bound holds uniformly with the system size, when the Dirichlet boundary condition is replaced by the periodic one.

\vspace{0.1cm}

We begin by stating a set of assumptions for this Section (only):

\begin{enumerate}[{\rm b}1.]

\item The Hamiltonian $h \in \Aa_d$ is of finite range:
\begin{equation}\label{Eq-HAssumption2}
h=\sum_{|x| \leq \Rr} h_x u_x \;, \quad \Rr < \infty \; .
\end{equation}

\vspace{0.1cm}

\item The fluctuations in the Fourier coefficients are small. That is, there exists $\bar \omega \in \Omega$ invariant under translations, $\tau_x \bar \omega= \bar \omega$ for all $x \in \ZM^d$, such that
\begin{equation} 
\sup_{\omega \in \Omega} \| \delta h_x(\omega) \| \; , \quad \delta h_x(\omega) =h_x(\omega) - h_x(\bar \omega) \; , \quad x \in \ZM^d \; , 
\end{equation}
are small (to be specified later how small). 

\end{enumerate}

In the following, we consider the reference Hamiltonian $h_0 \in \Aa_d$ whose Fourier coefficients are given by $h_{0,x}(\omega) = h_x(\bar \omega)$.

\begin{proposition}\label{Pro-H0ExpDecay} Recall the epimorphism $\tilde{\mathfrak p}$ from Proposition~\ref{Pro-OmegaPer} and the epimorphism $\hat{\mathfrak p}$ from Proposition~\ref{Lem-PerFinite}. We define $\hat h_0 \in \hat A_d$ as $\hat h_0 = (\hat{\mathfrak p} \circ \tilde{\mathfrak p})(h_0)$.  Let $z \in \CM$ and $\delta = {\rm dist}\big ({\rm Re}\, z,\sigma(h_0) \big)>0$. Then the resolvent of $\hat h_0$ satisfies: 
\begin{equation}\label{Eq-H0Exp}
\|(\hat h_0-z)^{-1}_x \| \leq A(\delta) e^{-\gamma(\delta)|x|}\;, \quad 0< A(\delta), \, \gamma(\delta) < \infty \;.
\end{equation}
As the notation suggests, the parameters on the righthand side depend only on $\delta$, hence \eqref{Eq-H0Exp} holds uniformly in $L$. Furthermore, $\gamma(\delta)$ has a sub-linear behavior for $\delta$ small.
\end{proposition} 

\proof From Proposition~\ref{Pro-PerFiniteF}, we have:
\begin{equation}
(\hat h_0-z)^{-1}_x = \sum_{y \in \ZM^d} (h_0 -z)^{-1}_{x + y(2L+1)} \;.
\end{equation}
The Fourier coefficients on the right are known to obey \cite{8Pro4}:
\begin{equation}
\|(h_0-z)^{-1}_x \| \leq B_\delta e^{-\gamma(\delta)|x|} \; , \quad \forall x \in \ZM^d \; ,
\end{equation}
where the parameters are as in the statement. Then, for $x \in V_L$:
\begin{align}
\|(\hat h_0-z)^{-1}_x\| & \leq B_{\delta}\sum_{y \in \ZM^d} e^{-\gamma(\delta)|x + y(2L+1)|} \; .
\end{align}
Now note that $|x + y(2L+1)|\geq |x|$ and $
|x + y(2L+1)|\geq L|y|$ hold for all $y\in \ZM^d$ if $x\in V_L$, hence:
\begin{align}
\sum_{y \in \ZM^d} e^{-\gamma(\delta)|x+y(2L+1)|} & \leq e^{-(1-\xi)\gamma(\delta)|x| }\sum_{y \in \ZM^d} e^{-\xi \gamma(\delta) L|y|} \; , \quad \forall \ \xi \in (0,1) \; ,
\end{align}
and the statement follows if we choose $A_\delta = 2 B_\delta$, since the remaining sum is uniformly bounded in $L$. In fact sum is continuous of $L$ and:
\begin{equation}
\sum_{y \in \ZM^d} e^{-\xi \gamma(\delta) L|y|} \rightarrow 1 \; , \quad \xi >0\;,
\end{equation}
as $L \rightarrow \infty$. Regarding the sub-linear behavior of $\gamma(d)$, see \cite{8BCH} and \cite{8PGP}. \qed

\begin{proposition}\label{Pro-AMbound} Assume b1 and b2 and define $\hat h \in \hat \Aa_d$ as $\hat h = (\hat{\mathfrak p} \circ \tilde{\mathfrak p})(h)$. Let $z \in \CM$ with $\delta = {\rm dist}\big ({\rm Re}\, z, \sigma(h_0) \big )>0$ fixed and $|{\rm Im} \, z | >0$. Then the Aizenman-Molchanov bound holds uniformly in $L$ and ${\rm Im}\, z$:
\begin{equation}\label{Eq-AM7}
\int_{\widehat \Omega} {\rm d} \widehat \PM(\hat \omega) \, \|( \hat h - z)_x^{-1}(\hat \omega) \|^s \leq A_s e^{-\gamma_s |x|} \; , \quad s \in (0,1) \; ,   
\end{equation}
for some strictly positive and finite constants $A_s$ and $\gamma_s$. 
\end{proposition}

\proof The resolvent identity and \eqref{Eq-FiniteAlgProd} give:
\begin{align}\label{Eq-RezId}
 (\hat h -z)_x^{-1}(\hat \omega) = (\hat h_0 -z)_x^{-1} 
+ \sum_{^{\ \ x_i \in V_L}_{s(\sum \hat x_i)=x}} e^{\I \phi}(\hat h_0 -z)_{x_1}^{-1} \,  \delta \hat h_{x_2}(\hat \omega_2) \, (\hat h -z)_{x_3}^{-1}(\hat \omega_3) \; ,
\end{align}
where $\hat \omega_i$'s are cyclic shifts of $\hat \omega$, $e^{\I \phi}$ is the appropriate phase factor and $\delta \hat h = \hat h_0 - \hat h$. Let us introduce the notation:
\begin{equation}
\overline{\delta h}_x = \sup_{\omega \in \Omega} \|\delta h_x(\omega)\| \;,
\end{equation}
and note that $\| \delta \hat h_x(\hat \omega) \| \leq \overline{\delta h}_x$ for all $\hat \omega \in \widehat \Omega$. Taking the norm of \eqref{Eq-RezId} and then raising the result to the power $s$, and using the fact that $(\sum \alpha_i)^s \leq \sum \alpha_i^s$ for any $\alpha_i >0$ and $s \in (0,1)$, gives:
\begin{align}
& \big \| (\hat h -z)_x^{-1}(\hat \omega) \big \|^s \leq \big \|(\hat h_0 -z)_x^{-1} \big \|^s \\
\nonumber + \sum_{^{\ \ x_i \in V_L}_{s(\sum \hat x_i)=x}} & \big \| (\hat h_0 -z)_{x_1}^{-1} \big \|^s \,  \big (\overline{\delta h}_{x_2} \big )^s \, \big \| (\hat h -z)_{x_3}^{-1}(\hat \omega_3) \big \|^s \; .
\end{align}
Taking the average over $\hat \omega$ and using the invariance of the measure w.r.t. translations:
\begin{align}\label{Eq-Inter5}
& \int_{\widehat \Omega} {\rm d} \widehat \PM(\hat \omega) \, \big \|(\hat h -z)_x^{-1}(\hat \omega) \big \|^s \leq \| (\hat h_0 -z)_x^{-1}\|^s  \\
\nonumber + \sum_{^{x_i \in V_L}_{s(\sum \hat x_i)=x}} & \big \| (\hat h_0 -z)_{x_1}^{-1} \big \|^s \,  \big (\overline{\delta h}_{x_2}\big )^s \, \int_{\widehat \Omega} {\rm d} \widehat \PM(\hat \omega) \, \big \| (\hat h -z)_{x_3}^{-1}(\hat \omega)\big \|^s \; .
\end{align} 
Next, we multiply \eqref{Eq-Inter5} by $e^{\xi \gamma_s(\delta) |x|}$, with $\xi \in (0,1)$, $\gamma_s(\delta)=s\gamma(\delta)$ and $\gamma(\delta)$ from Proposition~\ref{Pro-H0ExpDecay} Then we sum over $x$ and use the fact that $|x| \leq |x_1|+|x_2|+|x_3|$ to properly distribute the exponential factor.  Note that the summations over $x_i$ on the right-hand side can now be decoupled. At this point is useful to introduce the notation:
\begin{equation}
Q_0(\xi) = \sum_{x \in V_L} \big \| (\hat h_0 -z)_{x}^{-1} \big \|^s \, e^{\xi \gamma_s(\delta) |x|} \; ,
\end{equation}
and
\begin{equation}
Q(\xi) = \sum_{x \in V_L}\int_{\widehat \Omega} {\rm d} \widehat \PM(\hat \omega) \, \big \| (\hat h -z)_{x}^{-1}(\hat \omega) \big \|^s \, e^{\xi \gamma_s(\delta) |x|} \; ,
\end{equation}
as well as:
\begin{equation}
\overline{\delta h}(\xi) = \sum_{x \in V_L} \big ( \overline{\delta h}_x \big )^s \, e^{\xi \gamma_s(\delta) |x|} \; .
\end{equation}
Then:
\begin{equation}\label{Eq-CentralAM1}
Q(\xi) \leq Q_0(\xi) + \overline{\delta h}(\xi) Q_0(\xi) Q(\xi) \; ,
\end{equation}
which implies:
\begin{equation}
\quad Q(\xi) \leq \frac{Q_0(\xi)}{1- \overline{\delta h}(\xi) Q_0(\xi)} \; ,
\end{equation}
provided that:
\begin{equation}\label{Eq-CentralAM2}
\overline{\delta h}(\xi) Q_0(\xi) <1 \; .
\end{equation}
From Proposition~\ref{Pro-H0ExpDecay}, we know that $Q_0(\xi)$ is uniformly bounded w.r.t. $L$ and ${\rm Im}\, z$, and, due to the finite-range assumption b1, $\overline{\delta h}(\xi)$ becomes independent of $L$ when the latter becomes large. Therefore, \eqref{Eq-CentralAM2} holds if $\overline{\delta h}(\xi)$ is small enough. The latter can now be formulated more precise. Indeed, since both $\overline{\delta h}(\xi)$ and $Q_0(\xi)$ are continuous and decreasing functions of $\xi$, if we require:
\begin{equation}\label{Eq-SmallFlcut}
\overline{\delta h}(0) Q_0(0) <1 \; \Leftrightarrow \Big ( \sum_{x \in V_L} \big \| (\hat h_0 -z)_{x}^{-1} \big \|^s \Big ) \Big ( \sum_{x \in V_L} \big ( \overline{\delta h}_x \big )^s \Big ) <1 \; ,
\end{equation}
then we can be sure that condition \eqref{Eq-CentralAM2} is satisfied if we take $\xi$ small enough.
The conclusion is that, with assumptions b1 and \eqref{Eq-SmallFlcut} in place, we can always find $\xi>0$ such that $Q(\xi)$ becomes uniformly bounded w.r.t. $L$ and ${\rm Im}\, z$, and the statement follows. \qed 

\vspace{0.1cm}

In the original work \cite{8Aiz}, Aizenman arrives at \eqref{Eq-CentralAM1} by using the so called decoupling inequalities. In the present context, this will only give a more optimized expression for $\overline{\delta h}(\xi)$. Now, with this optimization or not, one can determine that \eqref{Eq-SmallFlcut}, hence \eqref{Eq-AM7} too, hold true even when ${\rm Re} \, z$ is located inside the spectrum of $h$. This can be seen by examining the behavior of the spectral edges w.r.t. the strength of the fluctuations, as it was done in the original work \cite{8Aiz}. The fact that $\gamma$ in Proposition~\ref{Pro-H0ExpDecay} depends sub-linearly on ${\rm dist}\big ({\rm Re}\, z, \sigma(h_0) \big )$ is important for this argument \cite{8BCH}. Lastly, let us recall that the physical implications of the Aizenman-Molchanov bound was already discussed in Section~\ref{Sec-DisRegimes}.

\section{Assumptions} Here we state a set of sufficient assumptions which ensure a fast convergence of the canonical finite-volume algorithm presented in Chapter~\ref{Cha-CanAlg}, when applied to computing correlation involving piecewise continuous functions. 

\begin{enumerate}[{\rm c}1.]

\item The Hamiltonian $h \in \Aa_d$ is of finite range and takes the linearized form from Example~\ref{Ex-LinDM}.

\vspace{0.2cm}

\item The Hamiltonian has a mobility gap $\Delta$ as introduced in Definition~\ref{Def-LocElement}.

\vspace{0.2cm}

\item The Aizenman-Molchanov bound for $\epsilon_F \in \Delta$ holds uniformly in $L$ for the finite-volume approximations. 

\end{enumerate}

\begin{remark}{Proposition~\ref{Pro-AMbound} shows that c3 holds with the simplifying assumptions b1 and b2 but, of course, c3 can hold in much more generic conditions. Hence, the assumption b2 will be removed from now on. 
\hfill $\diamond$
}
\end{remark}

\section{Derivation of Error Bounds}

\begin{theorem}\label{Th-ZeroTErrors} Let $h \in \Aa_d$ be a Hamiltonian in the infinite-volume setting and assume c1-c3 from above. Recall the epimorphism $\tilde{\mathfrak p}$ from Proposition~\ref{Pro-OmegaPer} and the epimorphism $\hat{\mathfrak p}$ from Proposition~\ref{Lem-PerFinite} and define $\hat h \in \hat \Aa_d$ as $\hat h = (\hat{\mathfrak p} \circ \tilde{\mathfrak p})(h)$. Then, for any $K \in \NM$, there exists the finite positive constant $A_K$ such that:
\begin{equation}\label{Eq-ErrorBound4}
\Big | \Tt \big (\partial^{\alpha_1} G_1(h) \ldots \partial^{\alpha_n}G_n(h) \big )- \widehat \Tt \big (\hat \partial^{\alpha_1} G_1(\hat h) \ldots \hat \partial^{\alpha_n}G_n(\hat h) \big ) \Big | \leq \frac{A_K}{(1+ |V_L|)^K} \; ,
\end{equation}
where $G_i$'s are Borel functions that are smooth away from the mobility gaps of $h$.
\end{theorem}

\proof We recall Corollary~\ref{Cor-BoreFC}, which assures us that the argument of $\Tt$ in \eqref{Eq-ErrorBound4} belongs to $\bar \Aa_d^\infty$. We appeal again to the ``derivation" from \eqref{Eq-FakeDeriv} and notice that:
\begin{equation}
\hat \partial^{\alpha_1} G_1(\hat h) \ldots \hat \partial^{\alpha_n}G_n(\hat h) = (\hat{\mathfrak p} \circ \tilde{\mathfrak p}) \big (\hat \partial^{\alpha_1} G_1(h) \ldots \hat \partial^{\alpha_n}G_n(h) \big ) \; .
\end{equation}
Furthermore, note that $\hat \partial^{\alpha_1} G_1(h) \ldots \hat \partial^{\alpha_n}G_n(h) \in \bar \Aa_d^\infty$ and the Fourier coefficients of $\partial^{\alpha_j} G_j(h)$ and $\hat \partial^{\alpha_j} G_j(h)$ coincide over $V_L$. Hence the error committed when replacing the argument of $\Tt$ in \eqref{Eq-ErrorBound4} by $\hat \partial^{\alpha_1} G_1(h) \ldots \hat \partial^{\alpha_n}G_n(h)$ can be bounded by terms as in the righthand side of  \eqref{Eq-ErrorBound4}. 

\vspace{0.1cm}

Now, let $a \in L^\infty(\Aa_d,\Tt_0)$ and $\hat a = (\hat{\mathfrak p} \circ \tilde{\mathfrak p})(a)$. Then, from the definition of the measure ${\rm d}\widehat \PM$:
\begin{equation}
\widehat \Tt(\hat a) = \widehat \Tt_0 (\hat a_0) = \int_\Omega {\rm d} \PM(\omega) \, \hat a_0 \big ( (\hat{\mathfrak q} \circ \tilde{\mathfrak q})\omega \big ) \; .
\end{equation}
If we simplify the notation and use $\hat \omega$ for $(\hat{\mathfrak q} \circ \tilde{\mathfrak q})\omega \in \widehat \Omega$, then:
\begin{align}
\Tt(a) - \widehat \Tt(\hat a) = \int_\Omega {\rm d} \PM(\omega) \, \big ( a_0(\omega) - \hat a_0 ( \hat\omega ) \big ) \; .
\end{align}
The Fourier coefficients can be compared using the representations on the physical spaces and, by treating $\ell^2(V_L)$ as a subspace of $\ell^2(\ZM^d)$, we arrive at:
\begin{equation}
\Tt(a) - \widehat \Tt(\hat a) = \int_\Omega {\rm d} \PM(\omega) \, \langle 0 | \pi_\omega (a) - \hat \pi_{\hat \omega}(\hat a) | 0 \rangle \; .
\end{equation}
We now introduce the notations:
\begin{equation}
a^{(k)} = \hat \partial^{\alpha_1} G_1(h) \ldots \hat \partial^{\alpha_{k-1}} G_{k-1}(h) \in L^\infty(\Aa_d, \Tt_0)\; , \quad k>1 \; ,
\end{equation}
and $a^{(k)} =1$ if $k=1$, as well as:
\begin{equation} 
\hat a^{(k)} = \hat \partial^{\alpha_{k+1}} G_{k+1}(\hat h) \ldots \hat \partial_n G_n(\hat h) \; \in L^\infty(\widehat \Aa_d, \widehat \Tt_0) \; , \quad k <n \;,
\end{equation}
and $\hat a^{(k)} =1$ if $k =n$. Then the following expansion holds:
\begin{align}
 & \big \langle 0 \big | \pi_\omega  \Big ( \prod_{j=1}^n \hat \partial^{\alpha_j} G_j(h) \Big ) -\hat \pi_{\hat \omega} \Big ( \prod_{j=1}^n \hat \partial^{\alpha_j} G_j(\hat h)  \Big ) \big | 0 \big \rangle = \sum_{k=1}^n \sum_{y_1 \in \ZM^d} \sum_{y_2 \in V_L}  \\
 \quad \langle 0 |& \pi_\omega  \big (a^{(k)} \big )|y_1 \rangle
\langle y_1 |\pi_\omega\big (\hat \partial^{\alpha_k} G_k(h)\big ) - \hat \pi_{\hat \omega}\big (\hat \partial^{\alpha_k} G_k(\hat h)\big ) |y_2\rangle
\langle y_2 | \hat \pi_{\hat \omega} \big ( \hat a^{(k)} \big ) | 0 \rangle \;. \nonumber
\end{align}
Using the explicit formulas \eqref{Eq-Rep1} and \eqref{Eq-Rep3} for the representations and by applying the classical Holder inequality:
\begin{align}
  \Big | \int_\Omega {\rm d} & \PM(\omega) \big \langle 0 \big | \pi_\omega \Big ( \prod_{j=1}^n \hat \partial^{\alpha_j} G_j(h) \Big ) -\hat \pi_{\hat \omega} \Big ( \prod_{j=1}^n \hat \partial^{\alpha_j} G_j(\hat h)  \Big ) \big | 0 \big \rangle \Big | \\
\nonumber & \leq \sum_{y_1 \in \ZM^d} \sum_{y_2 \in V_L} \sum_{k=1}^n \Big [ \int_\Omega {\rm d} \PM(\omega) \big |a^{(k)}_{-y_1}(\omega)\big |^{3} \Big ]^\frac{1}{3}  \, \Big [ \int_{\widehat \Omega} {\rm d} \widehat \PM(\hat \omega) \big | \hat a^{(k)}_{y_2} (\hat \omega) \big |^3 \Big ]^\frac{1}{3} \\
\nonumber & \qquad \times \Big [ \int_\Omega {\rm d} \PM(\omega) \big |\langle y_1 |\pi_\omega\big (\hat \partial^{\alpha_k} G_k(h)\big ) - \hat \pi_{\hat \omega}\big (\hat \partial^{\alpha_k} G_k(\hat h)\big ) |y_2\rangle \big |^3 \Big ]^\frac{1}{3} \; .
\end{align}
From Corollary~\ref{Cor-BoreFC}, it follows that $a^{(k)} \in \bar \Aa_d^\infty$, hence for any $K \in \NM$:
\begin{equation}
\Big [ \int_\Omega {\rm d} \PM(\omega) \big |a^{(k)}_{-y_1}(\omega)\big |^{3} \Big ]^\frac{1}{3} \leq \frac{B_K}{(1+|y_1|)^{2K}}\; .
\end{equation}
Since the Aizenman-Molchanov bound holds uniformly w.r.t. $L$, a similar bound holds for the second term:
\begin{equation}
\Big [ \int_{\widehat \Omega} {\rm d} \widehat \PM(\hat \omega) \big | \hat a^{(k)}_{y_2} (\hat \omega) \big |^3 \Big ]^\frac{1}{3} \leq \frac{B_K}{(1+|y_2|)^{2K}}\; .
\end{equation}
It remains to resolve the last term. We have:
\begin{align}
& \langle y_1 |\pi_\omega\big (\hat \partial^{\alpha_k} G_k(h)\big ) - \hat \pi_{\hat \omega}\big (\hat \partial^{\alpha_k} G_k(\hat h)\big ) |y_2\rangle \\
\nonumber = (- \I)^{|\alpha_k|} s &(\widehat{y_1 - y_2})^{\alpha_k} \langle y_1 |\pi_\omega\big ( G_k(h)\big ) - \hat \pi_{\hat \omega}\big ( G_k(\hat h)\big ) |y_2\rangle \; .
\end{align}
In the following we specialize for the case when $G_i$'s are step functions. Then:
\begin{align}
& \qquad \qquad \langle y_1 |\pi_\omega\big ( G_k(h)\big ) - \hat \pi_{\hat \omega}\big ( G_k(\hat h)\big ) |y_2\rangle \\
\nonumber & = \frac{\I}{2 \pi} \int_\Cc {\rm d} z \,  \langle y_1 | \pi_\omega(h-z)^{-1} - \hat \pi_{\hat \omega}(\hat h-z)^{-1} |y_2\rangle\; .
\end{align}
The integrand can be bounded by $2 |{\rm Im} \, z|^{-1}$, or it can be expressed via the resolvent identity. Using a combination of both:
\begin{align}
& \Big | \frac{\I}{2 \pi} \int_\Cc {\rm d} z \, \langle y_1 | \pi_\omega(h-z)^{-1} - \hat \pi_{\hat \omega}(\hat h-z)^{-1} |y_2\rangle \Big | \\
\nonumber & \quad \leq \frac{1}{2^{1-s}\pi} \int_\Cc \frac{|{\rm d} z|}{|{\rm Im} \, z|^{1-s}} \sum_{y_3,y_4 \in \ZM^d}  \big | \langle y_1 | \pi_\omega(h-z)^{-1}| y_3 \rangle \big |^s \\
\nonumber & \qquad \times \big | \langle y_3 | \pi_\omega(h) - \hat \pi_{\hat \omega}(\hat h) | y_4 \rangle \big |^s \, \big | \langle y_4 | \hat \pi_{\hat \omega}(\hat h-z)^{-1} | y_2 \rangle \big |^s \;.
\end{align}
Now, recall \eqref{Eq-Bloch1}. Since the Hamiltonian is of finite range, if we take $L \gg \Rr$, we have $\hat h_y(\hat \omega) = h_y(\omega)$ for $|y| \leq \Rr$ and $\hat h_y(\hat \omega) = 0$ in rest. Furthermore, $(\hat \tau_x \hat \omega)_{\hat 0} = (\tau_x \omega)_0$ if $x \in V_L$, hence, from assumption c1, $\hat h_y(\hat \tau_x \hat \omega) = h_y(\tau_x \omega)$ if $x \in V_L$. Then:
\begin{align}
\pi_\omega(h) - \hat \pi_{\hat \omega}(\hat h)  = \sum_{|y|\leq \Rr} & \Big ( \sum_{x \in \ZM^d} \chi(\{x,x-y\} \not\subset V_L) \, h_y(\tau_x \omega) \otimes |x \rangle \langle x | U_y \\
\nonumber & - \sum_{x \in V_L} \chi(x-y \notin V_L) \, \hat h_y(\hat \tau_x \hat \omega) \otimes |x \rangle \langle x | \hat U_y \Big )
 \; .
\end{align}
The important implication of the above is that $\langle x |\pi_\omega(h) - \hat \pi_{\hat \omega}(\hat h)|x'\rangle$ is identically zero if $x,x' \in V'_L$, where:
\begin{equation}
V'_L = \{ x \in V_L \, | \, x-y\in V_L, \ \forall y \in \ZM^d \ \mbox{with} \  |x-y| \leq \Rr \} \; .
\end{equation} 
Furthermore, for the rest of the cases, $\langle x |\pi_\omega(h) - \hat \pi_{\hat \omega}(\hat h)|x'\rangle$ can be easily seen to be uniformly bound. We conclude:
\begin{align}
& \Big | \frac{\I}{2 \pi} \int_\Cc {\rm d} z \, \langle y_1 | \pi_\omega(h-z)^{-1} - \hat \pi_{\hat \omega}(\hat h-z)^{-1} |y_2\rangle \Big |^3 \\
\nonumber & \quad \leq {\rm ct.} \int_\Cc \frac{|{\rm d} z_1|}{|{\rm Im} \, z_1|^{1-s}} \int_\Cc \frac{|{\rm d} z_2|}{|{\rm Im} \, z_2|^{1-s}} \int_\Cc \frac{|{\rm d} z_3|}{|{\rm Im} \, z_3 |^{1-s}}   \\
\nonumber & \sum_{y_3^\alpha,y_4^\alpha \in \ZM^d \setminus V'_L} \prod_{\alpha =1}^3\big | \langle y_1 |\pi_\omega(h-z_\alpha)^{-1}| y_3^\alpha \rangle \big |^s \, \big | \langle y_4^\alpha | \hat \pi_{\hat \omega}(\hat h-z_\alpha)^{-1} | y_2 \rangle \big |^s \; ,
\end{align}
therefore, if we apply Holder's inequality again:
\begin{align}
& \int_\Omega {\rm d} \PM(\omega)\big | \langle y_1 |\pi_\omega\big ( G_k(h)\big ) - \hat \pi_{\hat \omega}\big ( G_k(\hat h)\big ) |y_2\rangle \big |^3 \\
\nonumber & \quad \leq {\rm ct.} \int_\Cc \frac{{\rm d} z_1}{|{\rm Im} \, z_1|^{1-s}} \int_\Cc \frac{{\rm d} z_2}{|{\rm Im} \, z_2|^{1-s}} \int_\Cc \frac{{\rm d} z_3}{|{\rm Im} \, z_3 |^{1-s}}   \\
\nonumber & \sum_{y_3^\alpha,y_4^\alpha \in \ZM^d \setminus V'_L} \ \prod_{\alpha =1}^3 \Big [ \int_\Omega {\rm d} \PM(\omega) \big | \langle y_1 |\pi_\omega(h-z_\alpha)^{-1}| y_3^\alpha \rangle \big |^{6s}\Big ]^\frac{1}{6} \\
& \qquad \qquad \qquad \times \int_\Omega {\rm d} \PM(\omega) \big | 
\nonumber \langle y_4^\alpha | \hat \pi_{\hat \omega}(\hat h-z_\alpha)^{-1} | y_2 \rangle \big |^{6s} \Big ]^\frac{1}{6}\; .
\end{align}
By taking $s < \frac{1}{6}$, we can apply the Aizenman-Molchanov bound and, by noticing that singularities in the contour integrals are integrable, we obtain:
\begin{align}
& \int_\Omega {\rm d} \PM(\omega)\big | \langle y_1 |\pi_\omega\big ( G_k(h)\big ) - \hat \pi_{\hat \omega}\big ( G_k(\hat h)\big ) |y_2\rangle \big |^3 \\
\nonumber & \quad \leq {\rm ct.}  \sum_{y_3^\alpha,y_4^\alpha \in \ZM^d \setminus V'_L}  e^{-\beta\sum_{\alpha =1}^3(|y_1 - y_3^\alpha| + |y_2 - y_4^\alpha|)}\; ,
\end{align}
with $\beta$ strictly positive. At the end:
\begin{align}
 & \Big | \int_\Omega {\rm d}  \PM(\omega) \big \langle 0 \big | \pi_\omega \Big ( \prod_{j=1}^n \hat \partial^{\alpha_j} G_j(h) \Big ) -\hat \pi_{\hat \omega} \Big ( \prod_{j=1}^n \hat \partial^{\alpha_j} G_j(\hat h)  \Big ) \big | 0 \big \rangle \Big | \\
 \nonumber & \leq {\rm ct.} \sum_{y_1 \in \ZM^d} \sum_{y_2 \in V_L} \sum_{y_3^\alpha,y_4^\alpha \in \ZM^d \setminus V'_L} \big | s (\widehat{y_1 - y_2})^{\alpha_k} \big | \frac{e^{-\beta\sum_{\alpha =1}^3(|y_1 - y_3^\alpha| + |y_2 - y_4^\alpha|)}}{(1+|y_1|)^{2K} (1+ |y_2|)^{2K}}  \\
 \nonumber & \leq {\rm ct.} \sum_{y_1 \in \ZM^d} \sum_{y_2 \in V_L} \sum_{y_3^\alpha,y_4^\alpha \in \ZM^d \setminus V'_L}  \frac{e^{-\beta\sum_{\alpha =1}^3(|y_1 - y_3^\alpha| + |y_2 - y_4^\alpha|)}}{(1+|y_1|)^{2K'} (1+ |y_2|)^{2K'}} \; ,
\end{align}
with $2K' < 2K - \max|\alpha_k|$ (here we take $K$ large enough). By exploiting the behavior of the summand over various domains, the remaining sum can be bounded as follows:
\begin{align} 
&  \leq {\rm ct.} \, e^{-3\frac{\beta}{2}|\frac{L}{2}-\Rr|}\sum_{y_1,y_2 \in \ZM^d}  \sum_{y_3^\alpha,y_4^\alpha \in \ZM^d} \chi(y_1 \in V_\frac{L}{2})   \frac{e^{-\beta\sum_{\alpha =1}^3(\frac{1}{2}|y_1 - y_3^\alpha| + |y_2 - y_4^\alpha|)}}{(1+|y_1|)^{2K'} (1+ |y_2|)^{2K'}} \\
\nonumber & + \frac{{\rm ct.}}{(1+\frac{L}{2})^{K'}}\sum_{y_1,y_2 \in \ZM^d}  \sum_{y_3^\alpha,y_4^\alpha \in \ZM^d} \chi(y_1 \notin V_\frac{L}{2})   \frac{e^{-\beta\sum_{\alpha =1}^3(|y_1 - y_3^\alpha| + |y_2 - y_4^\alpha|)}}{(1+|y_1|)^{K'} (1+ |y_2|)^{2K'}} \\
\nonumber & \leq \frac{{\rm ct.}}{(1+L)^{K'}}\sum_{y_1,y_2 \in \ZM^d}  \sum_{y_3^\alpha,y_4^\alpha \in \ZM^d} \frac{e^{-\beta\sum_{\alpha =1}^3(\frac{1}{2}|y_1 - y_3^\alpha| + |y_2 - y_4^\alpha|)}}{(1+|y_1|)^{K'} (1+ |y_2|)^{2K'}} \; ,
\end{align}
and the remaining sum is evidently finite if we take $K'$ large enough. \qed

\chapter{Applications II: Topological Invariants}
\label{Cha-TopInv}

\abstract{This Chapter reports computer assisted calculations of the topological invariants for strongly disordered models of topological insulators from class A and AIII in the classification table~\ref{Table1}. The computations are based on the canonical finite-volume algorithm presented in Chapter~\ref{Cha-CanAlg}. The expressions of the topological invariants involve functional calculus with discontinuous step-wise functions, hence their numerical simulations are covered by the analysis in Chapter~\ref{Cha-NSCorrelations}, provided the discontinuities occur inside the Anderson localized spectrum. The performance and the accuracy of the algorithms will be quantified and analyzed in various ways. 
}

\section{Class AIII in $d=1$} We introduce here a disordered model whose phase diagram can be computed analytically. As we shall see, this phase diagram can be reproduced with high fidelity by simply mapping the topological invariant ${\rm Ch}_1(U_F)$ introduced in Section~\ref{Sec-TopInv}. The analysis and the numerical simulations are reproduced from \cite{9MSHP}.

\vspace{0.1cm}

The model is defined on the Hilbert space $\CM^2 \otimes \ell^2(\ZM)$ and takes the form: 
\begin{align}
\label{Eq-Hamdis}
& H_\omega
\;=\; 
\tfrac{1}{2}\sum_{x\in\ZM}
\Big [ m_x \,\sigma_2 \otimes |x\rangle\langle x| \\ 
\nonumber + \, t_x\big (  (\sigma_1+\I \sigma_2) & \otimes |x\rangle\langle x+1| \;+\; (\sigma_1-\I \sigma_2)\otimes |x+1\rangle\langle x|
\big)\; \Big ]
\;,
\end{align}
where $\sigma_i$'s represent Pauli's matrices. The fluctuating parameters are defined as
\begin{equation}
t_x = 1+W_1 \, \omega_x \;, \quad m_x =  m+W_2 \, \omega'_x \; ,
\end{equation}
where $\omega_x,\omega'_x\in[-\frac{1}{2},\frac{1}{2}]$ are independent and uniformly distributed random variables. As such, the disorder configuration space is the Tychonov space $\Omega=\big ([-\frac{1}{2},\frac{1}{2}]\times[-\frac{1}{2},\frac{1}{2}]\big )^{\ZM}$ and the probability measure on $\Omega$ is just the product measure:
\begin{equation}
{\rm d} \PM(\omega) = \prod_{x \in \ZM} {\rm d} \omega_x \, {\rm d} \omega'_x \; .
\end{equation}
Using the standard commutation relations of the Pauli's matrices, one can see immediately that the model does have the chiral symmetry:
\begin{equation}
\sigma_3 \, H_\omega \, \sigma_3 = - H_\omega \;, \quad \forall \omega \in \Omega \;.
\end{equation}
The parameter space of the model is 3-dimensional and defined by $(m,W_1,W_2)$.

\vspace{0.1cm}

For $\omega=0$ or $W_1 = W_2=0$, the Hamiltonian $H_\omega$ is similar, but not identical, with the Hamiltonian from Example~\eqref{Ex-ChiralModel}. In this clean limit, the Bloch Hamiltonian and the Fermi unitary operator take the form:
\begin{equation}
H_0(k) = \begin{pmatrix} 0 & e^{-\I k} +\I m \\ e^{\I k} - \I m & 0 \end{pmatrix} \; , \quad U_F(k) = \frac{e^{\I k}-\I m}{|e^{\I k}-\I m|} \;.
\end{equation}
Then:
\begin{equation}\label{KWindingNr}
{\rm Ch}_1(U_F) = {\rm Winding}\Big (\frac{e^{\I k}-\I m}{|e^{\I k}-\I m|} \Big )= \left \{
\begin{array}{l}
1 \ \mathrm{if} \ m\in (-1,1) \;, \\
0 \ \mathrm{otherwise} \;.
\end{array}
\right .
\end{equation}
Let us point out that the bulk energy-gap closes precisely at $m=\pm 1$, where the topological phase transitions occur. 

\begin{figure}
\center
  \includegraphics[width=0.6\textwidth]{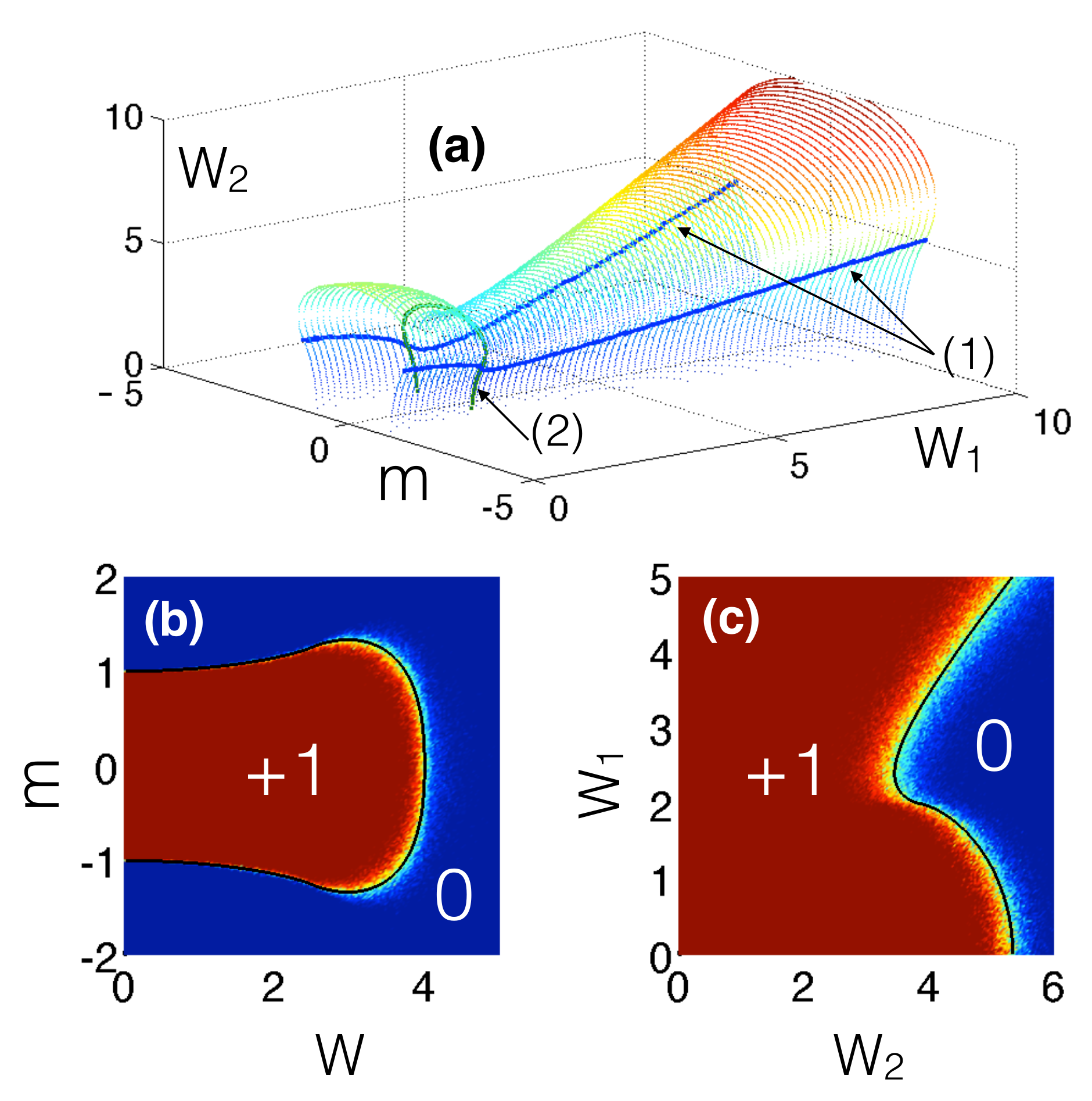}\\
  \caption{\small (Adapted with permission from \cite{9MSHP}) (a) The critical surface, where $\Lambda \rightarrow \infty$, plotted in the 3-dimensional phase space $(m,W_1,W_2)$ as derived from \eqref{Eq-LocLength1}. The lines (1) and (2) represent the singular points where the scaling is anomalous (see text). The next panels report maps of ${\rm Ch}_1(U_F)$, for two sections of the phase space defined by the constraints $W_2=2W_1=W$ (b) and $m=0.5$ (c). The analytic critical curves are shown as black lines in panels (b,c). The computations of ${\rm Ch}_1(U_F)$ were done on a lattice with $1000$ unit cells and were averaged over 10 disorder configurations.}
  \label{CriticalSurface}
\end{figure} 

\begin{figure}
\center
\includegraphics[width=0.55\textwidth]{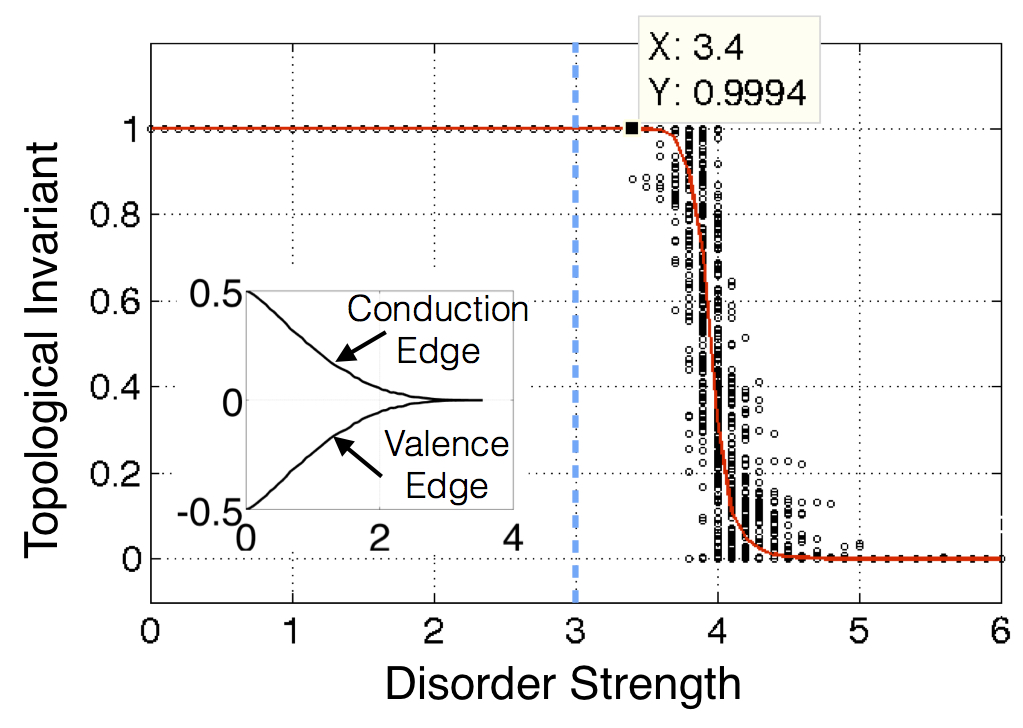}
\caption{\small (Adapted with permission from \cite{9MSHP}) Evolution of ${\rm Ch}_1(U_F)$ with disorder strength $W$, where $W_1=0.5W$ and $W_2=W$. The raw, un-averaged data for 200 disorder configurations is shown by the scattered points and the average by the solid line. Inset: the conduction and valence edges as functions of W, indicating a spectral gap closing at $W \approx 3$ (marked by the dashed line in the main panel). The marked data point reports a quantized $\nu=0.9994$, at disorder well beyond $W=3$.}\label{AIII1DInvariantVsW}
\end{figure}

 In the presence of disorder, the Schr\"odinger equation $H_\omega\psi = \epsilon \psi$ associated to \eqref{Eq-Hamdis} can be solved explicitly for $\epsilon = 0$. This is of course a special point of the energy axis but it is highly relevant for the systems from AIII class because the Fermi level is pinned at $\epsilon=0$. Now, the Schr\"{o}dinger equation reads:
\begin{equation}
\begin{pmatrix} 0 & t_x \\ 0 & 0 \end{pmatrix} \psi_{x+1} 
\;+\; 
\begin{pmatrix} 0 & 0 \\ t_x & 0 \end{pmatrix} \psi_{x-1} 
\;+\; \I \begin{pmatrix} 0 &  - m_x \\ m_x & 0 \end{pmatrix}  \psi_{x} 
\;=\; 0
\;,
\end{equation}
or, on the components:
\begin{equation}
t_x\psi_{x-\alpha,\alpha}+\I \alpha m_x \psi_{x,\alpha} =0\; , \quad \alpha = \pm 1 \; .
\end{equation}
The solution is:
\begin{equation}
\psi_{\xi_\alpha+x,\alpha}
\;=\;
\I^x \prod_{j=1}^{x} \left( \frac{t_j}{m_j}\right ) ^\alpha \psi_{\xi_\alpha,\alpha}
\;,
\end{equation}
where $\xi_\alpha = 0,1$ for $\alpha=\pm 1$, respectively. The inverse of the Anderson localization length is given by:
\begin{equation}
\Lambda^{-1} 
\;=\;
\max_{\alpha=\pm 1}\Big [- \lim\limits_{x\rightarrow \infty} \frac{1}{x}\log|\psi_{\xi_\alpha+x,\alpha}| \Big ] 
\;=\;
\Big|\lim\limits_{x\rightarrow \infty} \frac{1}{x}\sum_{j=1}^{x}(\ln |t_j|-\ln|m_j|)\Big|
\;.
\end{equation}
Using Birkhoff's ergodic theorem \cite{9Bir} on the last expression:  
\begin{equation}
\Lambda^{-1} 
\;=\;
\left | \ \int_{-1/2}^{1/2} {\rm d} \omega \int_{-1/2}^{1/2} {\rm d} \omega' \ \big (\ln|1+W_1 \, \omega| - \ln|m+W_2 \, \omega'| \big ) \right |
\;.
\end{equation}
The integrations can be performed explicitly and the result is:
\begin{equation}\label{Eq-LocLength1}
\Lambda^{-1}
\;=\;
\left | \ln \left [\frac{|2+W_1|^{\frac{1}{W_1}+\frac{1}{2}}}{|2-W_1|^{\frac{1}{W_1}-\frac{1}{2}}} \  \frac{| 2m-W_2|^{\frac{m}{W_2}-\frac{1}{2}}}{|2m+W_2|^{\frac{m}{W_2}+\frac{1}{2}}}  \right ] \right |
\;.
\end{equation}

\vspace{0.1cm}

A plot of the critical manifold where Anderson's localization length diverges, $\Lambda \rightarrow \infty$ is presented in Fig.~\ref{CriticalSurface}(a), where one can see two insulating phases fully separated by this critical manifold. In other words, the only way to cross from one insulating phase to the other is to go through an Anderson localization-delocalization quantum transition. The computation provides additional useful information, namely, the critical exponent at the crossing of the phase boundary which turns to be $\nu = 1$ for the entire critical boundary, except at the points marked by the lines (1) and (2) in Fig.~\ref{CriticalSurface}(a), where the exponent is anomalous (see \cite{9MSHP} for more details).  

\vspace{0.1cm}

Another interesting aspect is the behavior of the energy spectrum w.r.t. the disorder strength. Recall that the analysis of the previous paragraph applies strictly at $\epsilon = 0$, but the rest of the spectrum can be investigated by standard numerical means. The conclusion in \cite{9MSHP} was that the entire energy spectrum becomes Anderson localized as soon as the disorder is turned on. The spectrum remains localized upon increasing the disorder strength and, precisely when the phase boundary is reached, the spectrum at $\epsilon=0$ and only at $\epsilon=0$ becomes delocalized. This behavior is quite different from the one observed in the integer quantum Hall systems, where parts of the spectrum remain delocalized when the disorder is turned on.

\begin{figure}
\center
\includegraphics[width=0.9\textwidth]{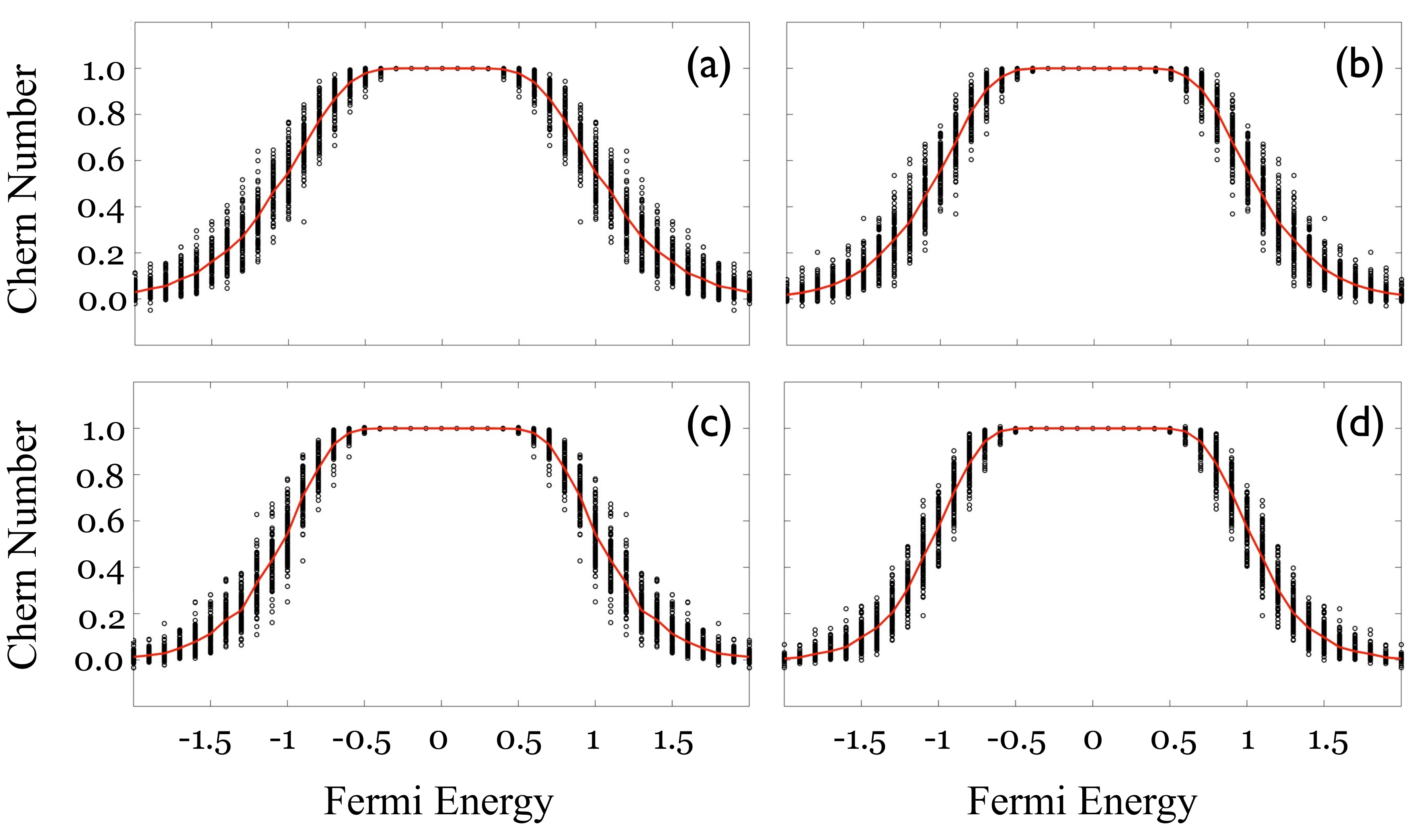}
\caption{\small Evolution of ${\rm Ch}_2(P_F)$ with the Fermi energy for model \eqref{Eq-KM}, as computed with the canonical algorithm on increasing lattice sizes of (a) $40 \times 40$, (b) $60 \times 60$, (c) $80 \times 80$ and (d) $100 \times 100$. The raw, un-averaged data for 100 disorder configurations is shown by the scattered points and the average by the solid line. The disorder strength is $\lambda=4$, large enough to completely close the spectral gap of the Hamiltonian.}
\label{Chern1}
\end{figure}

\vspace{0.1cm}
 In panels (b) and (c) of Fig.~\ref{CriticalSurface}, we report the maps of ${\rm Ch}_1(U_F)$ for two different sections of the phase space, as computed with the canonical finite-volume algorithm described in Chapter~\ref{Cha-CanAlg}. As one can see, the topological phase with ${\rm Ch}_1(U_F) =1$ extends all the way to the analytically computed critical boundary, in perfect agreement with the prediction of Theorem~\ref{Th-OddChernIndex}. Fig.~\ref{AIII1DInvariantVsW} zooms into a 1-dimensional section of the phase diagram, so that the evolution of the topological invariant can be observed in more detail. The inset of this figure reports the evolution of the spectral gap which disappears once the disorder strength exceeds $W=3$. The topological invariant, however, remains quantized to many digits of precision well beyond this value. 

\section{Class A in $d=2$} 

The Chern insulators were already introduced in Section~\ref{Sec-ChernTI} and here we continue the analysis of the spin-up sector of the Kane-Mele model with disorder introduced in \eqref{Eq-KM}, this time at zero temperature.  

\begin{figure}
\center
\includegraphics[width=0.8\textwidth]{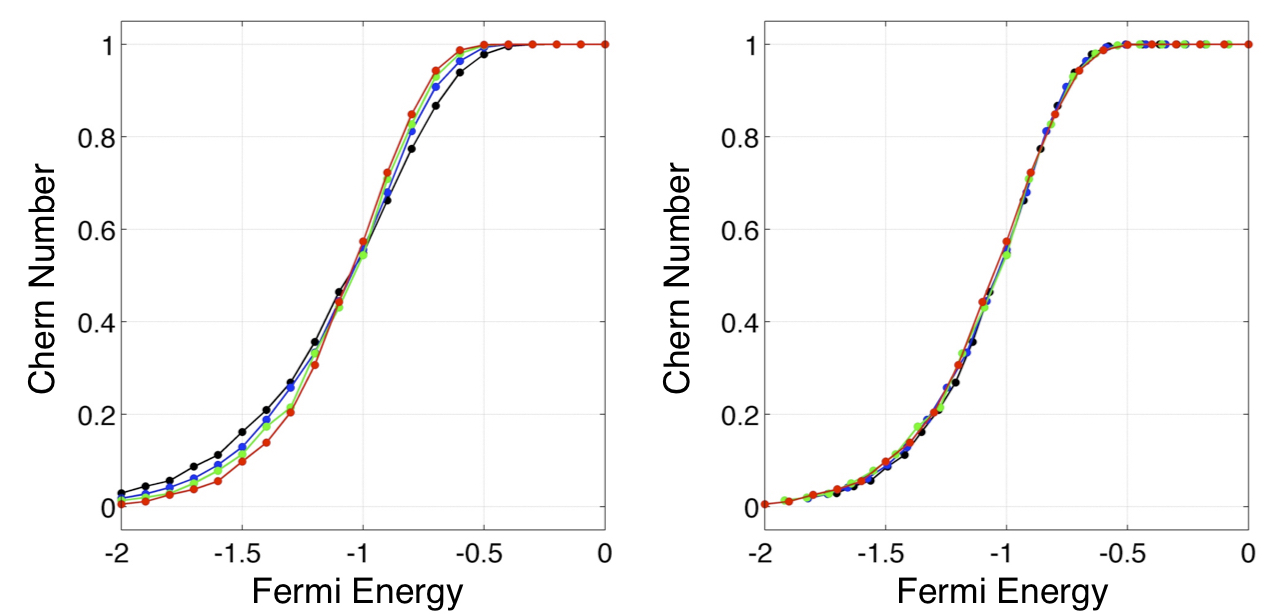}
\caption{\small (Left) Overlap of the averaged Chen number values from Fig.~\ref{Chern1}. (Right) Re-plot of the data after the horizontal axis was re-scaled as in \eqref{Eq-ReS}, with $\epsilon_F^c =-1.00$ and $\nu = 2.6$. }\label{Chern2}
\end{figure}

\vspace{0.1cm}
 The phase diagram of the model in the plane of $(\epsilon_F, \lambda)$ was computed in Fig.~6 of \cite{9Pro0}, using the classical method based on the level statistics. In the following, we consider the situation from panel (b) of this Fig.~6, corresponding to $\lambda=4$ (same as in Section~\ref{Sec-ChernTI}), where the spectral gap is closed. In Fig.~\ref{Chern1} we report the numerical computation of ${\rm Ch}_2(P_F)$ as function of $\epsilon_F$, obtained with the canonical finite-volume algorithm on lattices of increasing sizes. This figure shows disorder averaged and non-averaged data and a good quantization of the invariant at $+1$ is observed for both cases, at least for Fermi energies close to the origin. In fact, the energy interval where this quantization is observed is in good agreement with the mobility gap extracted from Fig.~6 of \cite{9Pro0}. On the opposite side, the topological invariant drops to zero and in between we expect a topological phase transition. According to Theorem~\ref{Th-EvenChernIndex}, this should be accompanied by an Anderson transition. Its signatures are the fluctuations w.r.t disorder in the data, which start to develop once the Anderson localization length becomes comparable to or larger than the size of the computational super-cell. These fluctuations persists when the size of the super-cell is increased, which provides the evidence of the existence of a critical point $\epsilon_F^c$ where the Anderson localization length diverges. Of course, we already know this from Section~\ref{Sec-ChernTI}.

\begin{table}
\scriptsize
\caption{The numerical values of averaged Chern numbers, for increasing lattice sizes.}
\begin{center}
\begin{tabular}{|c|c|c|c|c|}
\hline
       Fermi Energy & $40 \times 40$ & $60 \times 60$ & $80 \times 80$ & $100 \times 100$  \\
\hline             
      -2.0000000000000000   &    0.0293885304649968    &   0.0183147848896676   &    0.0134785966919230  &     0.0055726403061233 \\
      -1.8999999999999999    &   0.0442301583027775    &   0.0274502505545331    &   0.0200229343621875    &   0.0112501012411246 \\
      -1.8000000000000000   &    0.0563736772645283   &    0.0416811880195335   &    0.0285382576963500   &    0.0259995275657507 \\
      -1.7000000000000000    &   0.0868202901241971    &   0.0612803850743208  &     0.0506852078002088   &    0.0377798251819264 \\ 
      -1.6000000000000001    &   0.1121154018269069    &   0.0905166860071905    &   0.0781754600177580   &    0.0554182299457663 \\
      -1.5000000000000000   &    0.1617580454580226   &    0.1291516191502659    &   0.1133966598848624    &   0.0977984662347778 \\
      -1.3999999999999999    &   0.2093536896403097    &   0.1883311262238442    &   0.1733092018533850    &   0.1386844139850113 \\
      -1.3000000000000000   &    0.2687556358733589   &    0.2575144956897765   &    0.2146703753513447    &   0.2040079233029510 \\
      -1.2000000000000000   &    0.3565352143319771   &    0.3333569482253110    &   0.3319133571108642    &   0.3066419928551302 \\
      -1.1000000000000001   &    0.4646789224167249    &   0.4444784219466996   &    0.4310440221933989  &     0.4427699238861748 \\
      -1.0000000000000000   &    0.5479958396159215    &   0.5561471440680733   &    0.5442615536532044    &   0.5738596277941682 \\
      -0.9000000000000000   &    0.6624275864985472   &    0.6798953821199148   &    0.7086514094234754   &    0.7228749266484203 \\
      -0.8000000000000000   &    0.7742005453064691   &    0.8124137607528051   &    0.8270271100278364   &    0.8487923693232788 \\
      -0.7000000000000000    &   0.8672349391630054   &    0.9079791895178040   &    0.9301639459675241   &    0.9432234611493278 \\
      -0.6000000000000000   &    0.9392873717233425    &   0.9636994770114942   &    0.9802652381114992    &   0.9872940741308633 \\
      -0.5000000000000000   &    0.9784417158133359     &  0.9935074963179980   &    0.9974987656403326   &    0.9988846769813913 \\
      -0.4000000000000000   &    0.9958865415757685     &  0.9992024708366942   &    0.9998527876642247   &    0.9999656328302596 \\
      -0.3000000000000000    &   0.9998184404341747     &  0.9999824660477071    &   0.9999988087144891   &    0.9999996457562911 \\
      -0.2000000000000000    &   0.9999952917010211    &   0.9999977443008894    &   0.9999999997655000    &   0.9999999999862120 \\
      -0.1000000000000000    &   0.9999999046002306    &   0.9999999998972079   &    0.9999999999998473   &    0.9999999999999849 \\
       0.0000000000000000    &   0.9999999963422543    &   0.9999999999988873    &   0.9999999999999996    &   0.9999999999999999 \\
       \hline
       \end{tabular}
\end{center}
\label{Table2}
\end{table}

\vspace{0.1cm}
 To demonstrate the existence of such critical point and to detect its precise location, hence to locate the phase boundary using ${\rm Ch}_2(P_F)$ alone, we overlap in Fig.~\ref{Chern2} the disorder averaged Chern numbers computed on lattices of increasing sizes. In the left panel, one can see the curves intersecting at virtually one point and, in the right panel, the curves can be seen to overlap almost perfectly when the horizontal axis is rescaled as: 
\begin{equation}\label{Eq-ReS}
\epsilon_F \rightarrow \epsilon_F^c + (\epsilon_F - \epsilon_F^c) \Big ( \frac{L}{L_0} \Big )^\frac{1}{\nu} \; ,
\end{equation} 
with $\epsilon_F^c =-1.00 \pm 0.01$ and $\nu = 2.6 \pm 0.1$, in good agreement with the findings in Section~\ref{Sec-ChernTI}. We can conclude with great confidence that the system displays the single-parameter paradigm, which is indeed the hallmark signature of the Anderson transition. Note that the scaling exponent $\nu$ derived from our calculations is in good agreement with the latest numerical estimates \cite{9SO1,9KMO,9OSF,9FHA,9DET,9AMS,9SO2,9OGE}.

\vspace{0.1cm}
To conclude our analysis, let us present a table with the numerical values of the averaged Chern numbers. The data is reported in Table~\ref{Table2} and was compiled from Fig.~\ref{Chern1}. As expected, when the Fermi energy is inside the mobility gap, the quantization is reproduced with many digits of precision, and the number of exact digits rapidly increases with the size of the lattice. Particularly, when the Fermi energy is in the middle of mobility gap, the quantization is reproduce with machine precision for the largest system size. We recall that the Fermi energy is located in the essential spectrum and that the strength of the disorder potential is four times larger than the non-fluctuating coefficients of the Hamiltonian.

\begin{figure}
\center
\includegraphics[width=0.65\textwidth]{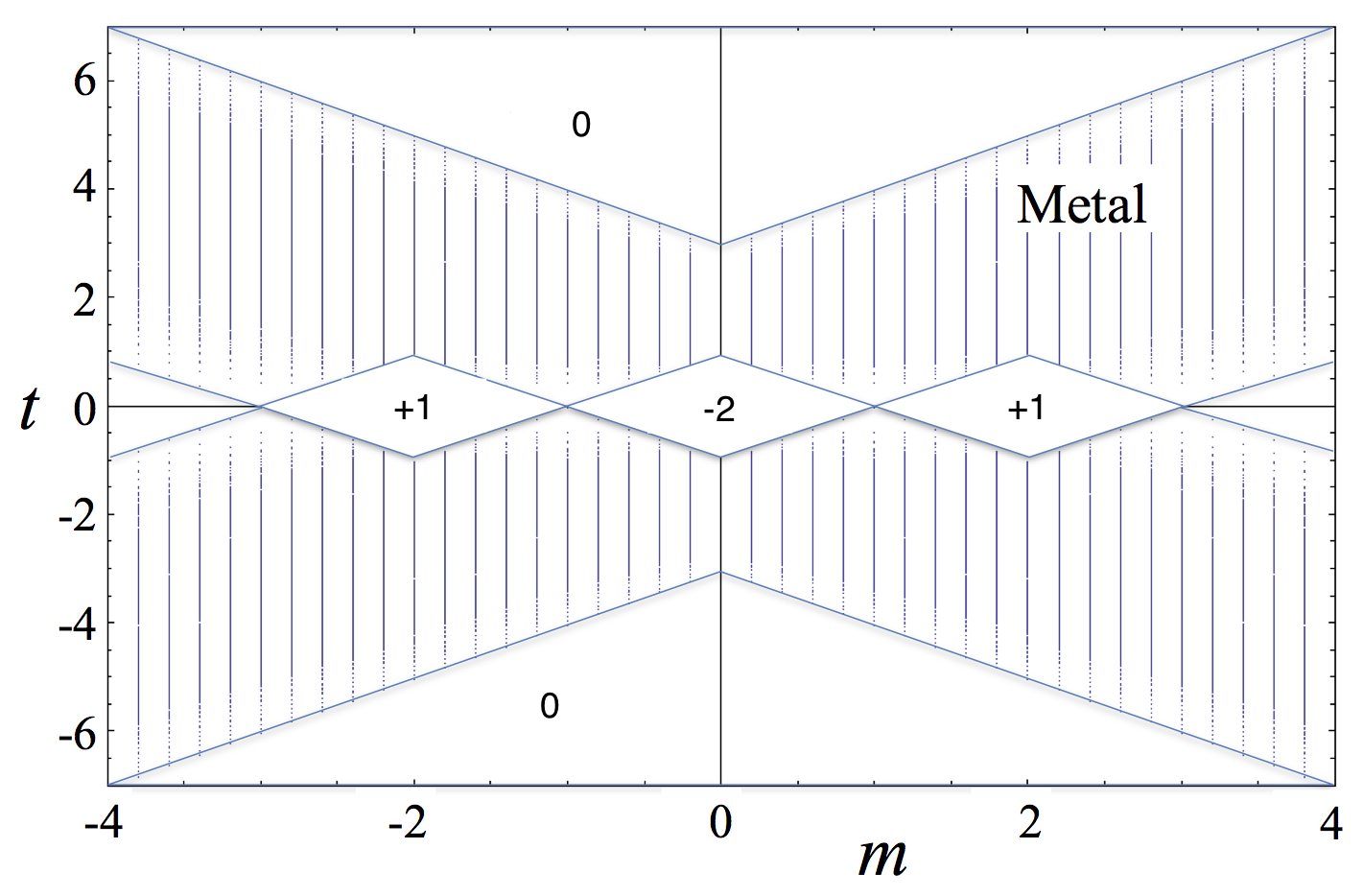}
\caption{\small (Adapted with permission from \cite{9SFP})The phase diagram of the clean model ($\lambda=0$) defined in \eqref{Eq-AIIID3Model}. Here, the reader can identify the topological phases with ${\rm Ch}_3(U_F)=-2$ and ${\rm Ch}_3(U_F)=+1$ (the rhombic domains), a large metallic phase (the shaded region) and the trivial topological phase ${\rm Ch}_3(U_F)=0$. }
\label{PhaseDiagramClean2}
\end{figure}

\section{Class AIII in $d=3$} Here we reproduce the simulations from \cite{9SFP} of the strong topological invariant ${\rm Ch}_3(U_F)$ for the following chiral-symmetric model defined on the Hilbert space $\CM^4 \otimes \ell^2(\ZM^3)$:
\begin{align}\label{Eq-AIIID3Model}
H_\omega \;  =\; 
\tfrac{1}{2\I} \sum_{j=1}^3 \gamma_j & \otimes (S_j-S_j^\ast) 
\;+\; 
\gamma_{4} \otimes \Big(m+\tfrac{1}{2}\sum_{j=1}^3  (S_j+S_j^\ast)\Big) \\
& + \I t \gamma_1 \gamma_3 \gamma_4 + \lambda \sum_{x \in \ZM^3} \omega_x |x\rangle \langle x|
\;,
\end{align} 
where $\{\gamma_j\}_{j=\overline{1,4}}$ provide an irreducible representation of the complex Clifford algebra ${\mathcal Cl}_4$ on $\CM^4$, and $\omega_x$ are drawn randomly and independently from the interval $[-\frac{1}{2},\frac{1}{2}]$. Without disorder, the model is almost identical with the on in \eqref{Eq-Model2}, the only difference being the presence of the additional term $\I t \gamma_1 \gamma_3 \gamma_4$. It has been introduced to break the time-reversal and particle-hole symmetries and make the model a true representative of the AIII class. Indeed, if $\gamma_0$ is the standard grading of even Clifford algebra ${\mathcal Cl}_4$, then:
\begin{equation}
\gamma_0 \, H_\omega \, \gamma_0 = - H_\omega
\end{equation}
remains the only symmetry of the Hamiltonian.

\begin{figure}
\center
\includegraphics[width=1\textwidth]{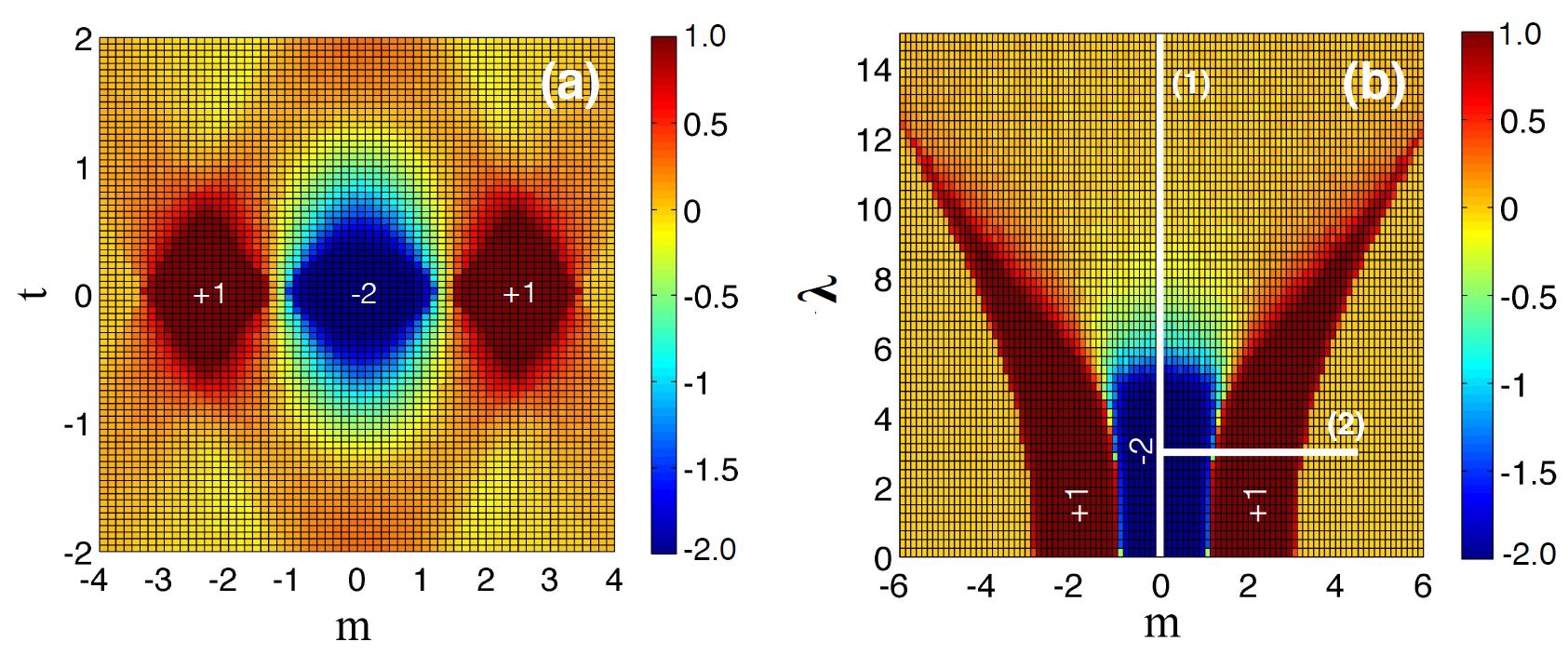}
\caption{\small (Adapted with permission from \cite{9SFP}) The phase diagram of the model \eqref{Eq-AIIID3Model} in (a) the plane $(m,t)$ at disorder strength $\lambda=4$ and (b) in the plane $(m,\lambda)$ at $t=0$. The computations were completed on a supercell containing $16\times 16 \times 16$ primitive cells.}
\label{3DAIIIPhaseDiagramDisorder}
\end{figure}

 The parameter space is 3-dimensional and given by $(m,t,\lambda)$. The phase diagram of the clean model ($\lambda=0$) is shown in Fig.~\ref{PhaseDiagramClean2}, where one can see the existence of topological phases with ${\rm Ch}_3(U_F) = -2$ and $+1$. Note that the phase diagram along the line $t=0$ is fully consistent with the analysis in Example~\ref{Ex-ChiralModel}.

\vspace{0.1cm}

Fig.~\ref{3DAIIIPhaseDiagramDisorder}(a) reports the map of the odd Chern number ${\rm Ch}_3(U_F)$ in the $(m,t)$ plane, computed at fixed disorder strength $\lambda=4$. Here one can see well defined regions where the odd Chern number remains quantized at non-trivial values, hence the topological phases seen in Fig.~\ref{PhaseDiagramClean2} remain clearly visible even at this large value of disorder strength. The boundary of the topological phases moved quite visibly when compared with Fig.~\ref{PhaseDiagramClean2}, with the topological phases actually occupying more volume after the disorder was turned on. Outside the topological regions, the odd Chern number does not drop sharply to zero, indicating the presence of a substantial metallic region where Anderson's localization length is infinite. This, together with the level spacings analysis, led us to conclude in \cite{9SFP} that the metallic phase present in  Fig.~\ref{PhaseDiagramClean2} survives the disorder. This is, of course, not a surprise in space dimension $d=3$. Fig.~\ref{3DAIIIPhaseDiagramDisorder}(b) reports the map of the odd Chern number in the plane $(m,\lambda)$, computed at $t=0$. As one can see, the phase boundaries are strongly affected by the disorder and the topological phases survive up to the extreme disorder strengths of $\lambda= 13$ for ${\rm Ch}_3(U_F)=1$ and $\lambda=6$ for ${\rm Ch}_3(U_F)=-2$. 

\vspace{0.1cm}

\begin{figure}
\center
\includegraphics[width=0.8\textwidth]{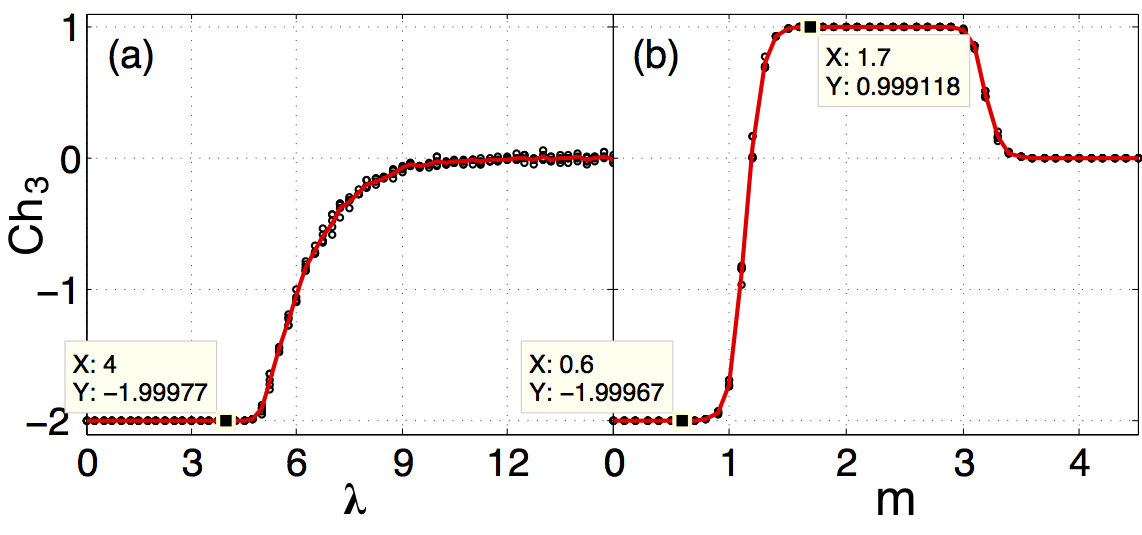}
\caption{\small (Adapted with permission from \cite{9SFP}) Evolution of the odd Chern number ${\rm Ch}_3(U_F)$ with (a) disorder strength $\lambda$  and (b) parameter $m$. The raw, un-averaged data for 5 disorder configurations is shown by the scattered points and the average by the solid line. The marked data points report quantized values with 3 digital precisions. The computations were done with a cubic lattice of $21\times 21 \times 21$ primitive cells.}
\label{WindingNumber}
\end{figure}

In order to illustrate the quality of the data, in Fig.~\ref{WindingNumber} we report the numerical values of ${\rm Ch}_3(U_F)$ along the paths (1) and (2) shown in Fig.~\ref{3DAIIIPhaseDiagramDisorder}(b). In this figure we show the results for five independent random configurations (the scattered markers), as well as the average over these five random configurations (the solid lines). The calculations have been completed on a larger lattice of size $21 \times 21 \times 21$. Several explicit numerical values of the averaged odd Chern numbers are displayed, showing a quantization with 3 digits of precision. The data also show the self-averaging property of the winding number, which can be deduced from the absence of fluctuations in the non-averaged data.

\end{document}